\documentclass[aps,pre,preprint,superscriptaddress]{revtex4-2}
\usepackage{amssymb}
\usepackage{graphicx}
\usepackage{subfigure}
\usepackage{threeparttable}
\usepackage{array}
\usepackage{multirow}
\usepackage{bm}
\usepackage{amsmath}
\usepackage{CJK}

\begin{document}


\title{Simplified Unified Wave-Particle Method with Quantified Model-Competition Mechanism for Numerical Calculation of Multi-Scale Flows}





\author{Sha Liu}
\email{shaliu@nwpu.edu.cn}
\affiliation{National Key Laboratory of Science and Technology on Aerodynamic Design and Research, Northwestern Polytechnical University, Xi'an, Shaanxi 710072, China}
\affiliation{School of Aeronautics, Northwestern Polytechnical University, Xi'an, Shaanxi 710072, China}

\author{Chengwen Zhong}
\email{Corresponding author: zhongcw@nwpu.edu.cn}
\affiliation{National Key Laboratory of Science and Technology on Aerodynamic Design and Research, Northwestern Polytechnical University, Xi'an, Shaanxi 710072, China}
\affiliation{School of Aeronautics, Northwestern Polytechnical University, Xi'an, Shaanxi 710072, China}

\author{Ming Fang}
\email{fangming\_cardc@163.com}
\affiliation{China Aerodynamics Research and Development Center, Mianyang, Sichuan 621000, China}
\date{\today}

\begin{abstract}
A Quantified Model-Competition (QMC) mechanism for multi-scale flows is extracted from the integral (analytical) solution of the Boltzmann-BGK model equation. In the QMC mechanism, the weight of the rarefied model and the weight of the continuum (aerodynamic/hydrodynamic) model are quantified. Then, a Simplified Unified Wave-Particle method (SUWP) is constructed based the on the QMC mechanism. In the SUWP, the stochastic particle method and the continuum Navier-Stokes method are combined together. Their weights are determined by the QMC mechanism quantitatively in every discrete cells of the computational domain. The validity and accuracy of the present numerical method are examined using a series of test cases including the high non-equilibrium shock wave structure case, the unsteady Sod shock-tube case with a wide range of Kn number, the hypersonic flow around the circular cylinder from the free-molecular regime to the near continuum regime, and the viscous boundary layer case. In the construction process of the present method, an anti-dissipation effect in the continuum mechanism is also discussed.
\end{abstract}


\maketitle

\section{Introduction}
For the flows around super/hyper-sonic aircrafts in near space, local rarefied regions often arise in the flow field due to the large gradients in the shock waves and the boundary layers. For micro-flows around Micro-Electro-Mechanical Systems (MEMS), rarefied flows often exist around the boundary of MEMS, since its characteristic length is comparable to the molecular mean free path (m.f.p.). In similar situations, the coexistence of the continuum flow and the rarefied flow in a single flow field makes the flow behavior and mechanism extremely complicated. In the reserches and simulations of these complex multi-scale flows, numerical methods that can cover the entire flow regime (including the continuum regime, slip regime, transitional regime and free-molecular regime) is in strong demand.

Since the Direct Simulation Monte Carlo (DSMC)~\cite{bird2013dsmc, Bird2003Molecular} and the discrete velocity method (DVM)~\cite{kolobov2007unified, wu2014solving} are able to simulate the rarefied flows, and the Navier-Stokes (N-S) solvers are able to simulate the continuum flows, hybrid methods are developed, in which a flow field is decomposed into continuum regions and rarefied regions with the corresponding solvers working on it. For example, the Modular Particle-Continuum (MPC) method~\cite{sun2004a, Schwartzentruber2007A} couples the Information Preservation (IP) DSMC method~\cite{fan2001statistical} with N-S solver, and the Unified Flow Solver (UFS)~\cite{kolobov2007unified} couples the DVM with the Gas-Kinetic Scheme (GKS)~\cite{xu2001a} which can be viewed as a N-S solver with better non-equilibrium performance. The hybrid methods should use empirical or semi-empirical criterion for domain decomposition. In hybrid methods, the continuum region and rarefied region are overlapped for better information exchange~\cite{sun2004a, Schwartzentruber2007A}. The model inaccuracy of N-S equation and insufficient particle number of DSMC in the overlap regions should be addressed well. When DVM method is used as the rarefied solver, the hybrid method should face the curse of dimensionality, which is adjoint to the discrete velocity space used in the DVM.

To enlarge the cell size and time step of particle methods such as the DSMC, the analytical solutions of homogenous Bhatnagar-Gross-Krook (BGK)-type model equations are used to categorize the particles into free-transport particles and particles participated in collisions~\cite{macrossan2001nu, gallis2011investigation}. Then, the velocities of the particles participated in collisions are sampled from the corresponding equilibrium distribution function. Up to now, ES-BGK, Shakhov, and Unified-BGK models are used~\cite{pfeiffer2018particle-based}, and real gas effect is considered~\cite{tumuklu2016particle, Pfeiffer2018Extending}. The homogenous treatment of particle collisions leads to a first order numerical scheme in which the extra numerical viscosity will harm the accuracy and the Asymptotic-Preserving (AP) property in prediction of continuum and near-continuum flows, especially in the boundary layer. In order to overcome this drawback, the Unified Stochastic Particle (USP) method is proposed recently~\cite{Zhang2019Particle, Fei2020A}, in which the extra relaxation terms toward a Grad distribution are added to both side of model equation, and the transport process is coupled with the extra relaxation process (this relaxation process can be viewed as some kind of particle collision), leading to a correct viscosity and AP property. On the other hand, using the Fokker-Planck (FP)-type Boltzmann model equation and the corresponding Langevin-type stochastic differential equation, particle FP method is proposed~\cite{jenny2010a} and further extended to real gas, gas mixture and dense gas~\cite{gorji2011fokker¨Cplanck, Gorji2012A, sadr2017a}. Further, the ES-FP~\cite{mathiaud2016a} and Cubic-FP models~\cite{gorji2011fokker¨Cplanck} are developed to achieve the right Pr number, and modifications are made to obtain the AP property~\cite{fei2017a}.

For simulating flows in entire flow regime, several unified methods are proposed based on the BGK-type model equation using the discrete velocity space in a deterministic way, such as Unified Gas-Kinetic Scheme (UGKS)~\cite{xu2010unified, Li2018A, ChenA}, Discrete Unified Gas-Kinetic Scheme (DUGKS)~\cite{guo2013discrete, wang2015a, zhu2016discrete}, Gas-Kinetic Unified Algorithm (GKUA)~\cite{li2009gas, peng2016implicit}, and the Improved Discrete Velocity Method (IDVM)~\cite{yang2018improved, yang2019an}. The free-transport process and collision process of particles are coupled together in UGKS and DUGKS, using the analytical solution of full BGK-type model equation (not simplified homogenous one), and the characteristic-line function, respectively. The coupled transport process is not only consistent with the physical nature, but also leads to the multi-scale property. Therefore, their cell size and time step are not limited by the mean free path and mean collision time anymore, and can be chosen according to the flow properties (such as the gradients of the flow field) and Courant-Friedrichs-Lewy (CFL) condition, respectively. Up to now, these methods are extended to real gas~\cite{liu2014unified}, plasma gas~\cite{Liu2017A}, phonon heat transfer~\cite{Guo2016Discrete}, radiation transfer~\cite{sun2017a}. Recently, a Unified Gas-Kinetic Wave-Particle (UGKWP) method is proposed based on the same philosophy of the UGKS method~\cite{Liu2018Unified, zhu2019unified}. In the UGKWP, the particles are categorized into free-transport particles and particles participated in collisions (named hydrodynamic particles in UGKWP) using the analytical solution of full BGK-type model equation. The particles participated in collisions are merged into the macroscopic variables, and their contribution to macroscopic flux are calculated from the time integral part of the analytical solution. Both the information of free-transport particles and macroscopic variables are updated in the UGKWP. After the updating, the particles participated in collisions are emerged from the macroscopic variables again. In continuum limit, there is almost no free-transport particles, then the UGKWP is equivalent to a N-S solver without the statistical fluctuation associated with particles methods, and the AP property is fulfilled directly.

In both the analytical solution of BGK-type equation and the time integral solution of the Langevin-type equation, there underlies a model-competition mechanism between the particle free-transport model and the continuum model, which directly leads to the multi-scale properties of the particle FP method, the UGKS method, and the recent UGKWP method. In this paper, enlightened from the construction process of the UGKWP method, a quantified model-competition mechanism is found by conducting a close investigation of the analytical solution of the BGK equation. With this quantified model-competition mechanism, a Simplified Unified Wave-Particle (SUWP) method is proposed, which combines the collisionless DSMC method (as the rarefied model) with the Navier-Stokes solver (as the continuum model). The weights of rarefied model and continuum model are determined from the quantified model-competition mechanism. Moreover, since the SUWP is not strictly based on the BGK-type model equation, it is flexible and can be extended to the gas mixture and chemical reaction easily in the future research.

The remaining of this paper is arranged as follows: Section~\ref{sec:GKT} is a quick review of the gas-kinetic theory and the BGK-type Boltzmann model equation. Section~\ref{sec:MCM} is an investigation of the analytical solution, from which the quantified model-competition mechanism is obtained. The SUWP method is in Section~\ref{sec:SUWP}. Section~\ref{sec:EXP} is the numerical experiments. The concluding remarks are in Section~\ref{sec:DIS}.

\section{Gas-Kinetic Theory and BGK-type model equation}\label{sec:GKT}
In the gas-kinetic theory, molecular motions are described in terms of the distribution function $f(\mathbf{x},\bm{\xi},t)$, which means the number density of molecules with the velocity $\bm{\xi}$ that arrive the location $\mathbf{x}$ at time $t$. For dilute gas, $f$ is governed by Boltzmann equation\cite{Kremer2010An}:
\begin{equation}\label{eq:Boltzmann}
\frac{{\partial f}}{{\partial t}} + {\bm{\xi }} \cdot \frac{{\partial f}}{{\partial {\mathbf{x}}}} + {\mathbf{a}} \cdot \frac{{\partial f}}{{\partial {\bm{\xi }}}} = C\left( {f,f} \right),
\end{equation}
where $\mathbf{a}$ is the acceleration of molecule. The left-hand side of Eq.~\ref{eq:Boltzmann} is the free-transport part, while the right-hand side is the five-fold nonlinear integral collision part. In most multi-scale methods, the BGK-type Boltzmann model equation is used in the following form:
\begin{equation}\label{eq:BGK}
\frac{{\partial f}}{{\partial t}} + {\bm{\xi }} \cdot \frac{{\partial f}}{{\partial {\mathbf{x}}}} + {\mathbf{a}} \cdot \frac{{\partial f}}{{\partial {\bm{\xi }}}} = \frac{\left( {g - f} \right)}{\tau},
\end{equation}
where the Boltzmann collision term in Eq.~\ref{eq:Boltzmann} is replaced by a simple relaxation term on the right-hand side of Eq.~\ref{eq:BGK}. Furthermore, $\tau$ is the relaxation time defined as $\mu/p$, where $\mu$ and $p$ are the temperature-dependent dynamic viscosity and the pressure, respectively. Moreover, $g$ is the Maxwellian distribution with the maximum local entropy, which is in the form below:
\begin{equation}\label{eq:Maxwell}
g = {\left( {\frac{m}{{2\pi kT}}} \right)^{3/2}}\exp \left( { - \frac{{m{\mathbf{c}} \cdot {\mathbf{c}}}}{{2kT}}} \right),
\end{equation}
where $n$, $\mathbf{c}$, $\mathbf{u}$, $T$, $k$ and $m$ are the number density, the peculiar velocity defined as $\bm{\xi}-\mathbf{u}$, the macroscopic velocity, the thermodynamic temperature, the Boltzmann constant, and the mass of molecule, respectively.

In the gas-kinetic theory, the macroscopic mass density $\rho$, momentum density $\rho \mathbf{u}$, energy density $\rho\left|{\bm{u}}\right|^{2}/2 + \rho e$ (here $e$ is inertial energy per unit mass), stress tensor $\bm{s}$ and heat flux $\mathbf{q}$ can be obtained from the distribution function $f$ with the following equations:
\begin{equation}\label{eq:constrain}
\begin{aligned}
&\rho  = \left\langle {mf} \right\rangle \\
&\rho {\mathbf{u}} = \left\langle {m{\bm{\xi }}f} \right\rangle \\
&\frac{1}{2}\rho\left|{\bm{u}}\right|^{2} + \rho e  = \frac{1}{2}\left\langle {m{\bm{\xi }} \cdot {\bm{\xi }}f} \right\rangle \\
&{\mathbf{s}} =  - \left\langle {m{\mathbf{cc}}f} \right\rangle  + p{\mathbf{I}}\\
&{\mathbf{q}}{ = }\frac{1}{2}\left\langle {m{\bm{c }}\left( {{\bm{c }} \cdot {\bm{c }}} \right)f} \right\rangle,
\end{aligned}
\end{equation}
where $\mathbf{I}$ is a indentity matrix, and the operator $\left\langle  \cdot  \right\rangle$ denotes an integral over of the whole velocity space as the following:
\begin{equation}\label{eq:operator}
\left\langle\cdot\right\rangle  = \int_{{R^3}} {\left(  \cdot  \right)} d{\bm{\xi }}.
\end{equation}

\section{Quantified Model-Competition Mechanism}\label{sec:MCM}
The analytical solution of BGK equation is in the following form
\begin{equation}\label{eq:analytical_solution}
f\left( {{\mathbf{x}},{\bm{\xi }},t} \right) = \frac{1}{\tau }\int_{\rm{0}}^t {g\left( {{\mathbf{x}} - {\bm{\xi }}t + {\bm{\xi }}t',{\bm{\xi }},t} \right){e^{\frac{{t' - t}}{\tau }}}dt'} {\rm{ + }}{e^{ - \frac{t}{\tau }}}f\left( {{\mathbf{x}} - {\bm{\xi }}t,{\bm{\xi }},0} \right).
\end{equation}
Here $f\left( {{\mathbf{x}} - {\bm{\xi }}t,{\bm{\xi }},0} \right)$ is the original distribution function at time $0$, and ${\mathbf{x}} - {\bm{\xi }}t$ is the original coordinate obtained by tracing the molecules (with velocity ${\bm{\xi }}$) back from $\mathbf{x}$. ${\mathbf{x}} - {\bm{\xi }}t + {\bm{\xi }}t'$ is the trace of molecules from time $0$ to time $t$, and $g\left( {{\mathbf{x}} - {\bm{\xi }}t + {\bm{\xi }}t',{\bm{\xi }},t} \right)$ is the equilibrium distribution function along this trace.

This analytical solution can be interpreted as:
\begin{enumerate}
  \item A cluster of particles with velocity ${\bm{\xi }}$  located at ${\mathbf{x}} - {\bm{\xi }}t$  at time $0$ transport in their velocity direction. Their initial number density is $f\left( {{\mathbf{x}} - {\bm{\xi }}t,{\bm{\xi }},0} \right)$.
  \item When they arrive at location $\mathbf{x}$ at time $t$, due to the intermolecular collisions, some molecules leave their original trace and do not belong to this cluster anymore. While, some molecules are not affected by intermolecular collisions. They are still in this trace, and their portion is $\exp\left(-t/\tau\right)$.
  \item On the other hand, intermolecular collisions also replenish this cluster with new molecules that emerge from other collisions with post-collision velocity ${\bm{\xi}}$. These post-collision molecules are determined from the equilibrium distribution $g\left( {{\mathbf{x}} - {\bm{\xi }}t + {\bm{\xi }}t',{\bm{\xi }},t} \right)$ along the trace.
\end{enumerate}

The analytical solution (Eq.~\ref{eq:analytical_solution}) explicitly shows that in a time interval $\left(0,t\right)$, there are a $\exp \left( {{{ - t} \mathord{\left/{\vphantom {{ - t} \tau }} \right.\kern-\nulldelimiterspace} \tau }} \right)$ portion of molecules are free-transport ones, and the others should experience at least one collision. As $t$ increases, the portion of free-transport molecules decreases. $t$ is actually the scale-dependent observation time. When $t$ is much larger than $\tau$, such as in the case of continuum regime, there is almost no free-transport molecule left. Since the portion of free-transport molecules is depended on the observation time, the analytical solution has a multi-scale property.

The molecules participated in collision are named hydrodynamic molecules in Ref.~\cite{LiuUnified2020}. This nomenclature is used in this paper. In order to conduct a close investigation of hydrodynamic molecules, a second order Taylor expansion is used for $g\left( {{\mathbf{x}} - {\bm{\xi }}t + {\bm{\xi }}t',{\bm{\xi }},t} \right)$ in the analytical solution (Eq.~\ref{eq:analytical_solution}), which is the equilibrium distribution along the trace. Denote the time integral term in the analytical solution by $h\left( {{\mathbf{x}},{\bm{\xi }},t} \right)$. With the second order Taylor expansion for $g\left( {{\mathbf{x}} - {\bm{\xi }}t + {\bm{\xi }}t',{\bm{\xi }},t} \right)$, it can be written as
\begin{equation}\label{eq:analytical_h1}
h\left( {{\mathbf{x}},{\bm{\xi }},t} \right) = \frac{1}{\tau }\int_{\rm{0}}^t {\left\{ {g\left( {{\mathbf{x}},{\bm{\xi }},0} \right){\rm{ + }}{{\left. {\left( { - {\bm{\xi }}t + {\bm{\xi }}t'} \right) \cdot \frac{{\partial g}}{{\partial {\mathbf{x}}}}} \right|}_{\left( {{\mathbf{x}},{\bm{\xi }},0} \right)}}{\rm{ + }}t{{\left. {\frac{{\partial g}}{{\partial t}}} \right|}_{\left( {{\mathbf{x}},{\bm{\xi }},0} \right)}}} \right\}{e^{\frac{{t' - t}}{\tau }}}dt'}.
\end{equation}
By calculating the integral in Eq.~\ref{eq:analytical_h1}, it becomes
\begin{equation}
h\left( {{\bf{x}},{\bm{\xi }},t} \right) = {\left\{ {\left( {1 - {e^{ - \frac{t}{\tau }}}} \right)g + \left( {{e^{ - \frac{t}{\tau }}}t + {e^{ - \frac{t}{\tau }}}\tau  - \tau } \right){\bm{\xi }} \cdot \frac{{\partial g}}{{\partial {\bf{x}}}} + \left( {t + {e^{ - \frac{t}{\tau }}}\tau  - \tau } \right)\frac{{\partial g}}{{\partial t}}} \right\}_{\left( {{\bf{x}},{\bm{\xi }},0} \right)}}.
\end{equation}
Here, all the information is located at $\left({\bf{x}},{\bm{\xi}},0\right)$. In order to get a clear physical picture, this equation is further rearranged as
\begin{equation}\label{eq:analytical_h2}
h\left( {{\bf{x}},{\bm{\xi }},t} \right) = {\left\{ {\left( {1 - {e^{ - \frac{t}{\tau }}}} \right)\left[ {g - \tau \left( {{\bm{\xi }} \cdot \frac{{\partial g}}{{\partial {\bf{x}}}} + \frac{{\partial g}}{{\partial t}}} \right)} \right] + {e^{ - \frac{t}{\tau }}}t\left( {{\bm{\xi }} \cdot \frac{{\partial g}}{{\partial {\bf{x}}}}{\rm{ + }}\frac{{\partial g}}{{\partial t}}} \right) + \left( {t - {e^{ - \frac{t}{\tau }}}t} \right)\frac{{\partial g}}{{\partial t}}} \right\}_{\left( {{\bf{x}},{\bm{\xi }},0} \right)}}.
\end{equation}
Here, the first term in the curly brackets are actually a distribution (in square brackets) multiplied by a scale factor, and this distribution corresponds to the 2nd order Chapman-Enskog (C-E) expansion of the BGK equation. The second term is an anti-dissipation term. The third term is a high order temporal term.

In the continuum regime, the relaxation time $\tau$ whose magnitude is in the same order with the mean collision time is greatly smaller than the observation time $t$ ($\tau \ll t$ and $t/\tau  \to \infty$). Then, Eq.~\ref{eq:analytical_h2} is reduced to:
\begin{equation}
h\left( {{\bf{x}},{\bm{\xi }},t} \right) = {\left\{ {\left[ g\emph{}{ - \tau \left( {{\bm{\xi }} \cdot \frac{{\partial g}}{{\partial {\bf{x}}}} + \frac{{\partial g}}{{\partial t}}} \right)} \right] + t\frac{{\partial g}}{{\partial t}}} \right\}_{\left( {{\bf{x}},{\bm{\xi }},0} \right)}}.
\end{equation}
Therefore, $h\left( {{\bf{x}},{\bm{\xi }},t} \right)$ becomes the 2nd order C-E distribution plus a high order temporal term.

In the free molecular regime, the relaxation time $\tau$ is greatly larger than the observation time $t$ ($\tau \gg t$ and $t/\tau  \to 0$). Then, $h\left( {{\bf{x}},{\bm{\xi }},t} \right)$ becomes
\begin{equation}
h\left( {{\bf{x}},{\bm{\xi }},t} \right) = {\left\{ {\left[ { - t\left( {{\bm{\xi }} \cdot \frac{{\partial g}}{{\partial {\bf{x}}}} + \frac{{\partial g}}{{\partial t}}} \right)} \right] + t\left( {{\bm{\xi }} \cdot \frac{{\partial g}}{{\partial {\bf{x}}}}{\rm{ + }}\frac{{\partial g}}{{\partial t}}} \right)} \right\}_{\left( {{\bf{x}},{\bm{\xi }},0} \right)}}=0.
\end{equation}
The coefficients of the transport term ($g$) of the C-E distribution and the high order temporal term in Eq.~\ref{eq:analytical_h2} are zero. Since the coefficient of the anti-dissipation term is opposite to that of the dissipation term in C-E distribution, these two terms are canceled, making a physical correct $h\left( {{\bf{x}},{\bm{\xi }},t} \right)=0$. This is consistent with the physical nature that there is no collision (hydrodynamic molecules) in the free molecular regime. Without the anti-dissipation term, a nonphysical dissipation proportional to the observation time $t$ will exist in the flow field permanently.

Combine the dissipation and anti-dissipation term and drop the high order temporal term, $h\left( {{\bf{x}},{\bm{\xi }},t} \right)$ can be finally written as
\begin{equation}\label{eq:analytical_h3}
h\left( {{\bf{x}},{\bm{\xi }},t} \right) = {\left\{ {\left( {1 - {e^{ - \frac{t}{\tau }}}} \right)\left[ {g - {c_{vis}}\tau \left( {{\bm{\xi }} \cdot \frac{{\partial g}}{{\partial {\bf{x}}}} + \frac{{\partial g}}{{\partial t}}} \right)} \right]} \right\}_{\left( {{\bf{x}},{\bm{\xi }},0} \right)}},
\end{equation}
where $c_{vis}$ (defined in Eq.~\ref{eq:analytical_h3a}) is the coefficient of the dissipation term after combined with the anti-dissipation term. The subscript ``vis'' stands for viscous since the dissipation term leads to the viscous flux in the N-S solver in the later analysis.
\begin{equation}\label{eq:analytical_h3a}
{c_{vis}} = 1 - \left( {\frac{t}{\tau }} \right)\frac{{{e^{ - \frac{t}{\tau }}}}}{{1 - {e^{ - \frac{t}{\tau }}}}}.
\end{equation}
The value of $c_{vis}$ is unity in continuum regime and zero in the free molecular regime.

So far, the mechanism can be extracted from the analytical solution of the BGK equation is quite clear. For time scale (observation time) $t$, $\exp\left({-t/\tau}\right)$ portion of molecules are free-transport molecules; $1-\exp\left({-t/\tau}\right)$ portion of molecules participate in collisions and follow a modified C-E expansion in Eq.~\ref{eq:analytical_h3}. The free transport molecules follow the free transport mechanism (rarefied model), and the hydrodynamic molecules follow a modified hydrodynamic mechanism (continuum model) since their distribution is a modified C-E distribution.

The macroscopic flux can be calculated from $\langle \left(\bm{\xi}\cdot\bm{n}\right)\bm{\psi}f \rangle$ (the operator is defined in Eq.~\ref{eq:operator}), where $\bm{\psi }=\left(m,m\bm{\xi},m\bm{\xi}\cdot\bm{\xi}\right/2)$ are the microscopic conservative variables, $\bm{n}$ is the normal direction of a cell interface, $f$ is defined at the central point of the interface. Therefore, set $\bm{x}$ the central point of the interface whose normal direction is $\bm{n}$, the macroscopic flux ${\bf{F}}_{hydro}$ caused by the hydrodynamic molecules is expressed as follows:
\begin{equation}\label{eq:fluxQCM1}
\begin{aligned}
&{{\bf{F}}_{hydro}} = \left\langle {\left( {{\bm{\xi }} \cdot {\bf{n}}} \right){\bf{\psi }}h} \right\rangle \\
&= \left( {1 - {e^{ - \frac{t}{\tau }}}} \right)\left\{ {\left\langle {\left( {{\bm{\xi }} \cdot {\bf{n}}} \right){\bf{\psi }}g} \right\rangle  + {c_{vis}}\left\langle {\left( {{\bm{\xi }} \cdot {\bf{n}}} \right){\bf{\psi }}\tau \left( {{\bm{\xi }} \cdot \frac{{\partial g}}{{\partial {\bf{x}}}} + \frac{{\partial g}}{{\partial t}}} \right)} \right\rangle } \right\}.
\end{aligned}
\end{equation}
Since the two integrals in Eq.~\ref{eq:fluxQCM1} are the inviscid flux and viscous flux of the N-S equation. ${{\bf{F}}_{hydro}}$ can be finally written as
\begin{equation}\label{eq:fluxQCM2}
{{\bf{F}}_{hydro}} = \left( {1 - {e^{ - \frac{t}{\tau }}}} \right)\left( {{{\bf{F}}_{NS,inv}} + {c_{vis}}{{\bf{F}}_{NS,vis}}} \right).
\end{equation}
Here, ``inv'' stands for inviscid. Since $\left( {1 - {\exp\left({ - t/\tau}\right)}} \right)$ is the portion of the hydrodynamic molecules, Eq.~\ref{eq:fluxQCM2} means that the flux caused by hydrodynamic molecules is in the form of N-S flux except a scale dependent coefficient $c_{vis}$ is multiplied to the viscous flux.

For numerical methods, the observation time $t$ is the numerical time step $\Delta t$. The weight of the rarefied model and the weight of the continuum model can be defined as follows, which are actually the proportions of free-transport and hydrodynamic molecules, respectively.
\begin{equation}
\begin{aligned}
&w_{free} = {e^{ - \frac{{\Delta t}}{\tau }}},\\
&w_{hydro} = 1 - {e^{ - \frac{{\Delta t}}{\tau }}}.
\end{aligned}
\end{equation}
The flux caused by the free-transport molecules ${\bf{F}}_{free}$ is directly obtained from their straight line motions. The flux cause by the hydrodynamic molecules ${\bf{F}}_{hydro}$ is in a modified N-S form:
\begin{equation}\label{eq:numericalFQMC}
{{\bf{F}}_{hydro}} = w_{hydro}\left( {{{\bf{F}}_{NS,inv}} + {c_{vis}}{{\bf{F}}_{NS,vis}}} \right),
\end{equation}
where the scale dependent coefficient $c_{vis}$ is multiplied to the viscous flux, and $c_{vis}$ is defined as
\begin{equation}
{c_{vis}} = 1 - \left( {\frac{\Delta t}{\tau }} \right)\frac{w_{free}}{w_{hydro}}.
\end{equation}
Finally, the philosophy of the Quantified Model-Competition (QMC) mechanism can be summarized explicitly as:
\begin{enumerate}
  \item During the time step $\Delta t$, $w_{free}$ portion of molecules are the free-transport ones. $w_{hydro}=1-w_{free}$ portion of molecules are hydrodynamic ones that participate in intermolecular collisions.
  \item The behavior the free-transport molecules is the straight free motion, and their macroscopic behavior is obtain by the direct summation of the molecular information (rarefied model). The hydrodynamic molecules follows the modified C-E expansion (Eq.~\ref{eq:analytical_h3} with $t$ replaced by $\Delta t$), and their macroscopic behavior is governed by a modified N-S mechanism (continuum model, Eq.~\ref{eq:numericalFQMC}).
  \item The weight of rarefied model and the weight of continuum model are the proportions of molecules governed by them, respectively.
\end{enumerate}

\section{Simplified Unified Wave-Particle Method}\label{sec:SUWP}
This section is about the Simplified Unified Wave-Particle Method (SUWP) that use the QMC mechanism for multi-scale flow simulations. Like other flow solvers, the physical space and time in the SUWP are the discrete ones. In discrete physical elements (cells), both the information of the free-transport molecules and the information of the macroscopic variables are recorded and renewed. The SUWP solver is composed by the following three parts:
\begin{enumerate}
  \item stochastic particle solver for free-transport molecules (the collisionless DSMC is used in this paper).
  \item N-S solver for the macroscopic behavior of hydrodynamic molecules.
  \item Functions for QMC mechanism (including information exchange between the two solvers).
\end{enumerate}

\subsection{Functions for QMC mechanism: categorization and supplementation of molecules}
\begin{figure}
\centering
\includegraphics[width=0.95\textwidth]{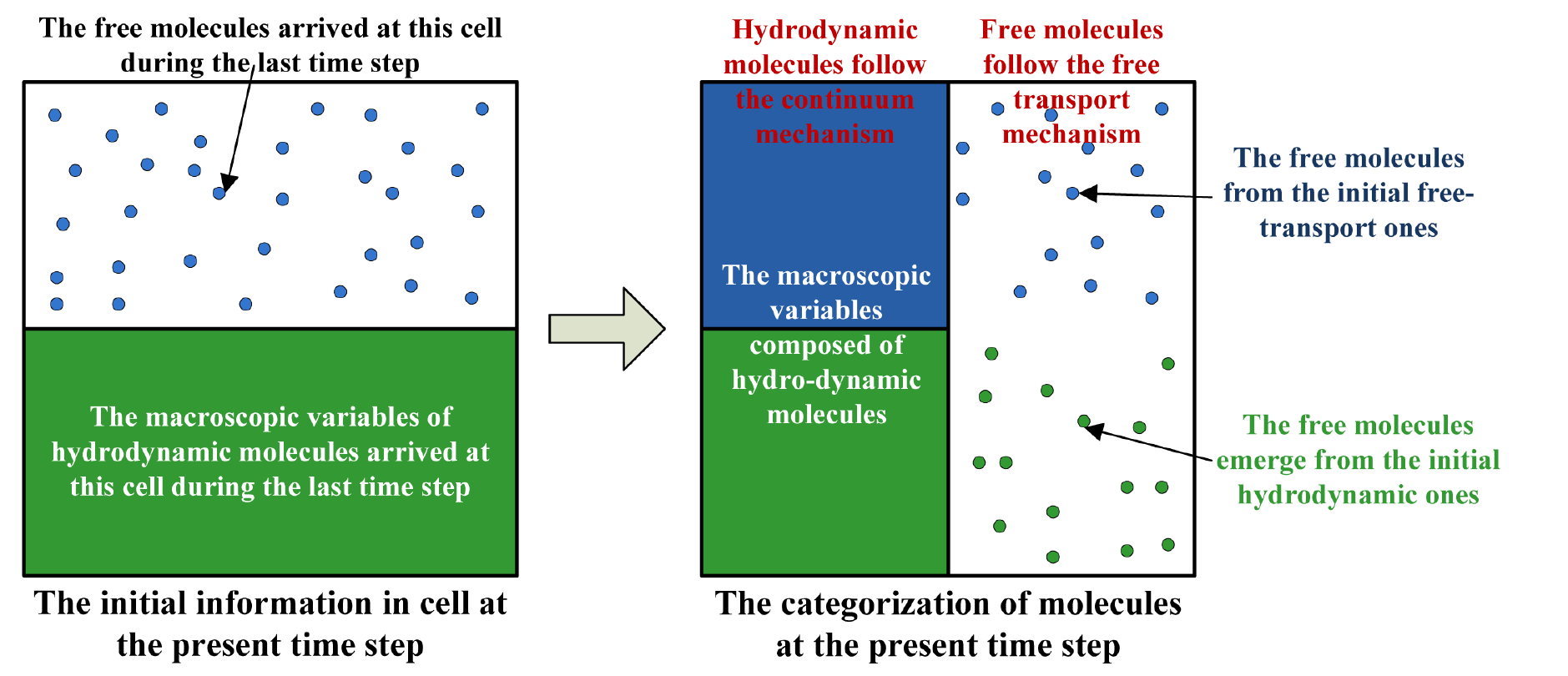}
\caption{\label{Fig:category} The categorization of molecules.}
\end{figure}

At the start of each time step (the left of Fig.~\ref{Fig:category}), there are two kinds of information included in discrete cells: the initial molecular information and the total macroscopic information. In the previous time step, some portion of molecules were categorized as free-transport molecules in each cell. When they finally arrived at certain cell at the end of this previous time step, they were recorded as molecules belonging to this cell. Therefore, the initial molecular information at the start of each time step is actually the individual information of the free molecules those were categorized in the previous time step. The individual molecular information includes its coordinate, mass and velocity. The total macroscopic information at the start of each time step are the macroscopic variables (mass, momentum and energy) for both free molecules and hydrodynamic molecules.

Therefore, the information at the start of each time step is incomplete, lacking of the individual information for the hydrodynamic molecules, since the transport of hydrodynamic molecules are modeled by their macroscopic (aggregate) behavior governed by N-S equation in the QMC mechanism, and their individual information are deleted for computational efficiency. In this paper, the macroscopic mass, momentum and energy in cell are defined as ${\bf{Q}}=\left(\rho\Omega, \rho{\bf{u}}\Omega, \rho\left(\left|{\bf{u}}\right|^2/2+e\right)\Omega\right)$, where $\Omega$ is the cell volume. The macroscopic variables of free-transport molecules, hydrodynamic molecules and total molecules are denoted by ${\mathbf{Q}}_{free}$, ${\mathbf{Q}}_{hydro}$ and ${\mathbf{Q}}_{total}$, respectively. Since ${\mathbf{Q}}_{total}$ is recorded in cell, and ${\mathbf{Q}}_{free}$ can be obtained by summing up the initial molecular information in cell, the macroscopic variables of hydrodynamic molecules can be obtained as ${\mathbf{Q}}_{hydro}={\mathbf{Q}}_{total}-{\mathbf{Q}}_{free}$. Then, the individual information of the hydrodynamic molecules are recovered by sampling them from the Maxwellian distribution determined by their macroscopic variables ${\mathbf{Q}}_{hydro}$ in a classic DSMC way~\cite{bird2013dsmc, Bird2003Molecular}.

After all the individual information of molecules are recovered, these molecules are categorized into new free-transport and hydrodynamic ones in this time step (the right of Fig.~\ref{Fig:category}). From QMC mechanism, there is $w_{free}$ portion of molecules are free-transport ones. In Ref.~\cite{Liu2018Unified, zhu2019unified}, the first collision time $t_{c} = \tau \ln \left( \eta  \right)$ is defined, where $\eta$ is a random number in $\left(0,1\right)$. For each molecule, $t_{c}$ is used to test whether it collides or not during the time step. If $t_{c}>\Delta t$, the molecule is categorized into free-transport one, else, it is categorized into  hydrodynamic one. The mathematical expectations of such test are $w_{free}$ and $w_{hyrdo}$. In continuum regime and near-continuum regime, there is almost no free-transport molecule existing in the flow field ($w_{free} \to 0$) since $\Delta t \gg \tau$. Therefore, there is no need to recover all the initial hydrodynamic molecules and only select a very small portion from them as the free-transport molecules in this time step. Instead, given ${\mathbf{Q}}_{hydro}$ and the weight $w_{hyrdo}$, free-transport molecules in this step can be directly sampled from $w_{hyrdo}{\mathbf{Q}}_{hydro}$ without $t_c$ test. The free-transport molecules in this step can also come from the initial free-transport molecules. Therefore, $w_{free}$ portion of free-transport molecules can be obtained by conducting the $t_{c}$ test to the initial free-transport molecules, while the portion of initial free-transport molecules that are categorized as hydrodynamic molecules in this time step, are deleted from computer memory.

After the categorization of molecules, the free-transport molecules transport freely in the flow filed, and their individual information is updated by the collisionless DSMC. The macroscopic flux caused by the hydrodynamic molecules are calculated by Eq.~\ref{eq:numericalFQMC}. And the total macroscopic variables are updated using the following equation:
\begin{equation}\label{eq:evolution}
{\bf{Q}}_{total}^{n + 1} = {\bf{Q}}_{total}^n + \left( {{\bf{Q}}_{free}^{n + \delta } - {\bf{Q}}_{free}^n} \right) - \Delta t\sum\limits_k^{{k_{\max }}} {\left({{\bf{F}}_{hydro}}\right){S_k}},
\end{equation}
where $n$ denotes the present time step, and $n+1$ denotes the next time step, $k$ is the index of cell interface, $k_{\max}$ is the number of interface in this cell. ${\bf{Q}}_{free}^n$ is the macroscopic variables of free-transport molecules at the present time step after the categorization. ${\bf{Q}}_{free}^{n+\delta}$ is the molecules belong to this cell at the end of the present time step. Without loss of generality, the normal directions of all interfaces are pointing outside.

Finally, the process of SUWP can be written as:
\begin{enumerate}
  \item Get ${\mathbf{Q}}_{free}$ by summing up the initial molecular information from the molecules which were categorized into free-transport ones and arrived at this cell during the last time step. Get the macroscopic variables of the initial hydrodynamic molecules from ${\mathbf{Q}}_{hydro}={\mathbf{Q}}_{total}-{\mathbf{Q}}_{free}$.
  \item Calculate $w_{free}$ and $w_{hydro}$.
  \item Sample free-transport molecules from $w_{hydro}{\mathbf{Q}}_{hydro}$ (one source of the free-transport molecules in this time step).
  \item Conduct the $t_{c}$ test on initial free-transport molecules, and only retain those pass this test ($t_{c}>\Delta t$, another source of the free-transport molecules in this time step).
  \item Now the information in cells includes the individual molecular information of free-transport molecules that are categorized in this time step, and the total macroscopic variables for all molecules in this cell (ready for transport).
  \item The motion of the free-transport molecules in this time step are calculated using a collisionless DSMC method. The transport of hydrodynamic molecules in this time step are modeled by the N-S mechanism, and their flux is calculated from Eq.~\ref{eq:numericalFQMC}.
  \item At the end of this time step, the coordinates of free-transport molecules are updated, and they are assigned to new cells if they have transported across the cell interfaces. The total macroscopic variables in cells are updated using Eq.~\ref{eq:evolution} by given the free molecular information and the macroscopic flux from hydrodynamic molecules.
\end{enumerate}

\subsection{collisionless DSMC solver}
For the free-transport molecules in SUWP, their transport processes are calculated by the collisionless DSMC solver. The particle tracing method on unstructured mesh is similar to that in Ref.~\cite{zhu2019unified}, expect a straddle test is used for computational efficiency in the present two-dimensional case. The tracing process is listed as follows:
\begin{enumerate}
  \item Given the iteration time $\Delta t$, the destination of a free-transport molecule (initially at ${\mathbf{x}}_{start}$ with velocity $\bm{\xi}$) can be directly obtained from ${\mathbf{x}}_{end} = {\mathbf{x}}_{start}+{\bm{\xi}}\Delta t$. (forming a line segment ${\mathbf{x}}_{start}{\mathbf{x}}_{end}$).
  \item loop all faces of the cell to test whether ${\mathbf{x}}_{start}{\mathbf{x}}_{end}$ intersects with one of them (Eq.~\ref{eq:straddle}). If there is no intersection between ${\mathbf{x}}_{start}{\mathbf{x}}_{end}$ and the faces, this molecule is in the present cell (process ends). If intersection is detected, break the loop and move to step 3
  \item Calculate the intersection point ${\mathbf{x}}_{inter}$ and move the molecule to it. Then, this molecule belongs to the adjacent cell. Calculate the remaining time $t_{remian}$ of the molecule (after been moved to ${\mathbf{x}}_{inter}$). Then move to step 4. Notice that if the intersection face is a solid wall, the molecular velocity is changed according to the wall boundary condition~\cite{Bird2003Molecular, bird2013dsmc}(after hitting the wall, its velocity is denoted by ${\bm{\xi}}_{wall}$), and its destination is changed to ${\mathbf{x}}_{end} = {\mathbf{x}}_{inter}+{\bm{\xi}}_{wall} t_{remain}$. Also notice that if the intersection face is other boundaries of the flow field, such inlet, outlet, farfield, the molecule can be deleted (process ends).
  \item loop all faces (expect the intersection face with ${\mathbf{x}}_{inter}$ on it) of the cell to test whether ${\mathbf{x}}_{inter}{\mathbf{x}}_{end}$ intersect with it (Eq.~\ref{eq:straddle}). if there is no intersection between ${\mathbf{x}}_{start}{\mathbf{x}}_{end}$ and the faces, the molecule is in the cell (process ends). If intersection is detected, break the loop, and move back to step 3.
\end{enumerate}

The straddle test is used in this paper to judge whether two segments intersect. Two segments (AB and CD) should fulfill the following two inequations in order to intersect with each other
\begin{equation}\label{eq:straddle}
\begin{aligned}
\left\{ {\left( {{{\mathbf{x}}_A} - {{\mathbf{x}}_C}} \right) \times \left( {{{\mathbf{x}}_D} - {{\mathbf{x}}_C}} \right)} \right\}\left\{ {\left( {{{\mathbf{x}}_B} - {{\mathbf{x}}_C}} \right) \times \left( {{{\mathbf{x}}_D} - {{\mathbf{x}}_C}} \right)} \right\} < 0\\
\left\{ {\left( {{{\mathbf{x}}_C} - {{\mathbf{x}}_A}} \right) \times \left( {{{\mathbf{x}}_B} - {{\mathbf{x}}_A}} \right)} \right\}\left\{ {\left( {{{\mathbf{x}}_D} - {{\mathbf{x}}_A}} \right) \times \left( {{{\mathbf{x}}_B} - {{\mathbf{x}}_A}} \right)} \right\} < 0
\end{aligned}
\end{equation}

In this paper, the unstructured rectangular meshes are adopted in the two-dimensional test cases. The sampling process into a rectangular mesh cell is translated into a sampling process into two triangular mesh cells denoted by $T_{A}$ and $T_{B}$, whose areas are $S_{A}$ and $S_{B}$ respectively. Given $\eta$ is the random real number sample in $\left(0,1\right)$. If $\eta \le S_{A}/\left(S_{A}+S_{B}\right)$, a molecule is sampled into $T_{A}$, else it is sampled into $T_{B}$. After the target triangle is chosen, the coordinate of this molecule is obtained by the direct sampling method~\cite{zhu2019unified} into the triangle.

Since the boundaries condition used for the present collisionless DSMC is exactly the same with the classical treatment~\cite{Bird2003Molecular, bird2013dsmc}, they are not discussed in this paper.

\subsection{Navier-Stokes solver}
In the SUWP, the classical N-S solver is used for calculating the macroscopic flux of the hydrodynamic molecules without change, except a scale-dependent coefficient is multiplied to the viscous flux (Eq.~\ref{eq:numericalFQMC}). Notice that during the entire N-S calculations, the physical variables (density, velocity and temperature) are from the total macroscopic variables ${\bf{Q}}_{total}$.

At the present stage, the inviscid flux in Ref.~\cite{SunA2016} is used. It is the weighted summation of the flux of Kinetic Flux-Vector Splitting (KFVS) method and the Totally Thermalized Transport (TTT) method~\cite{Xu1998report}. In this paper, both the mathematical forms of KFVS flux and the TTT process are rearranged in a classical Computational Fluid Dynamics (CFD) way. Here, the normal direction of the cell interface is from its left side to its right side. The macroscopic velocity in normal direction and two tangential directions are denoted by $u$, $v$, $w$ respectively. The KFVS flux $\mathbf{K}$ is rearrange as follows:
\begin{equation}\label{eq:KFVS}
\begin{aligned}
&{K_{mass}} = \frac{{\rm{1}}}{{\rm{2}}}\left( {{\rho _L}{u_L} + {\rho _R}{u_R}} \right) + \frac{1}{2}\left( {{\rho _L}{u_L}{\eta _L} - {\rho _R}{u_R}{\eta _R}} \right) + \frac{1}{2}\left( {{\rho _L}{\theta _L} - {\rho _R}{\theta _R}} \right),\\
&{K_{xmon}} = \frac{{\rm{1}}}{{\rm{2}}}\left[ {\left( {{\rho _L}u_L^2 + {p_L}} \right) + \left( {{\rho _R}u_R^2 + {p_R}} \right)} \right] + \frac{{\rm{1}}}{{\rm{2}}}\left[ {\left( {{\rho _L}u_L^2 + {p_L}} \right){\eta _L} - \left( {{\rho _R}u_R^2 + {p_R}} \right){\eta _R}} \right] + \frac{{\rm{1}}}{{\rm{2}}}\left( {{\rho _L}{u_L}{\theta _L} - {\rho _R}{u_R}{\theta _R}} \right),\\
&{K_{ymon}} = \frac{{\rm{1}}}{{\rm{2}}}\left( {{\rho _L}{u_L}{v_L} + {\rho _R}{u_R}{v_R}} \right) + \frac{1}{2}\left( {{\rho _L}{u_L}{v_L}{\eta _L} - {\rho _R}{u_R}{v_R}{\eta _R}} \right) + \frac{1}{2}\left( {{\rho _L}{v_L}{\theta _L} - {\rho _R}{v_R}{\theta _R}} \right),\\
&{K_{zmon}} = \frac{{\rm{1}}}{{\rm{2}}}\left( {{\rho _L}{u_L}{w_L} + {\rho _R}{u_R}{w_R}} \right) + \frac{1}{2}\left( {{\rho _L}{u_L}{w_L}{\eta _L} - {\rho _R}{u_R}{w_R}{\eta _R}} \right) + \frac{1}{2}\left( {{\rho _L}{w_L}{\theta _L} - {\rho _R}{w_R}{\theta _R}} \right),\\
&{K_{energy}} = \frac{{\rm{1}}}{{\rm{2}}}\left( {{\rho _L}{u_L}{h_L} + {\rho _R}{u_R}{h_R}} \right) + \frac{1}{2}\left( {{\rho _L}{u_L}{h_L}{\eta _L} - {\rho _R}{u_R}{h_R}{\eta _R}} \right) + \frac{1}{2}\left[ {\left( {{\rho _L}{h_L} - \frac{{{p_L}}}{2}} \right){\theta _L} - \left( {{\rho _R}{h_R} - \frac{{{p_R}}}{2}} \right){\theta _R}} \right],
\end{aligned}
\end{equation}
where the subscript ``L" and ``R" represent the left and right hand of the cell interface where the flux is calculated. $\eta$ and $\theta$ are defined as:
\begin{equation}\label{eq:coeff_KFVS}
\begin{aligned}
&{\eta _\alpha } = {\rm{erf}}\left( {\frac{{{U_\alpha }}}{{\sqrt {2R{T_\alpha }} }}} \right)\\
&{\theta _\alpha } = \sqrt {\frac{{2R{T_\alpha }}}{\pi }} \exp \left( { - \frac{{U_\alpha ^2}}{{2R{T_\alpha }}}} \right)
\end{aligned}
\end{equation}
where the subscript ``$\alpha$" can be ``L" or ``R".

The TTT flux $\mathbf{T}$ is the simple Euler flux using the averaged values
\begin{equation}\label{eq:Euler}
\begin{aligned}
&{T_{mass}} = \bar \rho \bar u,\\
&{T_{xmon}} = \bar \rho {{\bar u}^2} + \bar p,\\
&{T_{ymon}} = \bar \rho \bar u\bar v,\\
&{T_{zmon}} = \bar \rho \bar u\bar w,\\
&{T_{energy}} = \bar \rho \bar h,
\end{aligned}
\end{equation}
where the averaged values can be obtained from the following TTT process:
\begin{equation}\label{eq:TTT}
\begin{aligned}
&\overline \rho   = \frac{{\rm{1}}}{{\rm{2}}}\left( {{\rho _L}{\rm{ + }}{\rho _R}} \right) + \frac{{\rm{1}}}{{\rm{2}}}\left( {{\rho _L}{\eta _L} - {\rho _R}{\eta _R}} \right),\\
&\overline {\rho u}  = \frac{{\rm{1}}}{{\rm{2}}}\left( {{\rho _L}{u_L}{\rm{ + }}{\rho _R}{u_R}} \right) + \frac{1}{2}\left( {{\rho _L}{\theta _L} - {\rho _{\rm{R}}}{\theta _R}} \right) + \frac{1}{2}\left( {{\rho _L}{u_L}{\eta _L} - {\rho _R}{u_R}{\eta _R}} \right),\\
&\overline {\rho v}  = \frac{{\rm{1}}}{{\rm{2}}}\left( {{\rho _L}{v_L}{\rm{ + }}{\rho _R}{v_R}} \right) + \frac{{\rm{1}}}{{\rm{2}}}\left( {{\rho _L}{v_L}{\eta _L} - {\rho _R}{v_R}{\eta _R}} \right),\\
&\overline {\rho w}  = \frac{{\rm{1}}}{{\rm{2}}}\left( {{\rho _L}{w_L}{\rm{ + }}{\rho _R}{w_R}} \right) + \frac{{\rm{1}}}{{\rm{2}}}\left( {{\rho _L}{w_L}{\eta _L} - {\rho _R}{w_R}{\eta _R}} \right),\\
&\overline {\rho e}  = \frac{{\rm{1}}}{{\rm{2}}}\left( {{\rho _L}{e_L}{\rm{ + }}{\rho _R}{e_R}} \right) + \frac{1}{4}\left( {{\rho _L}{u_L}{\theta _L} - {\rho _{\rm{R}}}{u_R}{\theta _R}} \right) + \frac{1}{2}\left( {{\rho _L}{e_L}{\eta _L} - {\rho _R}{e_R}{\eta _R}} \right).
\end{aligned}
\end{equation}
with the aid of $\bar u=\overline {\rho u}/\bar \rho$, $\bar v=\overline {\rho v}/\bar \rho$, $\bar w=\overline {\rho w}/\bar \rho$, $\bar T=\overline {\rho e}/C_v$ and $\bar h=C_p T$.

So far, the inviscid flux on the cell interface can be calculated using:
\begin{equation}\label{eq:inviscid}
\mathbf{F}_{inv} = \beta\mathbf{K} + (1-\beta)\mathbf{T},
\end{equation}
where
\begin{equation}
\beta = {\rm{tanh}}\left(C\frac{\left|p_R-p_L\right|}{p_R-p_L}\right)
\end{equation}
Here ${\rm{tanh}}$ is the hyperbolic tangent function, $C$ is chosen as 10 according to Ref.~\cite{SunA2016}.

In this paper, the second order reconstruction is considered. The gradients of conservative variables $\rho, \rho{\bf{u}}, \rho\left(\left|{\bf{u}}\right|^2/2+e\right)$ are calculated by the least square method. The Venkatakrishnan slop limiter~\cite{Venkatakrishnan1993On} is adopted to the gradients.

The viscous flux ${\mathbf{F}}_{vis}$ is calculated by the central scheme. The physical variables and their gradients at the cell interface for calculating the viscous flux are obtained by conducting a weighted average of the central values of two neighbor cells. Up to now, the framework of flux calculation is the same with the classical CFD way. Finally, the modified N-S flux can be obtained from Eq.~\ref{eq:numericalFQMC}.

When $\tau$ approaches infinity (free molecular flow limit), $c_{vis}$ becomes zero, and the continuum mechanism has no contribution to the dissipation (it becomes the Euler mechanism). At the free molecular flow limit, the dissipation is totally from the free-transport mechanism. In earlier particle methods based on BGK-type equation, particles follow the Maxwellian distribution after collisions, or follow other equilibrium distributions depending on the chosen model equation. After collision, these particles transport freely. Since the free-transport of particles from Maxwellian distribution leads to the Euler mechanism without dissipation, the earlier particle methods can predict the rarefied flows with large Kn number well. While for flows with small Kn number, where the dissipation in continuum mechanism can not be neglected, they often have the problem of inaccurate viscous behavior. In USP method, this viscous problem is repaired by forcing some particles emerging from a Grad distribution which becomes second order C-E distribution in continuum limit with N-S viscous term in it. When $\tau$ approaches zero (continuum limit), the flow is totally dominated by continuum mechanism. At the same time, $c_{vis}$ is unity, and the dissipation mechanism recovers the N-S mechanism completely.

\section{Numerical Experiments}\label{sec:EXP}
\subsection{Shock wave structure}
The shock wave structure case is a benchmark case for the validity and accuracy of multi-scale numerical methods in non-equilibrium flow simulations. A normal shock wave is a discontinuity from macroscopic point of view, across which the physical properties change precipitously. While the profiles of a normal shock wave are actually smooth from microscopic point of view. Molecules in the shock wave are a mixture of the ones before the shock wave, where flow is supersonic or hypersonic with a relatively low temperature and the molecules after the shock wave, where the flow is subsonic with a high temperature. Since the molecular collisions in the thin shock wave (about twenty m.f.p.) are insufficient, the distribution function will be far from the equilibrium for large Ma numbers. Given the heat index $\omega$ and molecular scattering factor $\alpha$, the m.f.p. of Variable Soft Sphere (VSS) model can be written as
\begin{equation}
m.f.p. = \frac{1}{\beta }\sqrt {\frac{{RT}}{{2\pi }}} \frac{\mu }{p},
\end{equation}
where $\beta$ is a gas-model depended coefficient defined as
\begin{equation}\label{eq:beta}
\beta = \frac{{5\left( {\alpha {\rm{ + 1}}} \right)\left( {\alpha {\rm{ + 2}}} \right)}}{{{\rm{4}}\alpha \left( {5 - 2\omega } \right)\left( {7 - 2\omega } \right)}}.
\end{equation}
In this section, the Argon gas is chosen as the working gas. The benchmark solutions from UGKS~\cite{LiuUnified2020} chooses the Variable Hard Sphere (VHS) molecular model for Argon, so it is adopted in the present SUWP calculation. Since $\alpha=1$ will reduce the general VSS model to the VHS model, $\omega=0.81$ and $\alpha=1.0$ are chosen. The cell Kn number ${\rm{Kn}}_{\rm{cell}}$ (the reference length is chosen as the cell length) is set $2.0$ to resolve the profiles. The upstream and downstream boundaries are determined by Rankine-Hugoniot condition.

\begin{figure}
\centering
\subfigure[\label{Fig:Ma8Shock} Ma=8.0]{
\includegraphics[width=0.45\textwidth]{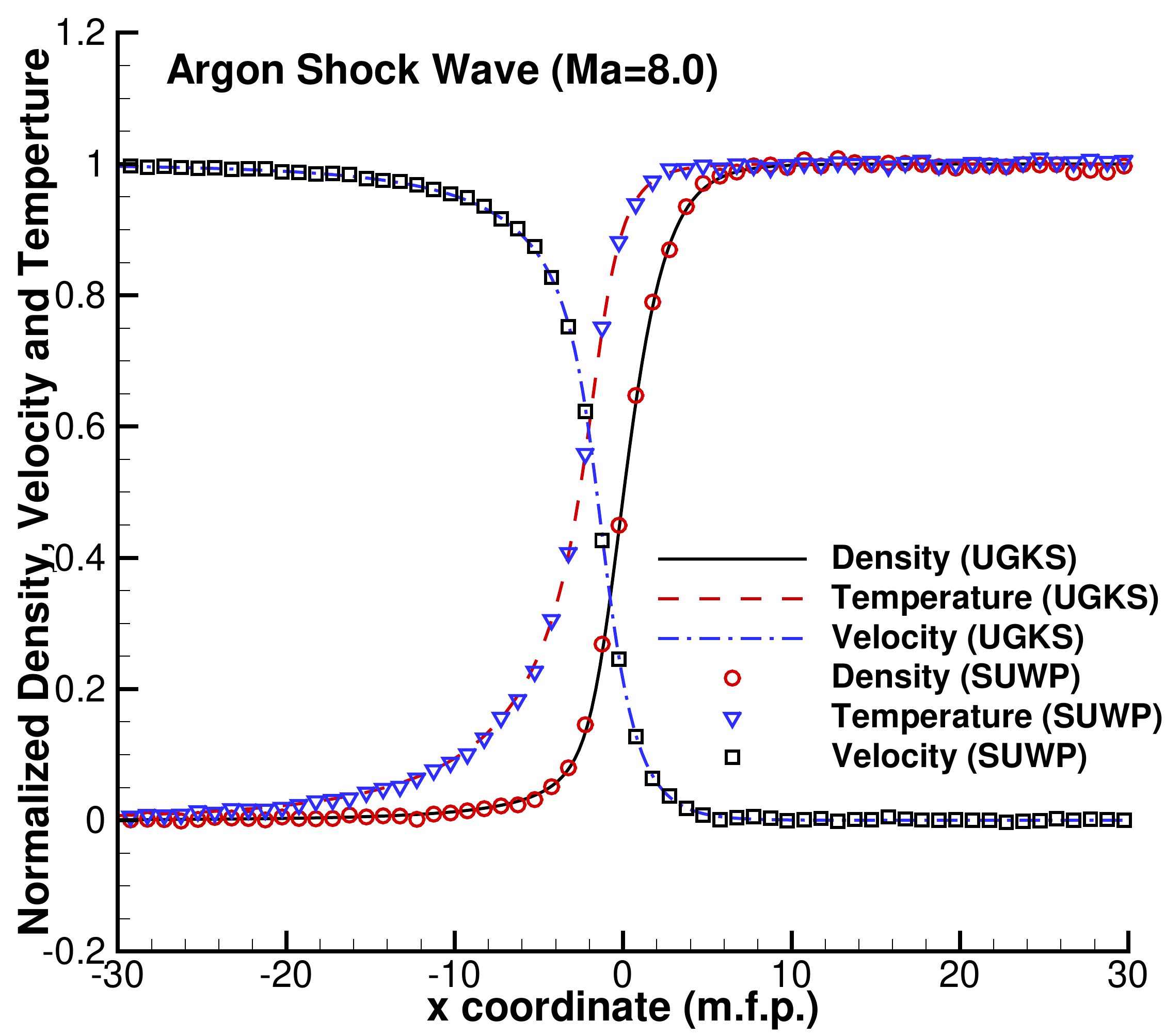}
}\hspace{0.05\textwidth}%
\subfigure[\label{Fig:Ma10Shock} Ma=10.0]{
\includegraphics[width=0.45\textwidth]{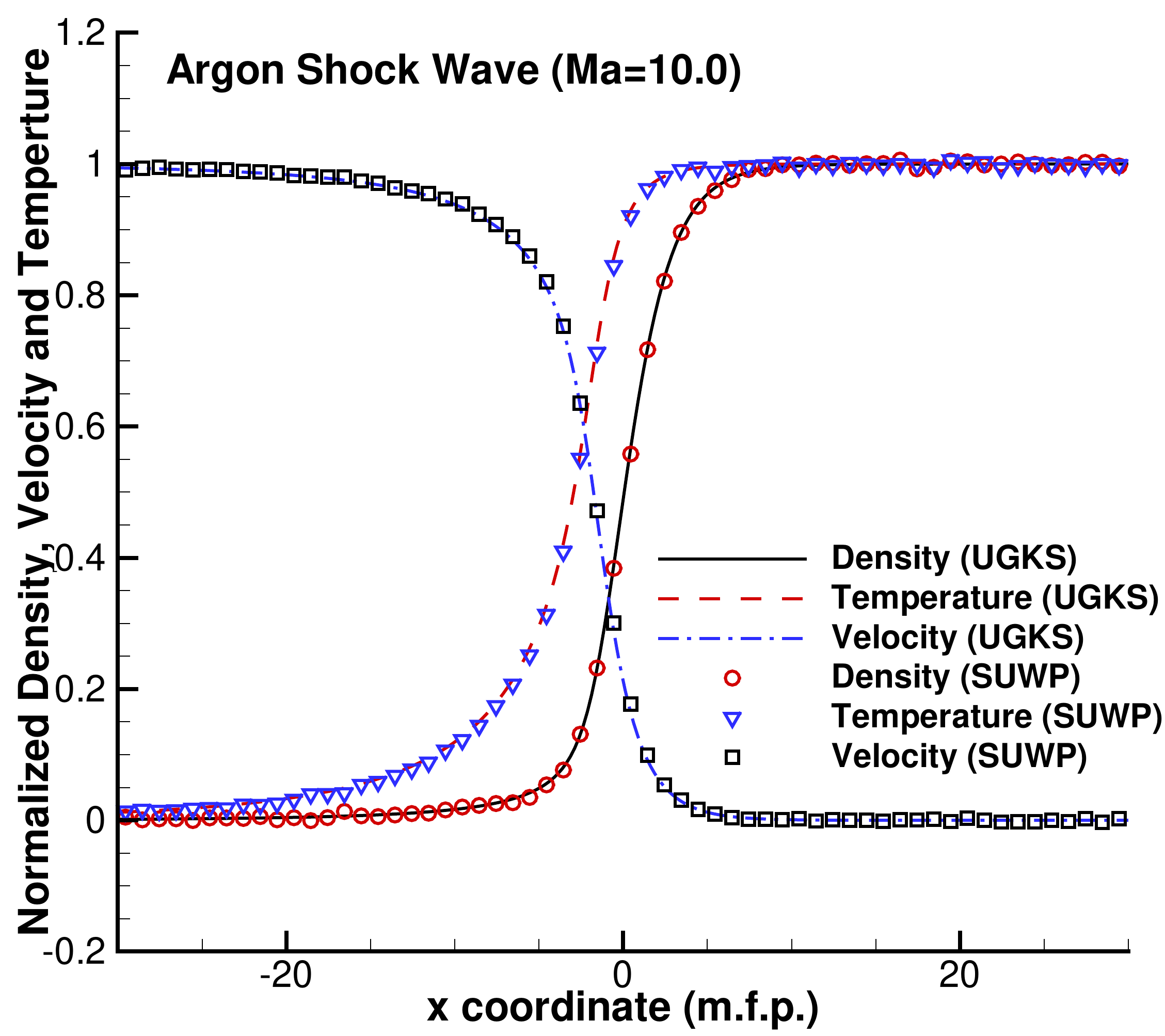}
}
\caption{\label{Fig:Shock} Argon shock wave structures (lines are the benchmark solution from UGKS, symbols are the present SUWP numerical solutions}
\end{figure}

Fig.~\ref{Fig:Shock} illustrates the profiles of the shock waves with Ma number 8 and 10, where the density, velocity and temperature are normalized as follows:
\begin{equation}\label{eq:rhoTnonD}
\begin{aligned}
&\hat \rho = \frac{{\rho  - {\rho _{up}}}}{{{\rho _{down}} - {\rho _{up}}}},\\
&\hat u = \frac{u-u_{down}}{u_{up}-u_{down}},\\
&\hat T = \frac{{T - {T_{up}}}}{{{T_{down}} - {T_{up}}}},
\end{aligned}
\end{equation}
where the subscript ``up'' and ``down'' stand for the upstream and downstream, respectively. The $x$ coordinate in the shock wave is normalized by the m.f.p. of upstream boundary. It can be seen from Fig.~\ref{Fig:Shock} that the numerical solutions obtained from SUWP match well with the benchmark solution from the UGKS method.

\subsection{Sod shock-tube}
\begin{figure}
\centering
\subfigure[\label{Fig:sodn1d} Density]{
\includegraphics[width=0.45\textwidth]{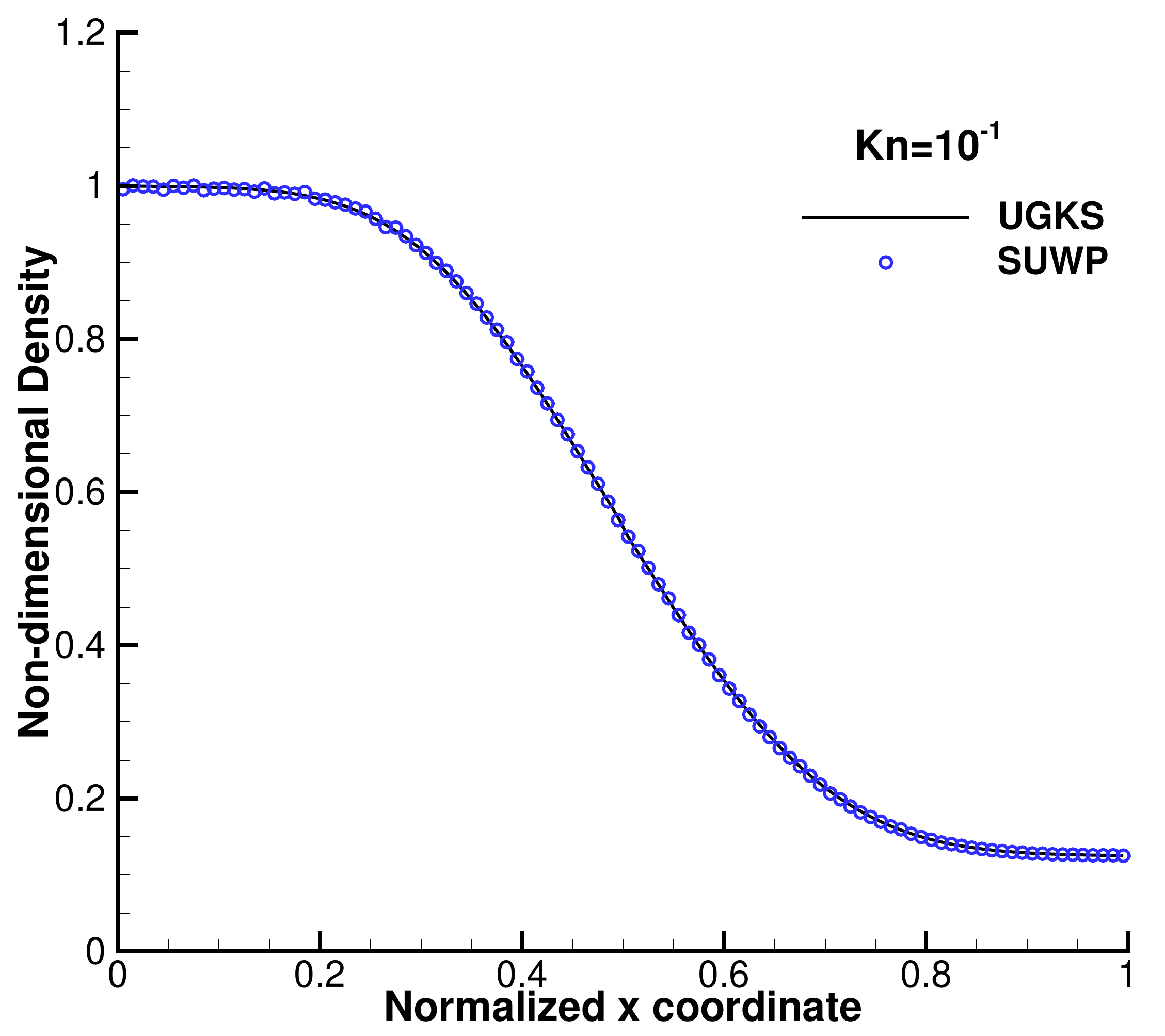}
}\hspace{0.05\textwidth}%
\subfigure[\label{Fig:sodn1u} Velocity]{
\includegraphics[width=0.45\textwidth]{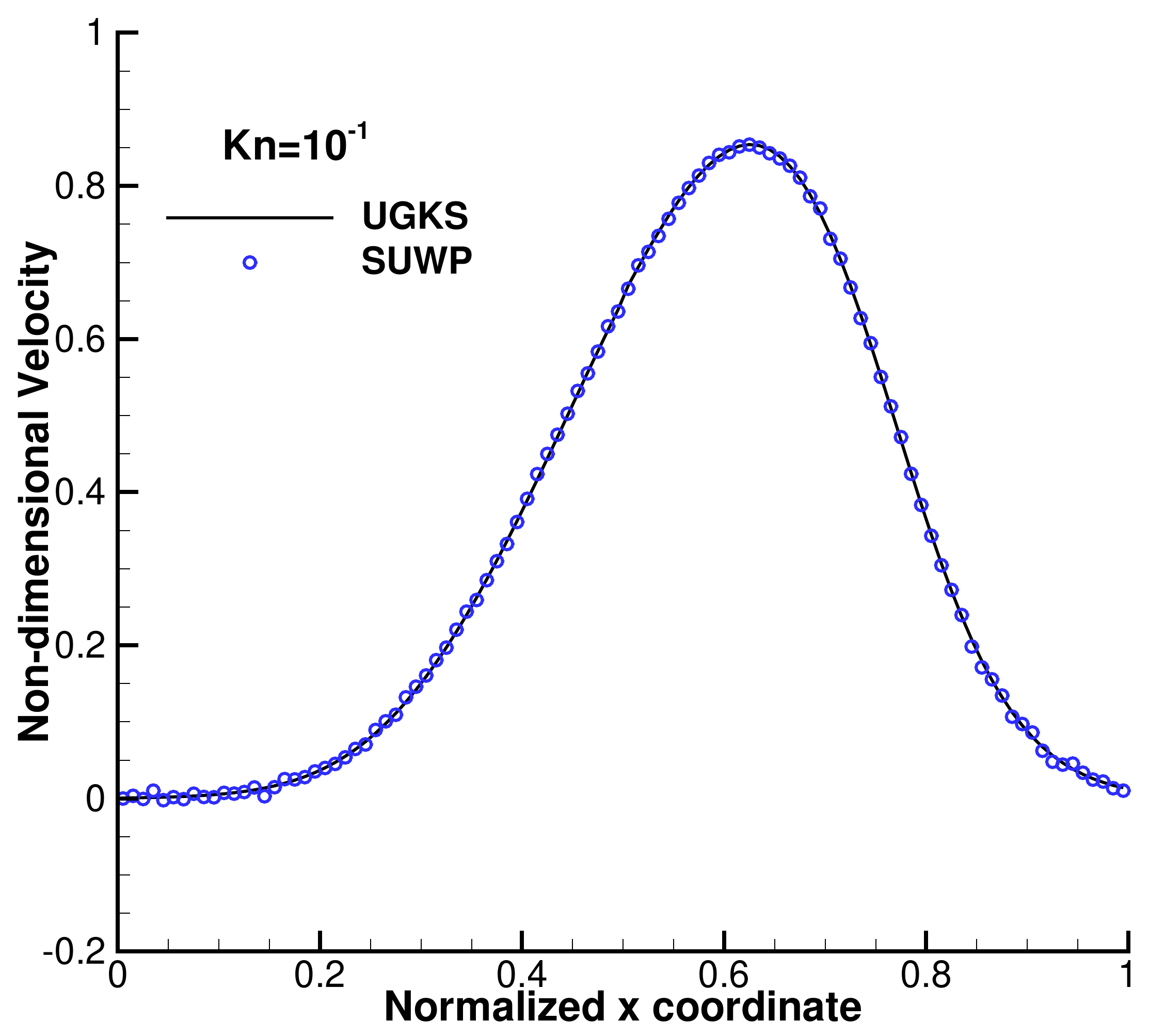}
}\\
\subfigure[\label{Fig:sodn1t} Temperature]{
\includegraphics[width=0.45\textwidth]{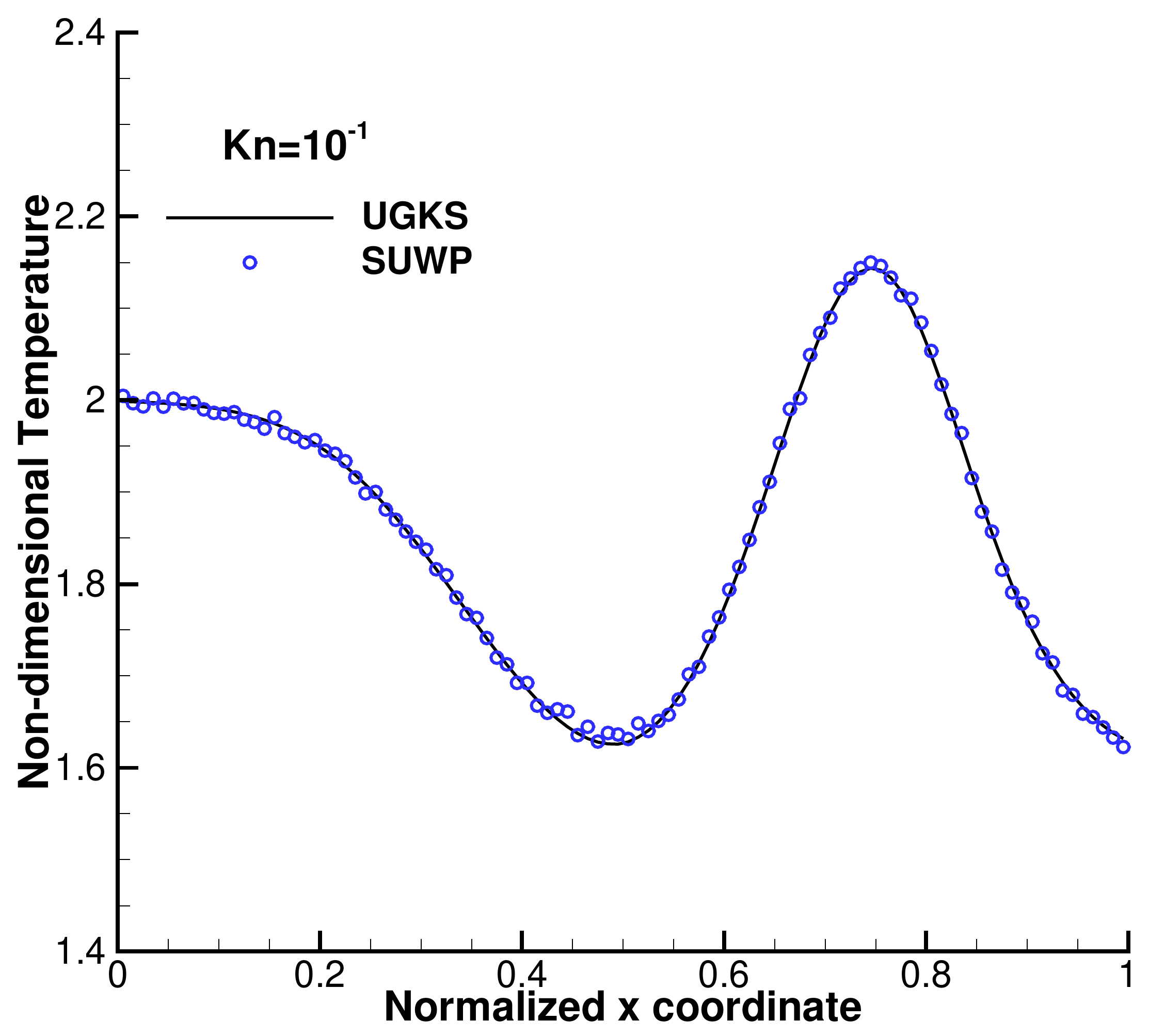}
}\hspace{0.05\textwidth}%
\subfigure[\label{Fig:sodn1p} Pressure]{
\includegraphics[width=0.45\textwidth]{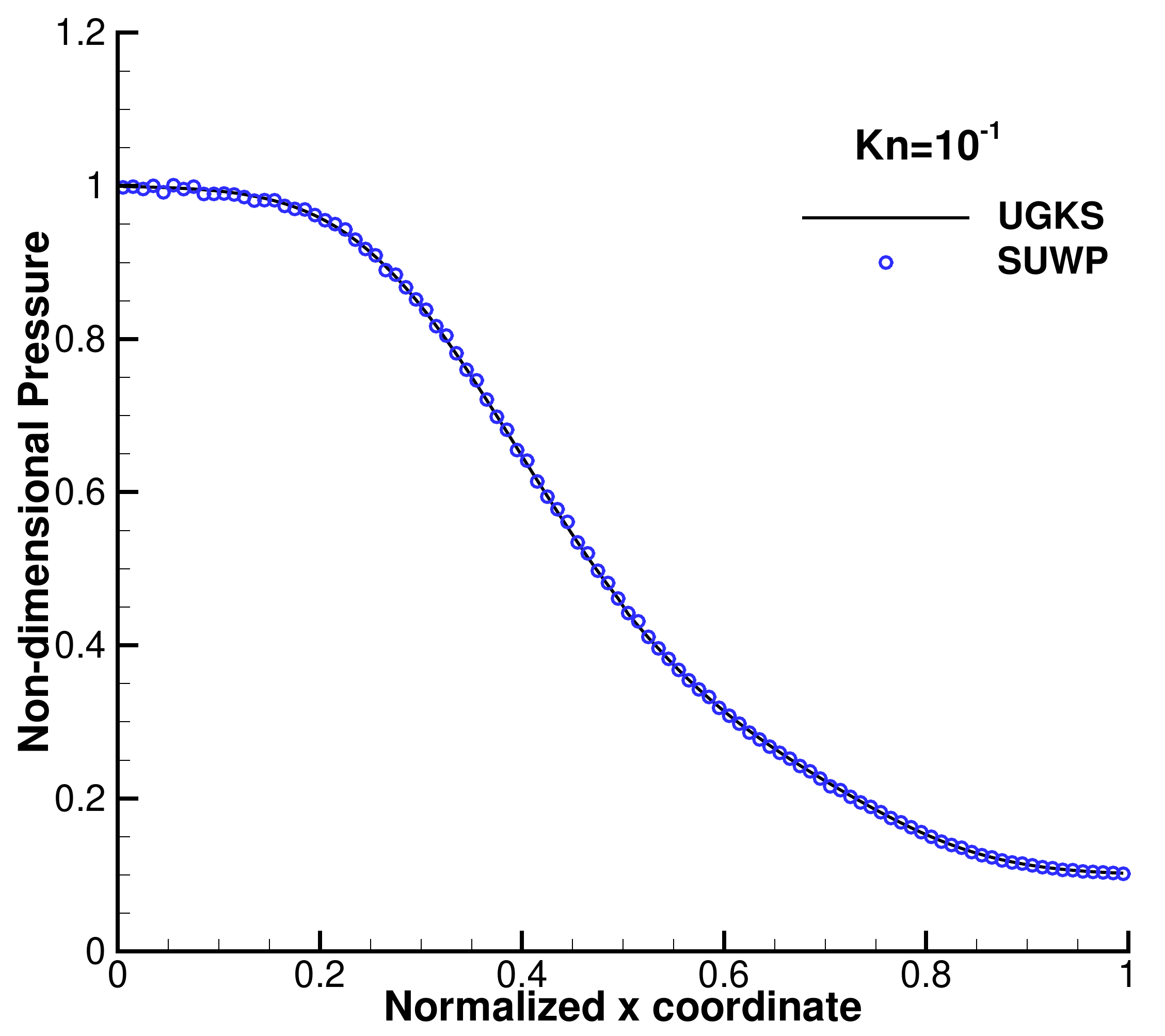}
}
\caption{\label{Fig:sodn1} Profiles of Sod shock-tube at $t=0.15$ with $\rm{Kn}=10^{-1}$. The SUWP solution is shown in symbols and the reference UGKS solution is shown in solid lines.}
\end{figure}

\begin{figure}
\centering
\subfigure[\label{Fig:sodn3d} Density]{
\includegraphics[width=0.45\textwidth]{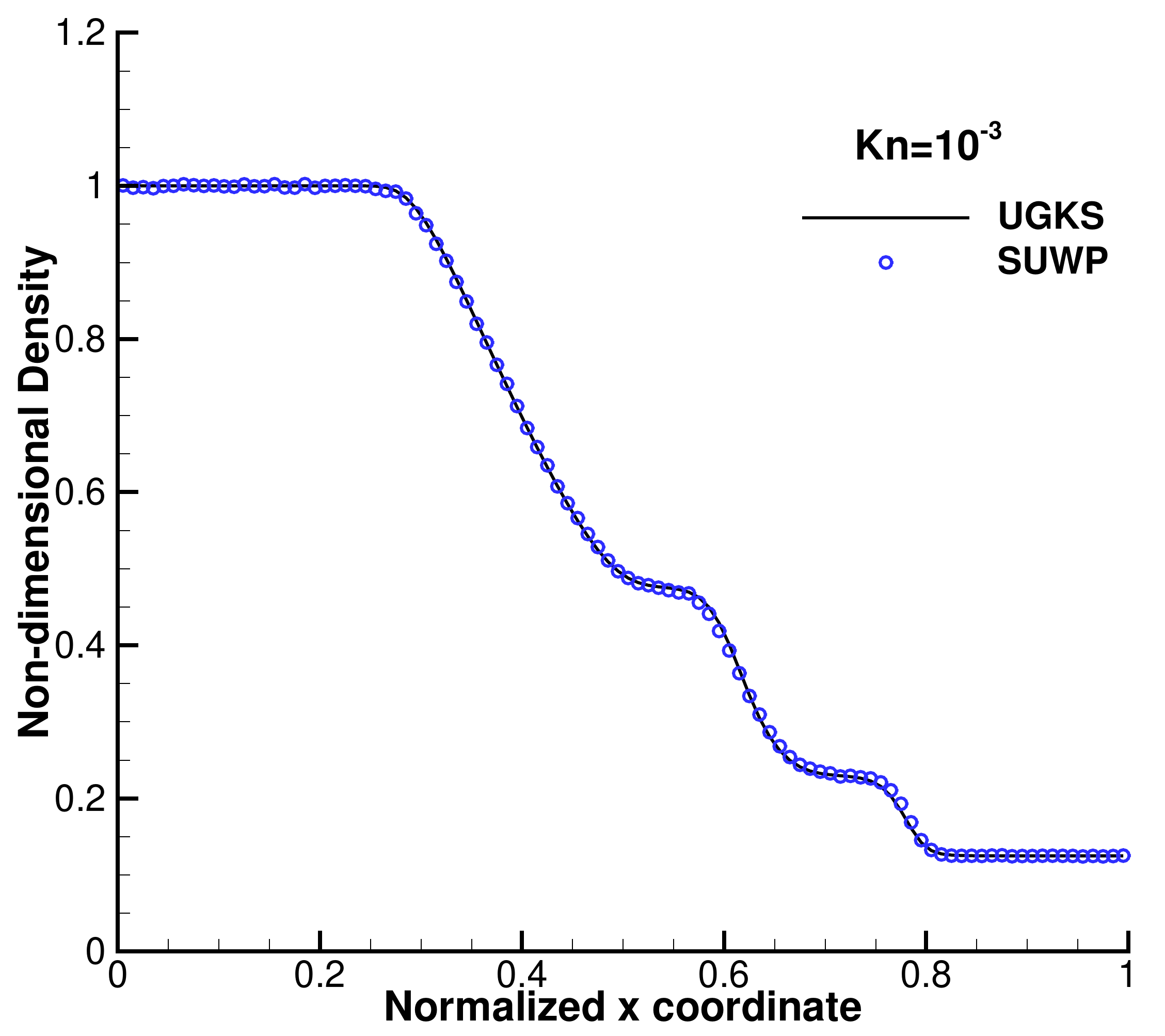}
}\hspace{0.05\textwidth}%
\subfigure[\label{Fig:sodn3u} Velocity]{
\includegraphics[width=0.45\textwidth]{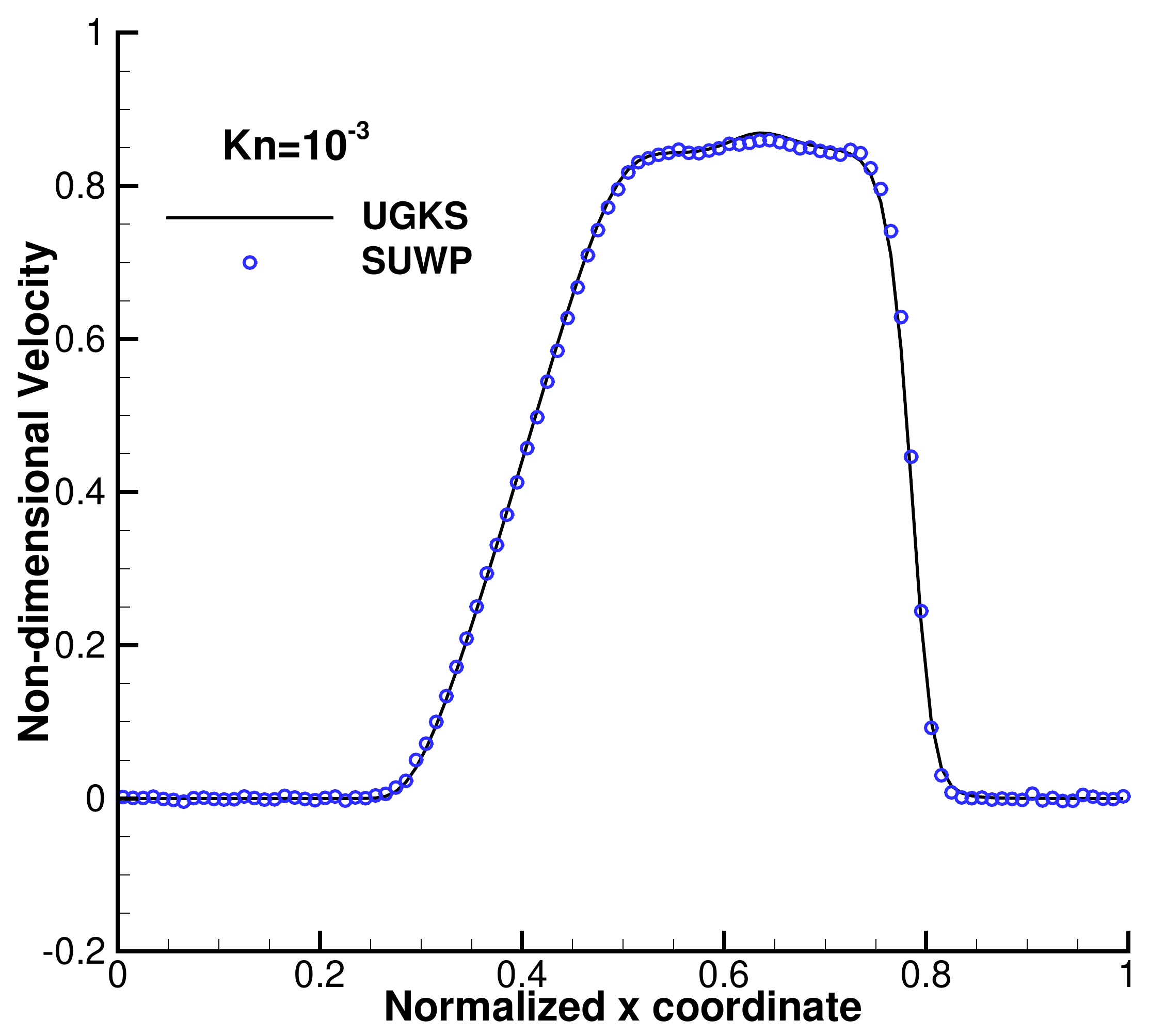}
}\\
\subfigure[\label{Fig:sodn3t} Temperature]{
\includegraphics[width=0.45\textwidth]{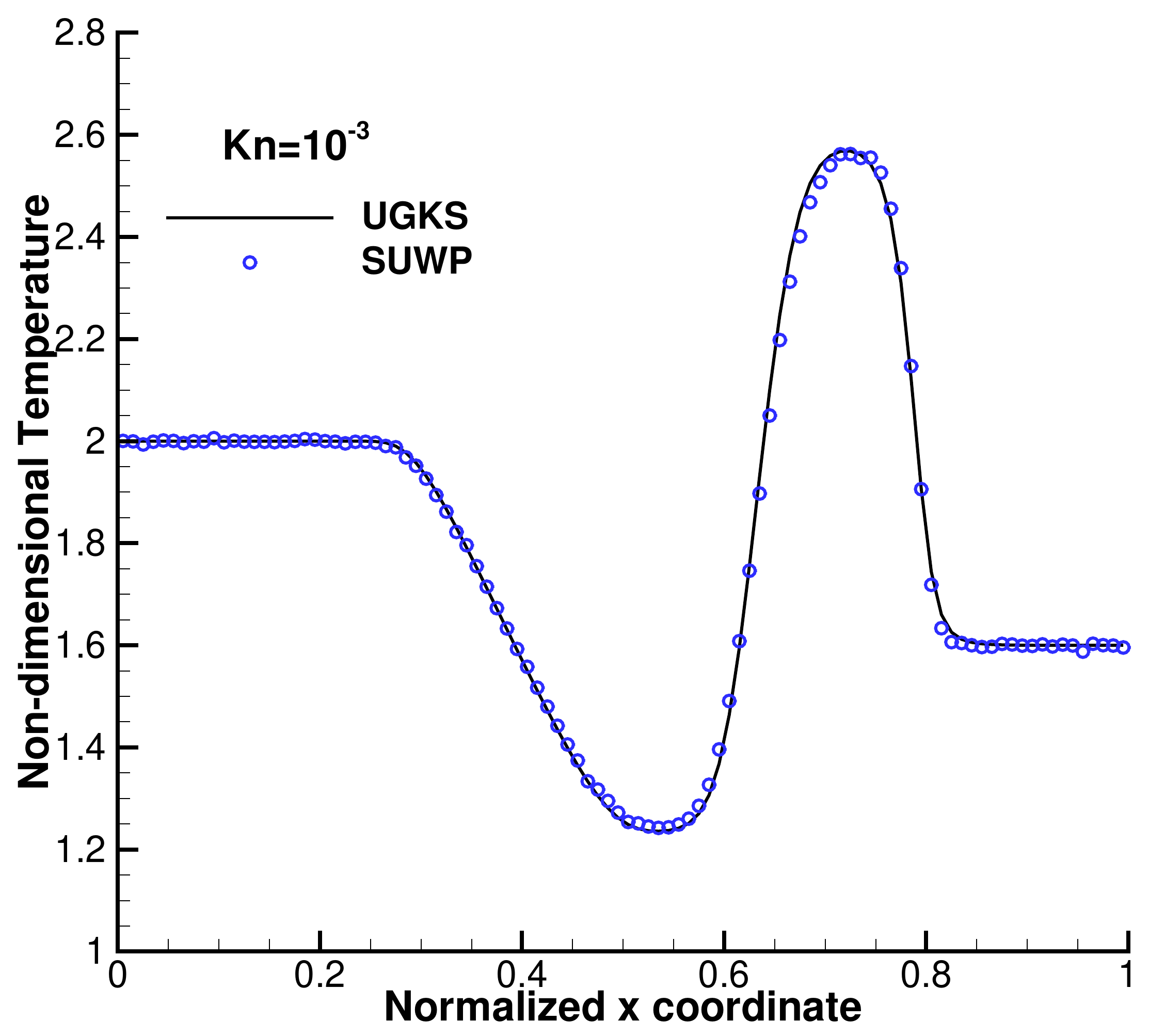}
}\hspace{0.05\textwidth}%
\subfigure[\label{Fig:sodn3p} Pressure]{
\includegraphics[width=0.45\textwidth]{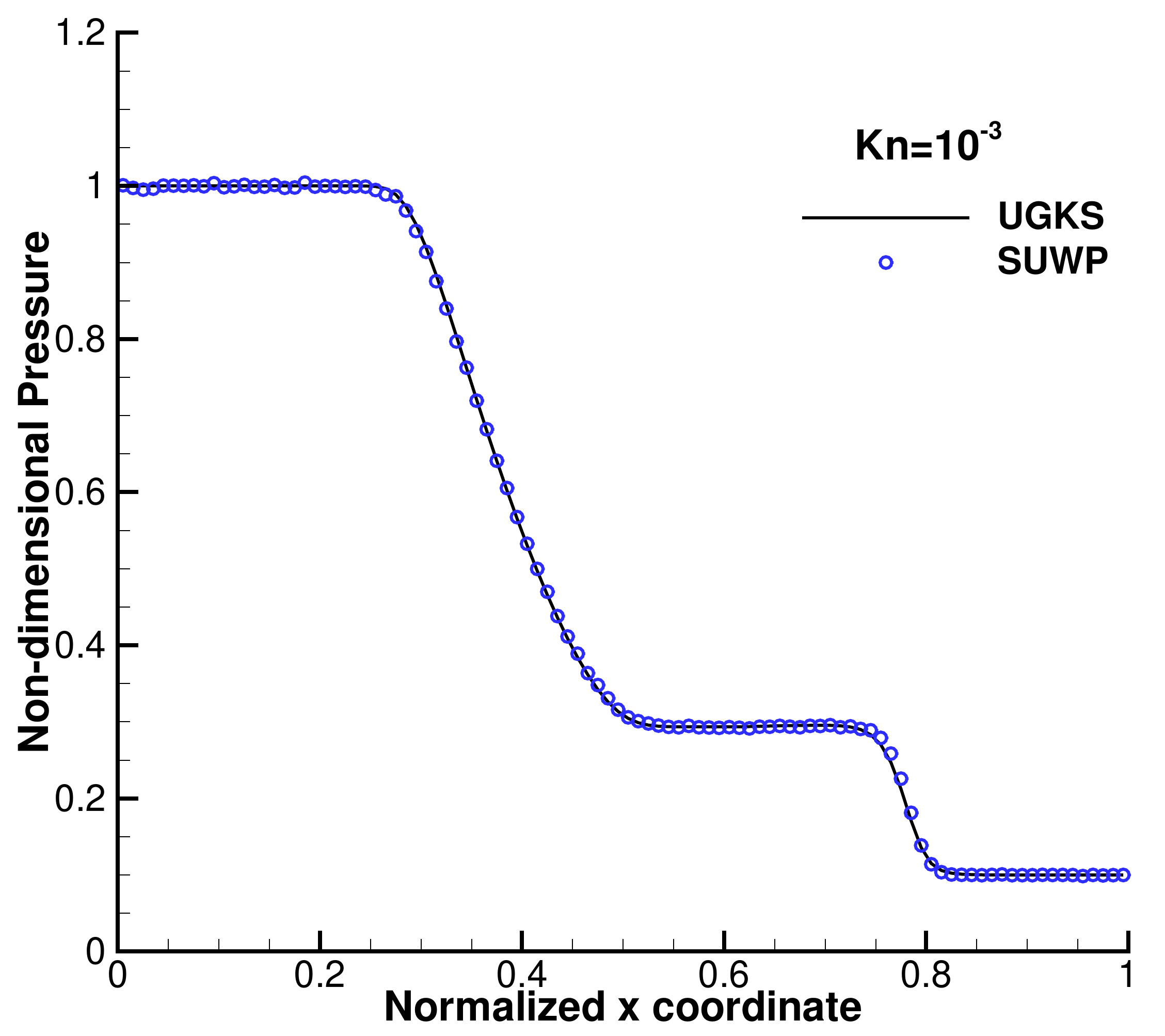}
}
\caption{\label{Fig:sodn3} Profiles of Sod shock-tube at $t=0.15$ with $\rm{Kn}=10^{-3}$. The SUWP solution is shown in symbols and the reference UGKS solution is shown in solid lines.}
\end{figure}

\begin{figure}
\centering
\subfigure[\label{Fig:sodn5d} Density]{
\includegraphics[width=0.45\textwidth]{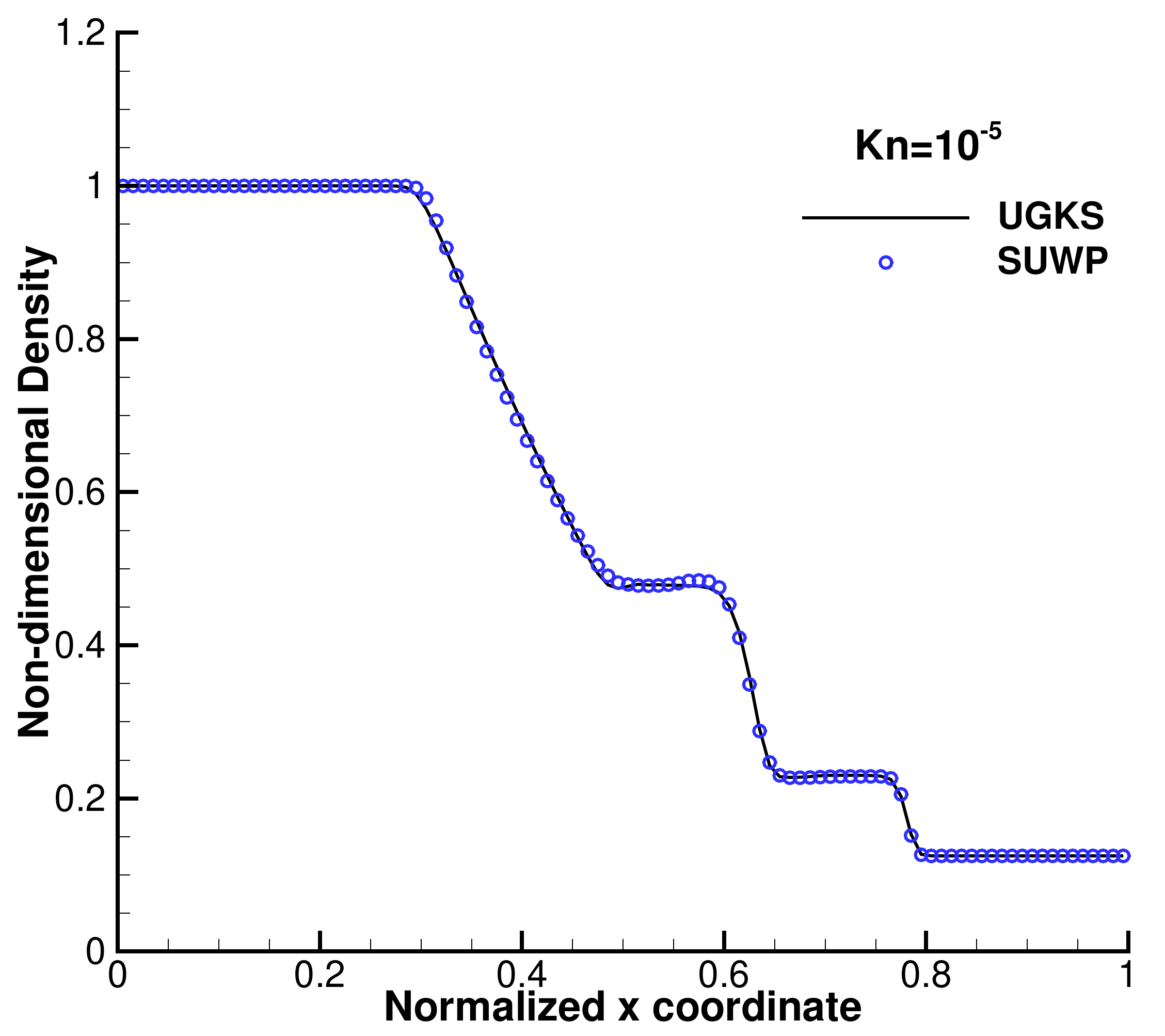}
}\hspace{0.05\textwidth}%
\subfigure[\label{Fig:sodn5u} Velocity]{
\includegraphics[width=0.45\textwidth]{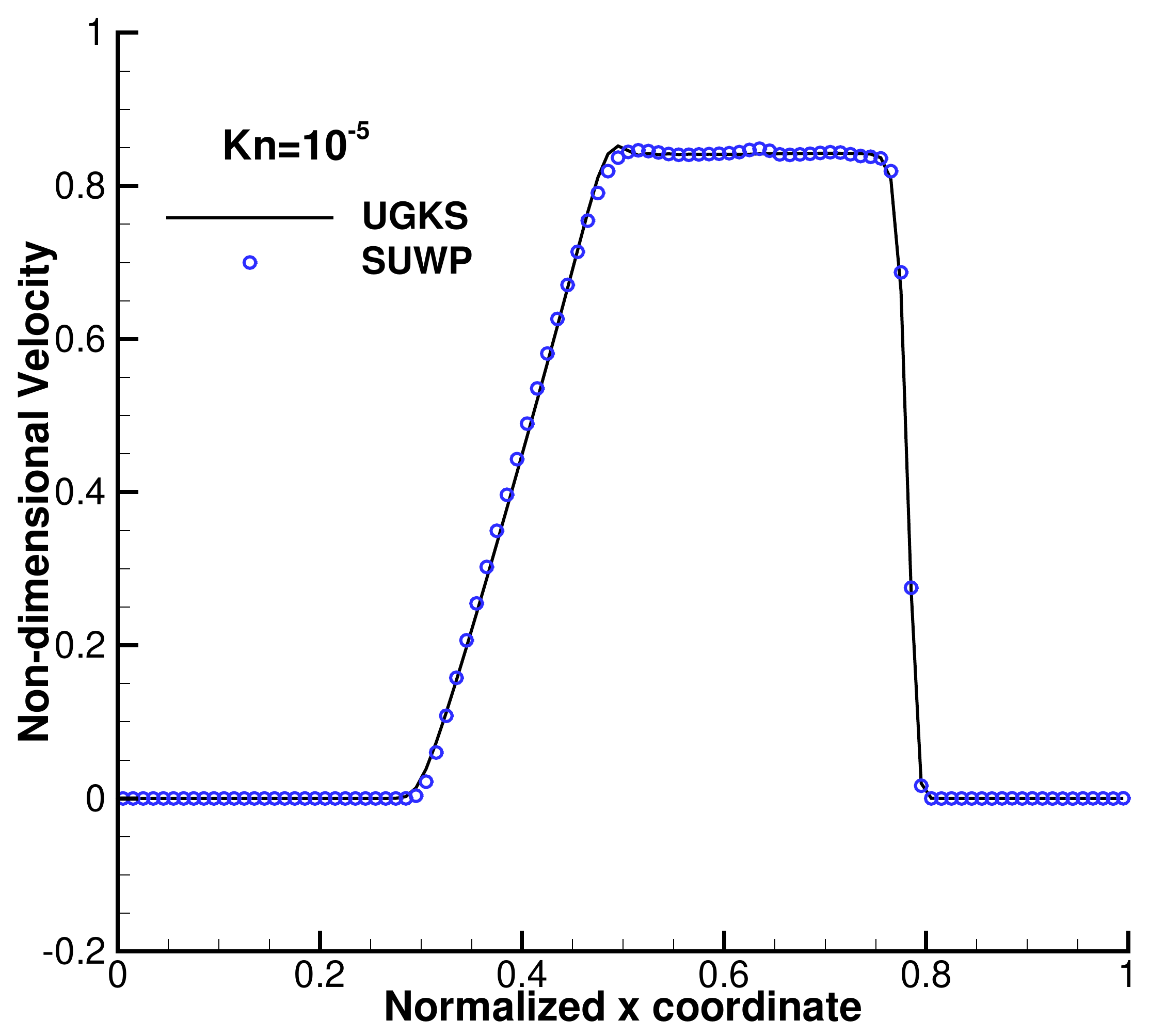}
}\\
\subfigure[\label{Fig:sodn5t} Temperature]{
\includegraphics[width=0.45\textwidth]{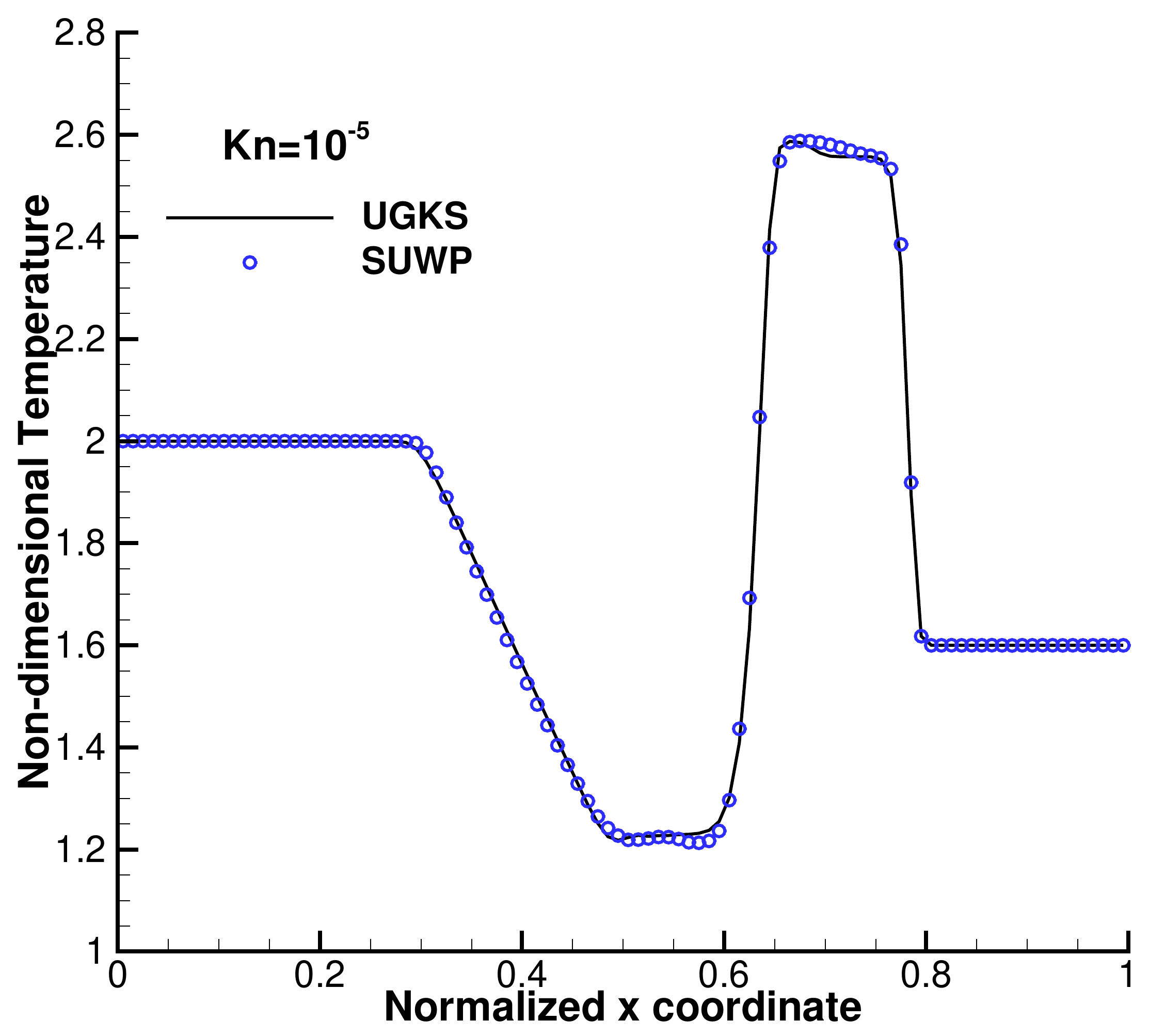}
}\hspace{0.05\textwidth}%
\subfigure[\label{Fig:sodn5p} Pressure]{
\includegraphics[width=0.45\textwidth]{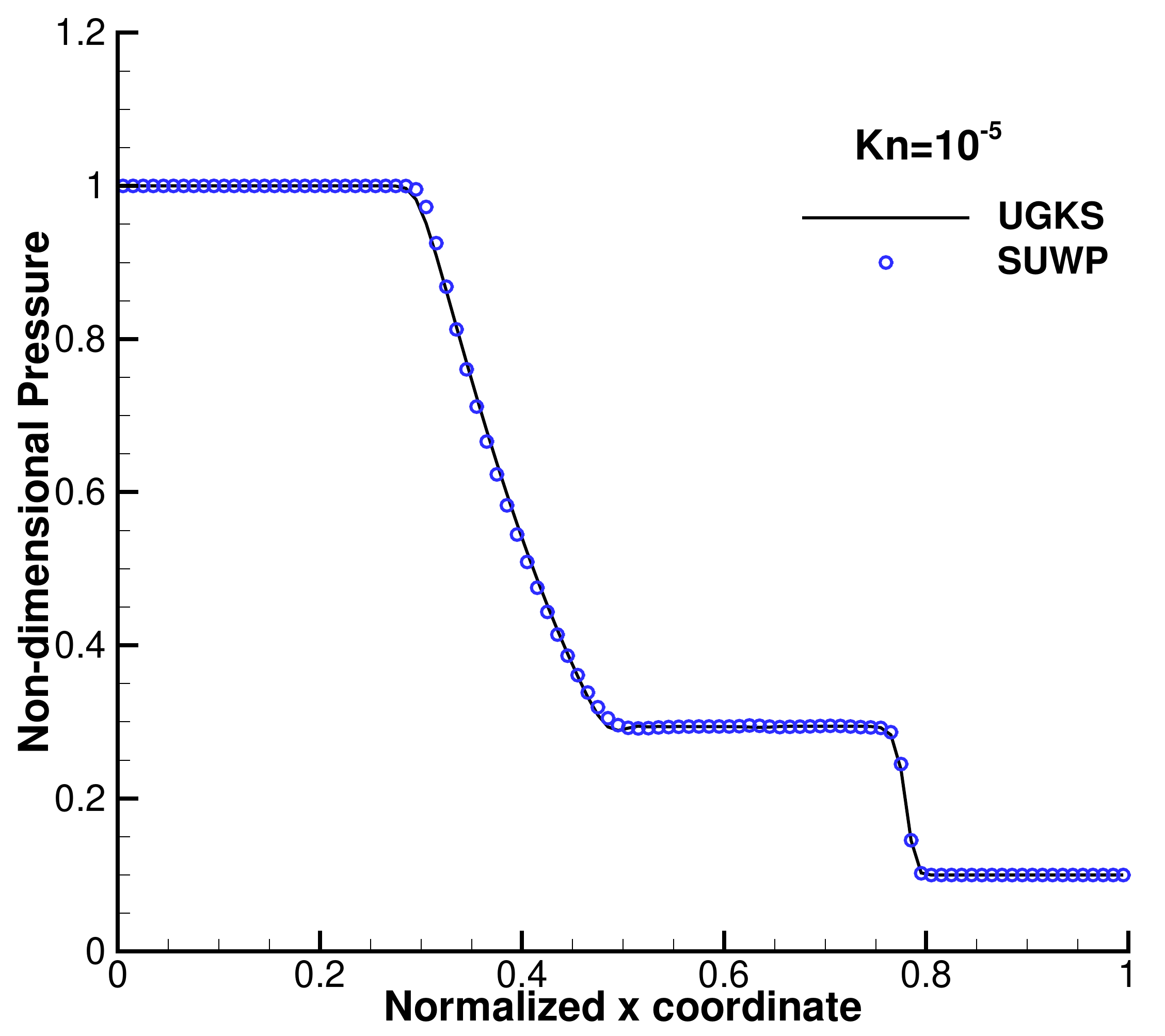}
}
\caption{\label{Fig:sodn5} Profiles of Sod shock-tube at $t=0.15$ with $\rm{Kn}=10^{-5}$. The SUWP solution is shown in symbols and the reference UGKS solution is shown in solid lines.}
\end{figure}

The sod shock-tube case with different Kn numbers ($10^{-1}$, $10^{-3}$ and $10^{-5}$) are used to examine the validity of SUWP in unsteady multi-scale flow simulations covering the transitional, slip and continuum flow regimes. The computational domain is $\left[-0.5,0.5\right]$. Being same with the benchmark solution from Ref.~\cite{LiuUnified2020}, the VHS model with $\omega=0.81$ is used. The reference length is chosen as the length of flow domain. The m.f.p. for Kn number is from the initial condition on the left half of the domain.
The initial condition is
\begin{equation}
\left( {\rho ,u,p} \right) = \left\{ {\begin{array}{*{20}{c}}
{\left( {1,0,1} \right),{\rm{   }}x \le 0}\\
{\left( {0.125,0,0.1} \right),{\rm{   }}x > 0}
\end{array}} \right.
\end{equation}
The density, velocity, temperature and pressure profiles with different Kn numbers at $t=0.15$ are illustrated in Fig.~\ref{Fig:sodn1}, Fig.~\ref{Fig:sodn3} and Fig.~\ref{Fig:sodn5}. The profiles calculated by the SUWP match well with those from the benchmark solution. In this case, the number of free-transport molecules emerging from the macroscopic variables is adjusted to make the total number of particles in the cell is around 300. For Kn$=0.1$ case, the weight of the particle method is large, and stochastic fluctuation can be seen in the profiles (Fig.~\ref{Fig:sodn1}). For Kn$=10^{-5}$ case (in continuum regime), the weight of the particle method can be neglected, and no obvious stochastic fluctuation can be seen in the profiles.

\subsection{Hypersonic flow around cylinder}
The hypersonic flow around cylinder with four Kn numbers (10, 1, 0.1, 0.01) is calculated using SUWP in this section. Since the Kn number is from 10 to 0.01, the free molecular, transitional, and slip flow regimes are covered by this test case.

The working gas is Argon with $\omega=0.81$ and $\gamma=5/3$. VHS model for intermolecular potential is used (VSS model with $\alpha=1$). The radius of the cylinder $L$ is chosen as the reference length ($L_{ref}=L$). The reference temperature is $T_{ref} = T_{\infty}$. Subscript $\infty$ represents the inflow physical variables in this case. The reference density and velocity are $\rho_{ref}=\rho_{\infty}$ and $u_{ref}=\sqrt{2RT_{ref}}$, respectively. The computational domain is enclosed by a circle with a radius of 15L. The domain is decomposed into $75 \times 62$ cells, where 75 cells are used in the radial direction and 62 cells are arranged along the wall of cylinder. The height of the cell adjacent to the wall boundary is 0.05, and its length is 0.1. The CFL number is chosen as 0.5 for all Kn numbers. The number of free molecules emerging from the hydrodynamic macroscopic variables is set $100$ times the rate of their mass to the total mass in cell.

\begin{figure}
\centering
\subfigure[\label{Fig:cylinderKn10D} Density]{
\includegraphics[width=0.45\textwidth]{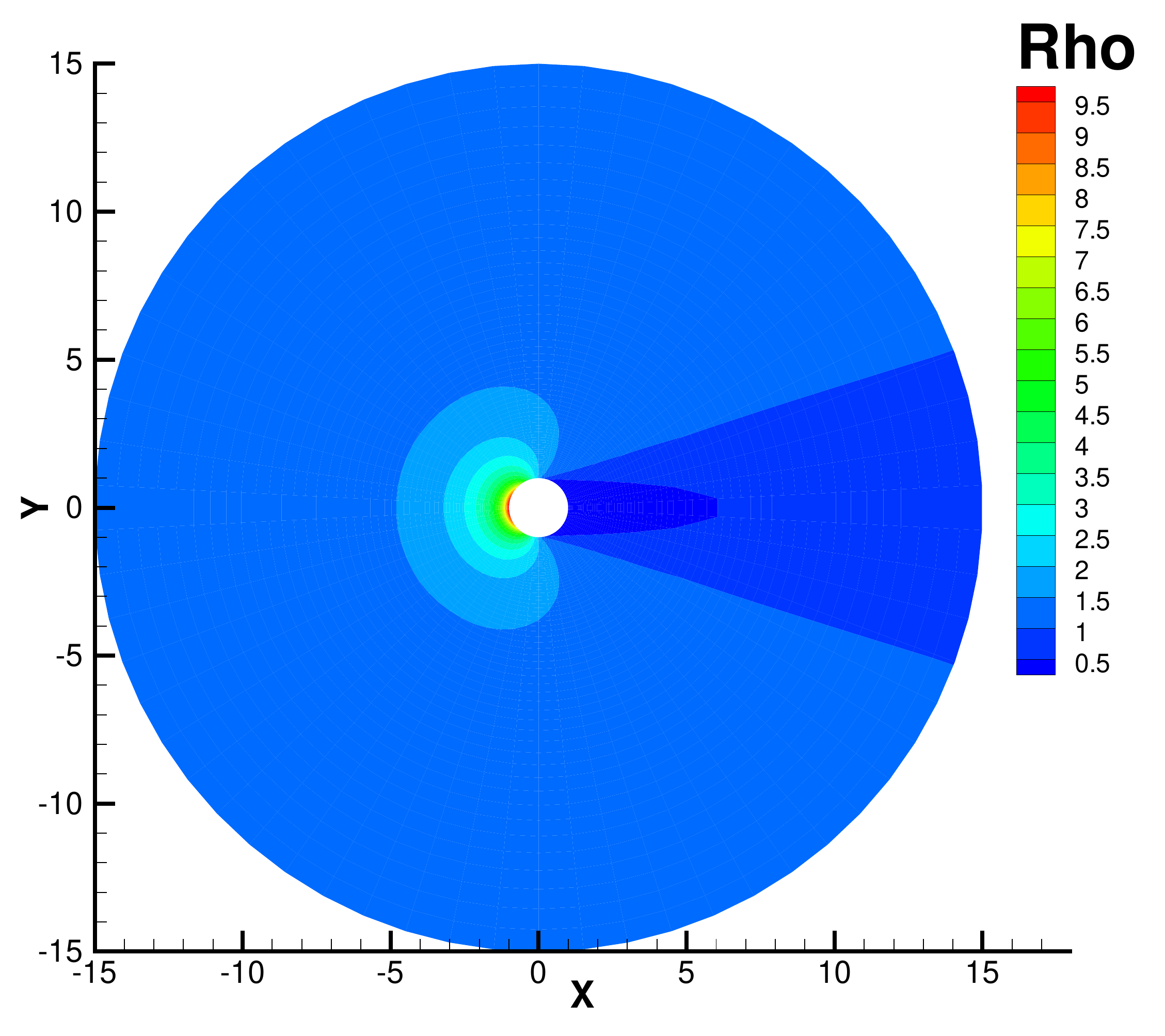}
}\hspace{0.05\textwidth}%
\subfigure[\label{Fig:cylinderKn10U} U-velocity]{
\includegraphics[width=0.45\textwidth]{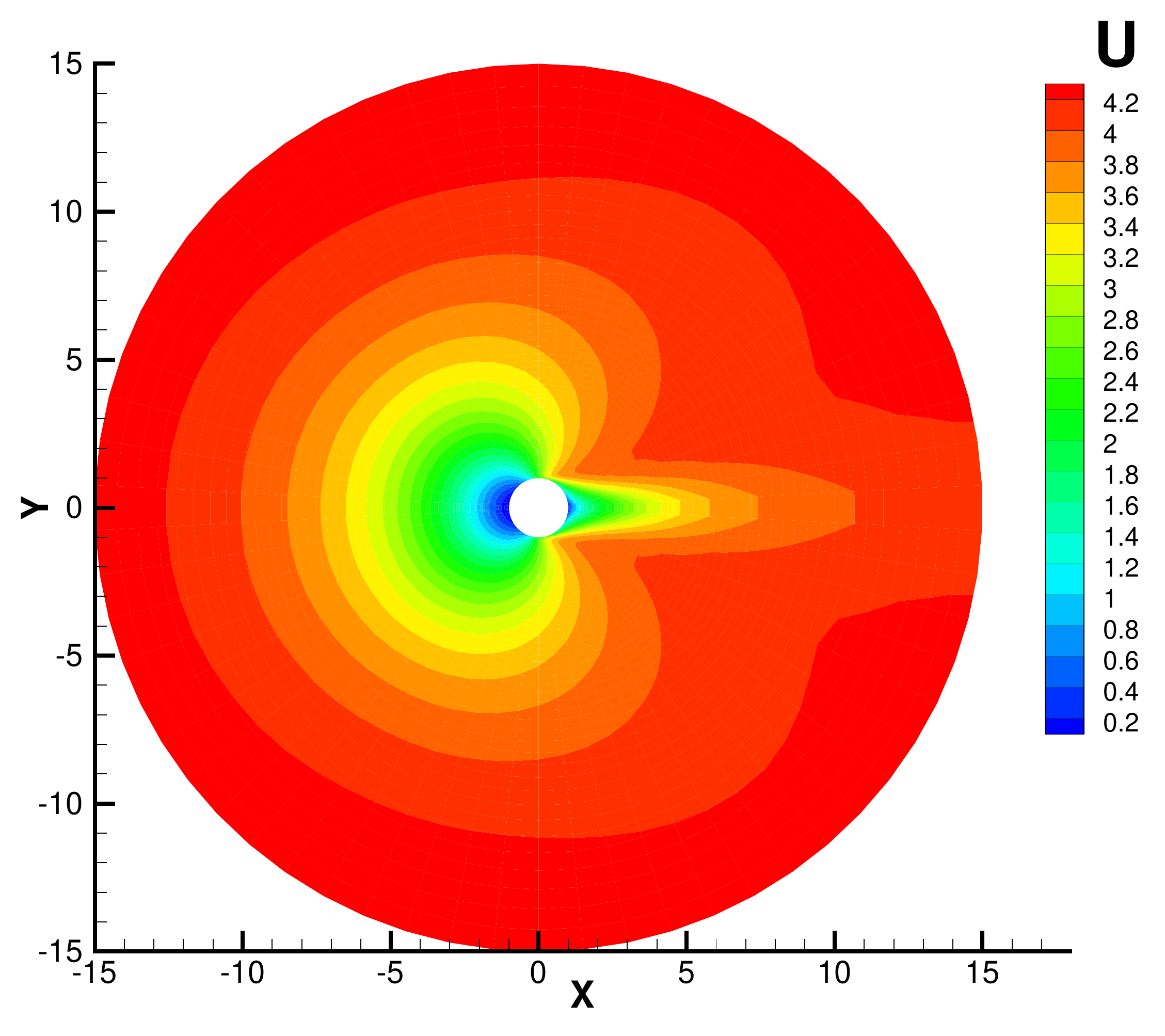}
}\\
\subfigure[\label{Fig:cylinderKn10V} V-velocity]{
\includegraphics[width=0.45\textwidth]{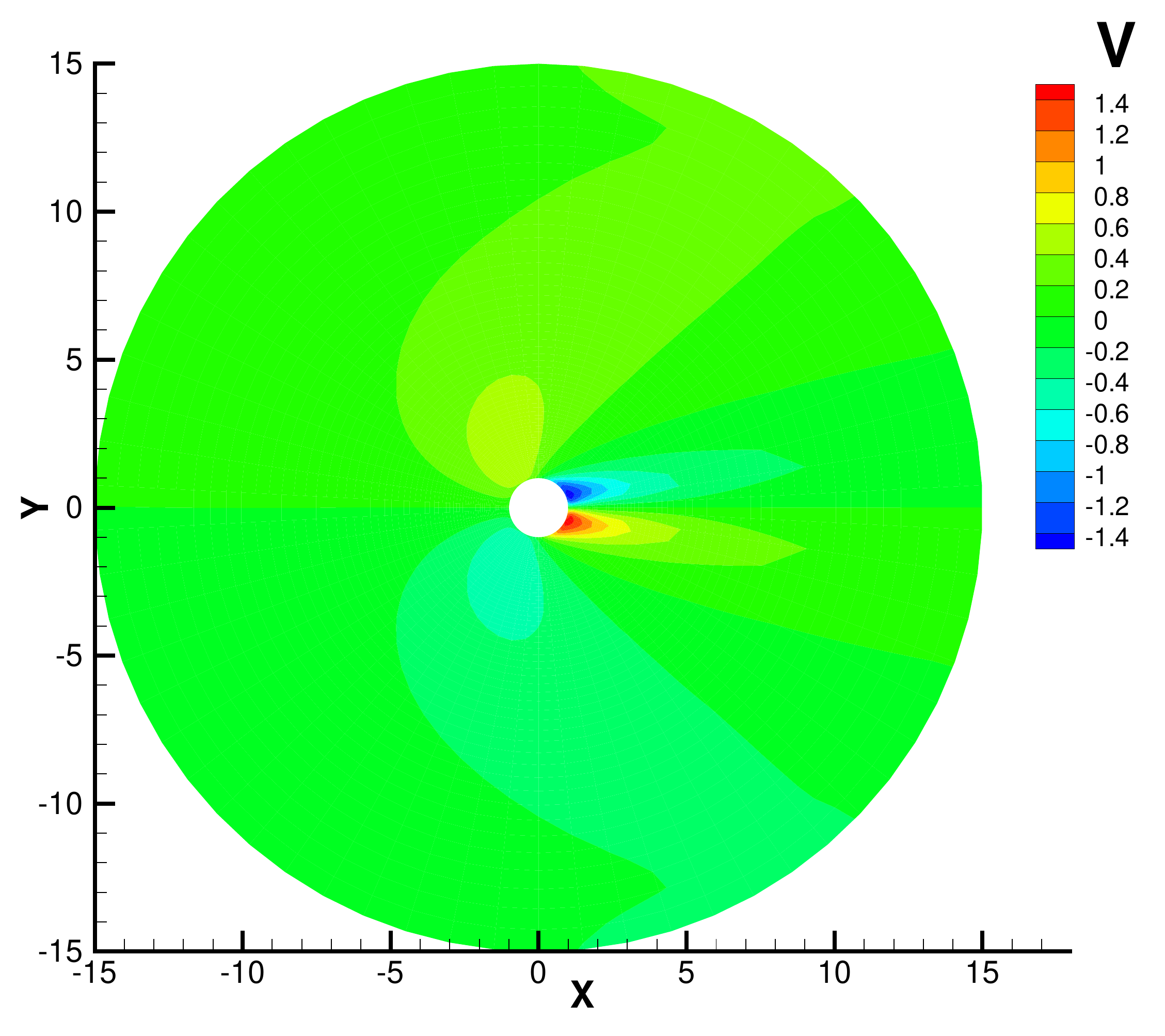}
}\hspace{0.05\textwidth}%
\subfigure[\label{Fig:cylinderKn10T} Temperature]{
\includegraphics[width=0.45\textwidth]{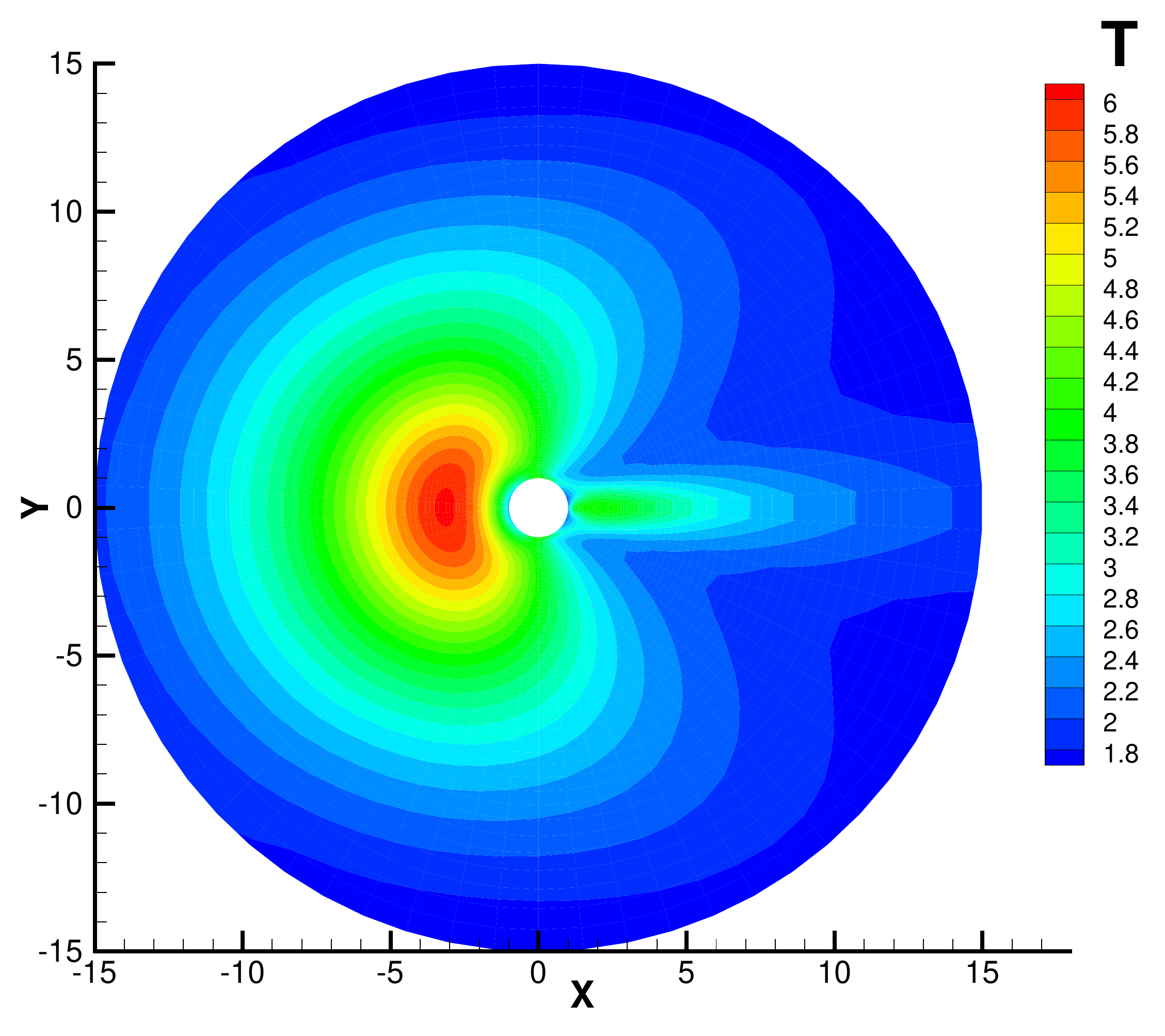}
}
\caption{\label{Fig:cylinderKn10} Ma$=5$ flow around cylinder at Kn$=10$}
\end{figure}

\begin{figure}
\centering
\subfigure[\label{Fig:cylinderKn1D} Density]{
\includegraphics[width=0.45\textwidth]{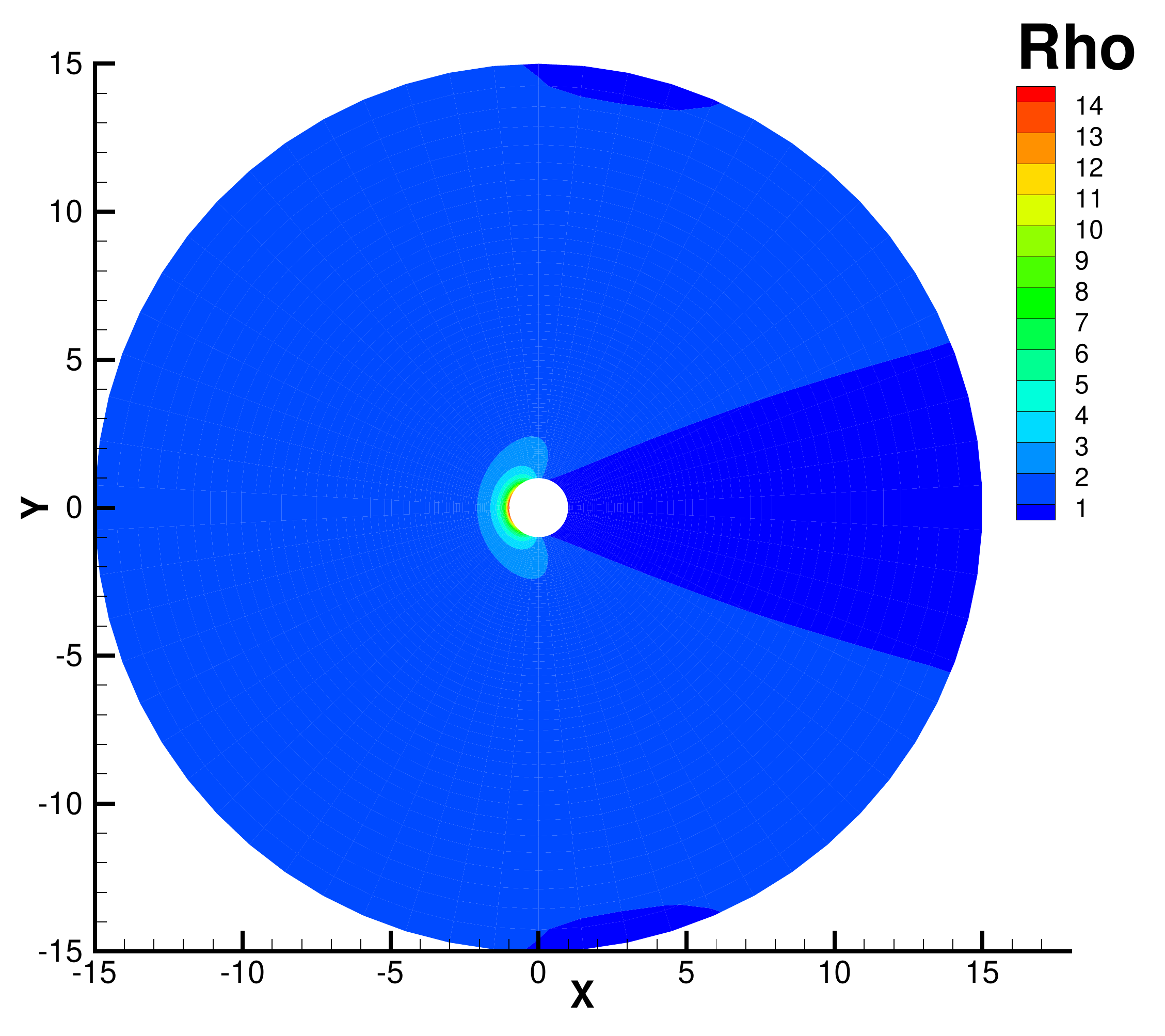}
}\hspace{0.05\textwidth}%
\subfigure[\label{Fig:cylinderKn1U} U-velocity]{
\includegraphics[width=0.45\textwidth]{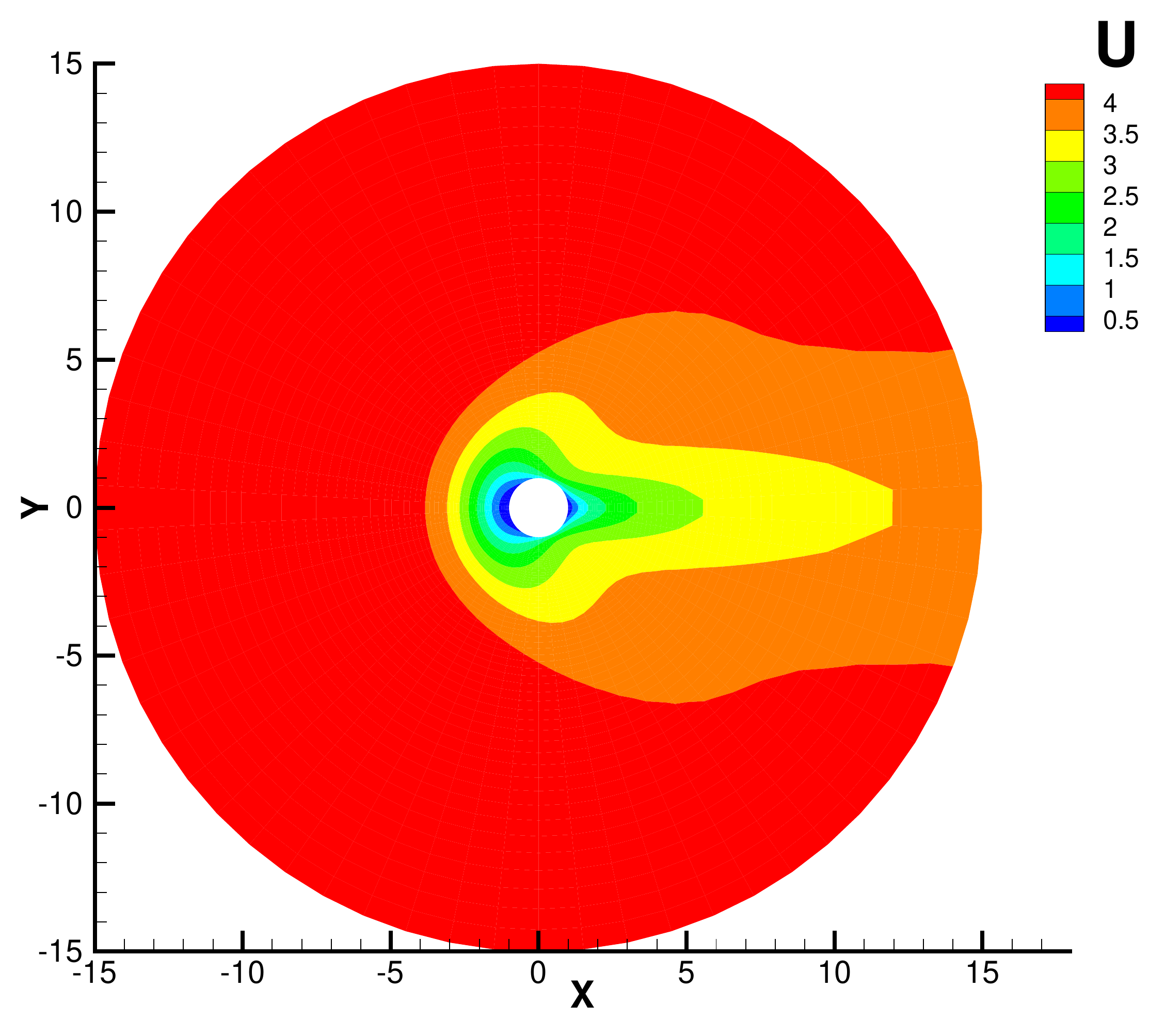}
}\\
\subfigure[\label{Fig:cylinderKn1V} V-velocity]{
\includegraphics[width=0.45\textwidth]{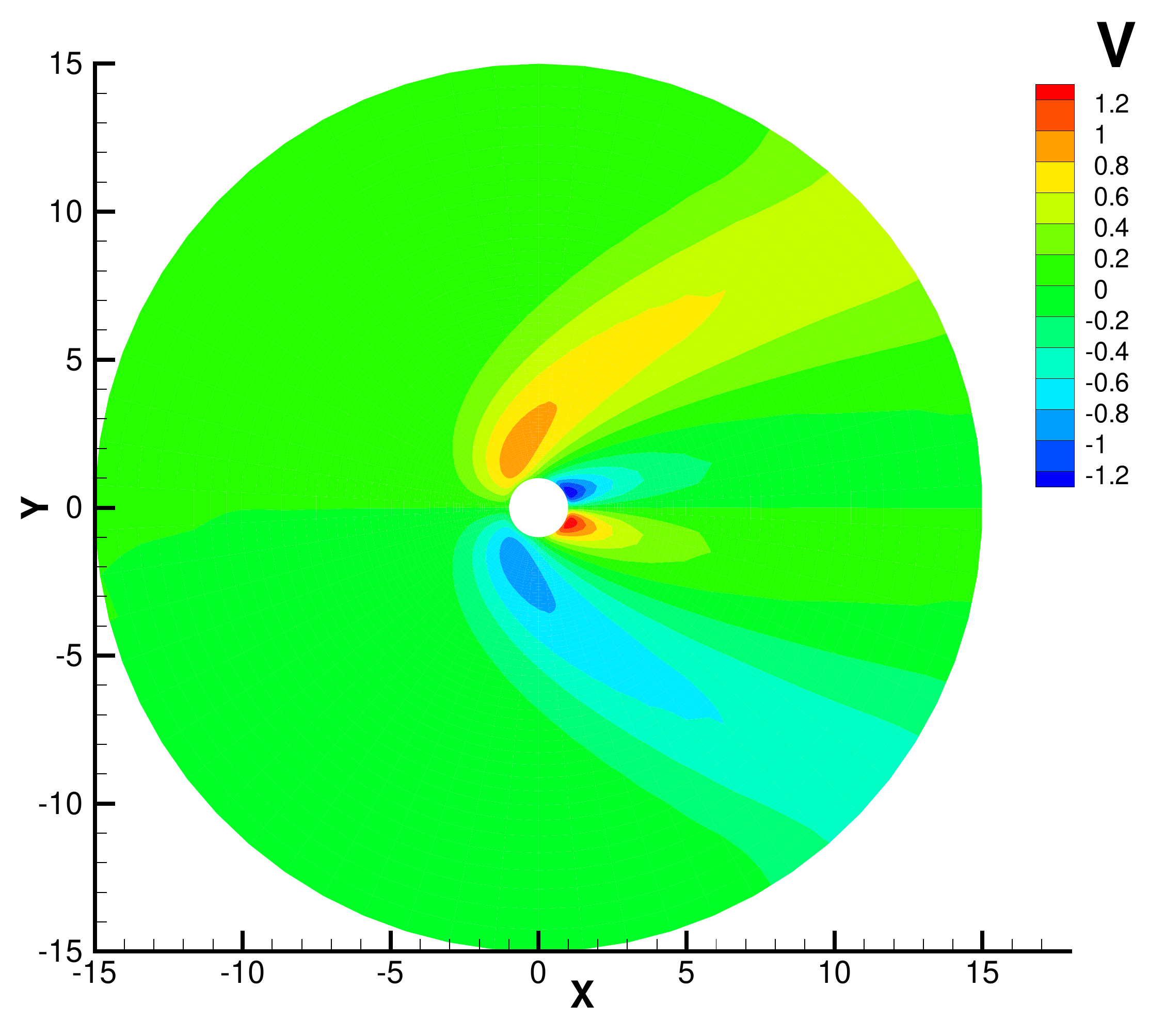}
}\hspace{0.05\textwidth}%
\subfigure[\label{Fig:cylinderKn1T} Temperature]{
\includegraphics[width=0.45\textwidth]{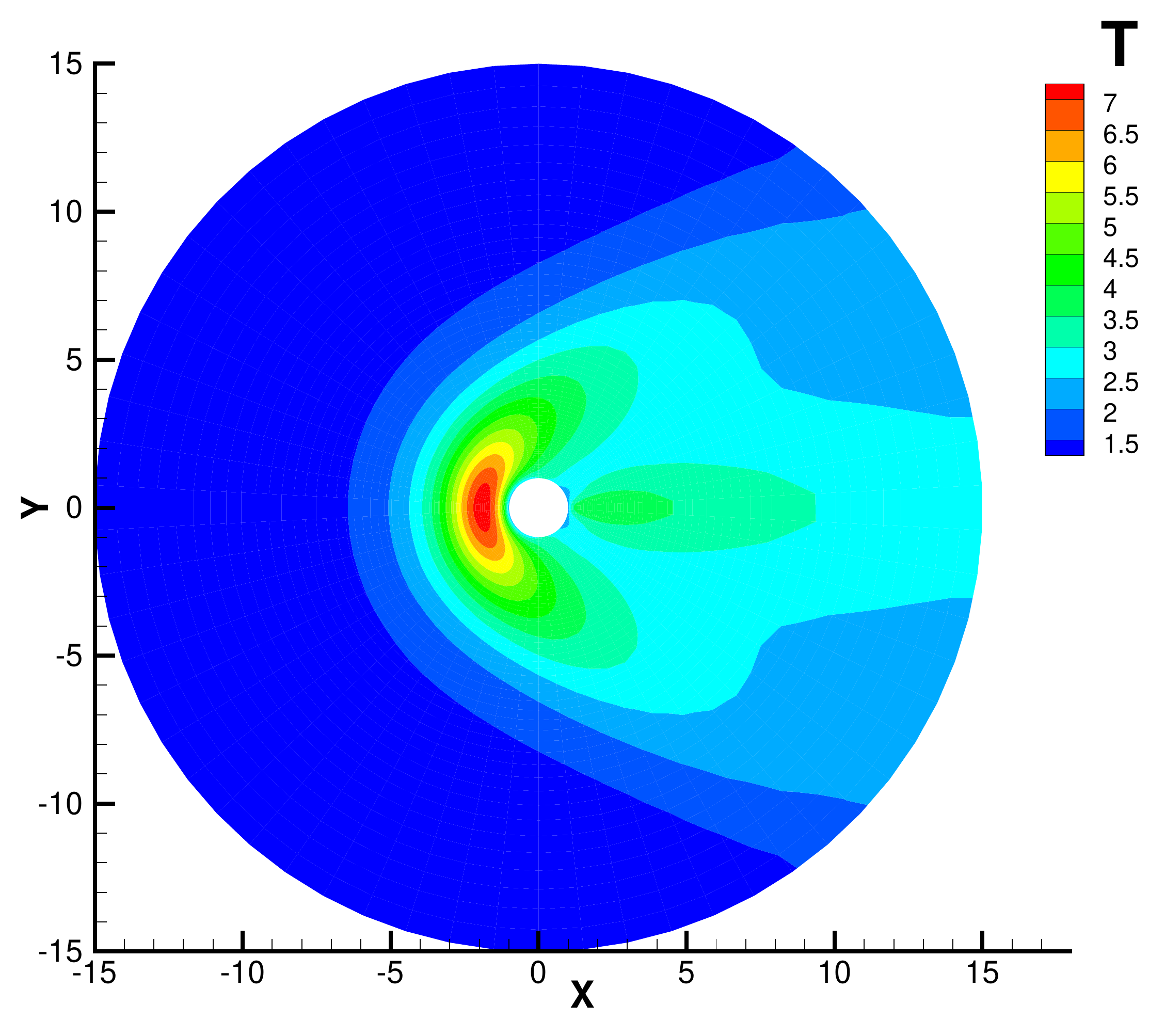}
}
\caption{\label{Fig:cylinderKn1} Ma$=5$ flow around the cylinder at Kn$=1$}
\end{figure}

\begin{figure}
\centering
\subfigure[\label{Fig:cylinderKn01D} Density]{
\includegraphics[width=0.45\textwidth]{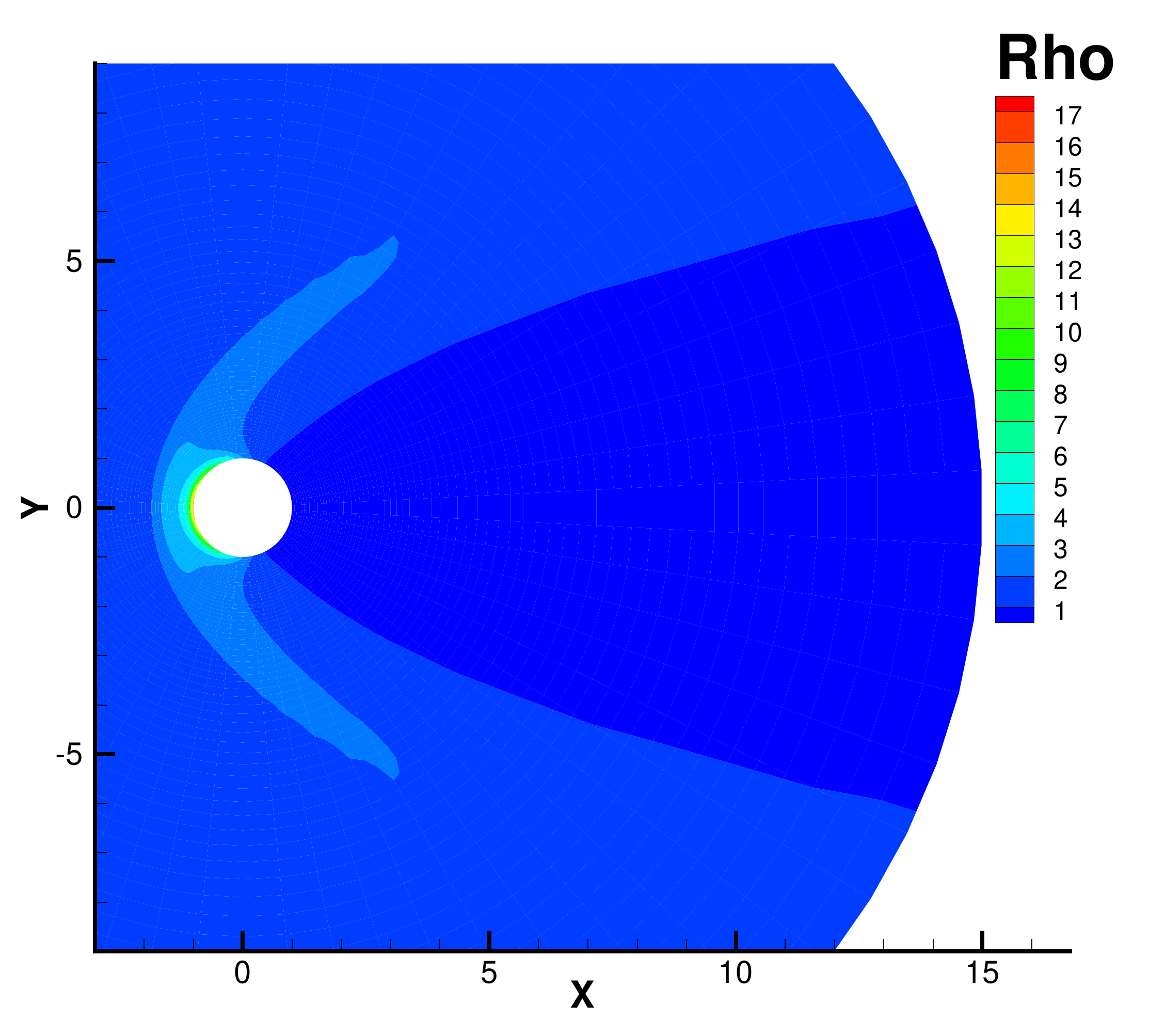}
}\hspace{0.05\textwidth}%
\subfigure[\label{Fig:cylinderKn01U} U-velocity]{
\includegraphics[width=0.45\textwidth]{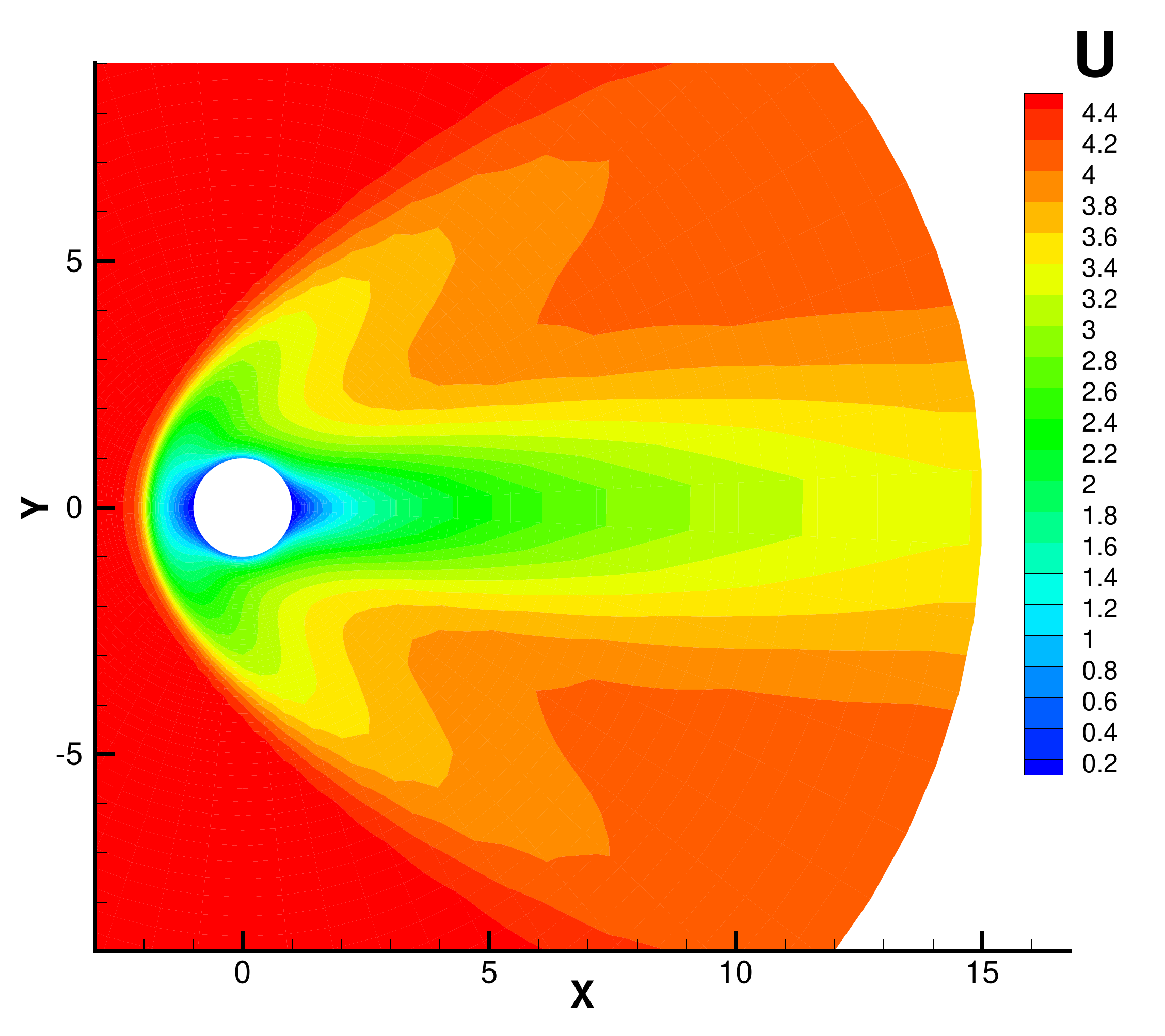}
}\\
\subfigure[\label{Fig:cylinderKn01V} V-velocity]{
\includegraphics[width=0.45\textwidth]{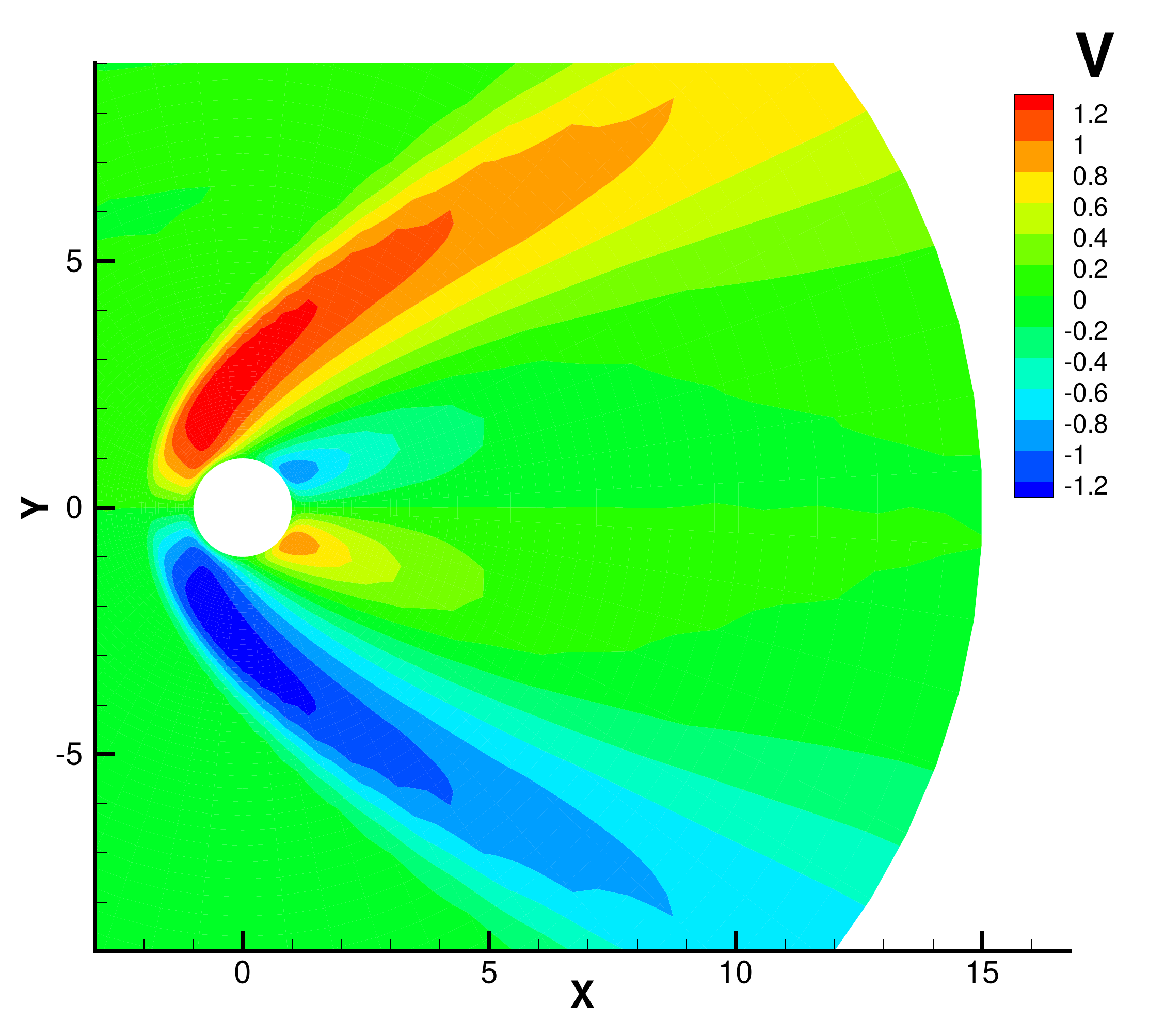}
}\hspace{0.05\textwidth}%
\subfigure[\label{Fig:cylinderKn01T} Temperature]{
\includegraphics[width=0.45\textwidth]{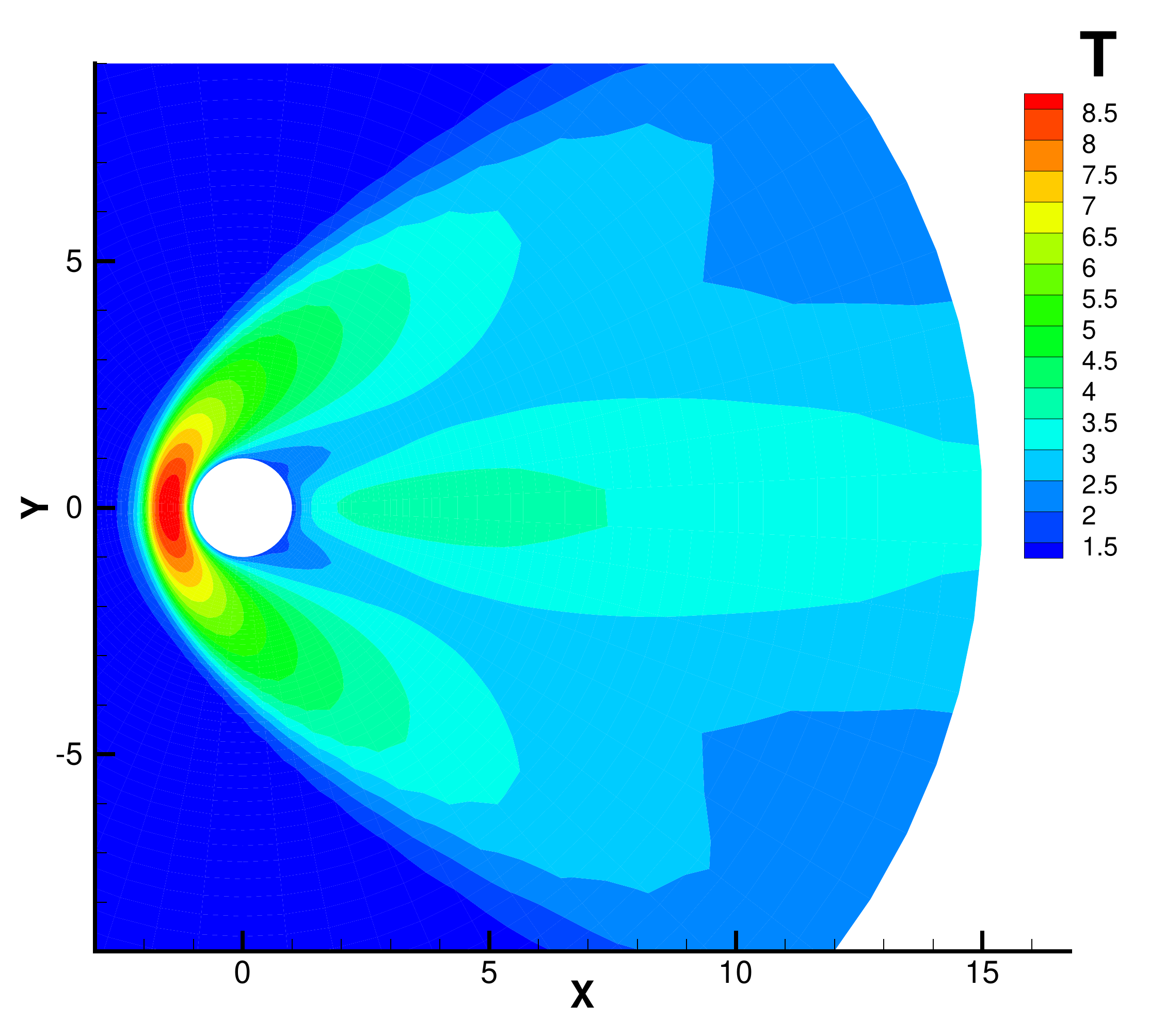}
}
\caption{\label{Fig:cylinderKn01} Ma$=5$ flow around the cylinder at Kn$=0.1$}
\end{figure}

\begin{figure}
\centering
\subfigure[\label{Fig:cylinderKn001D} Density]{
\includegraphics[width=0.45\textwidth]{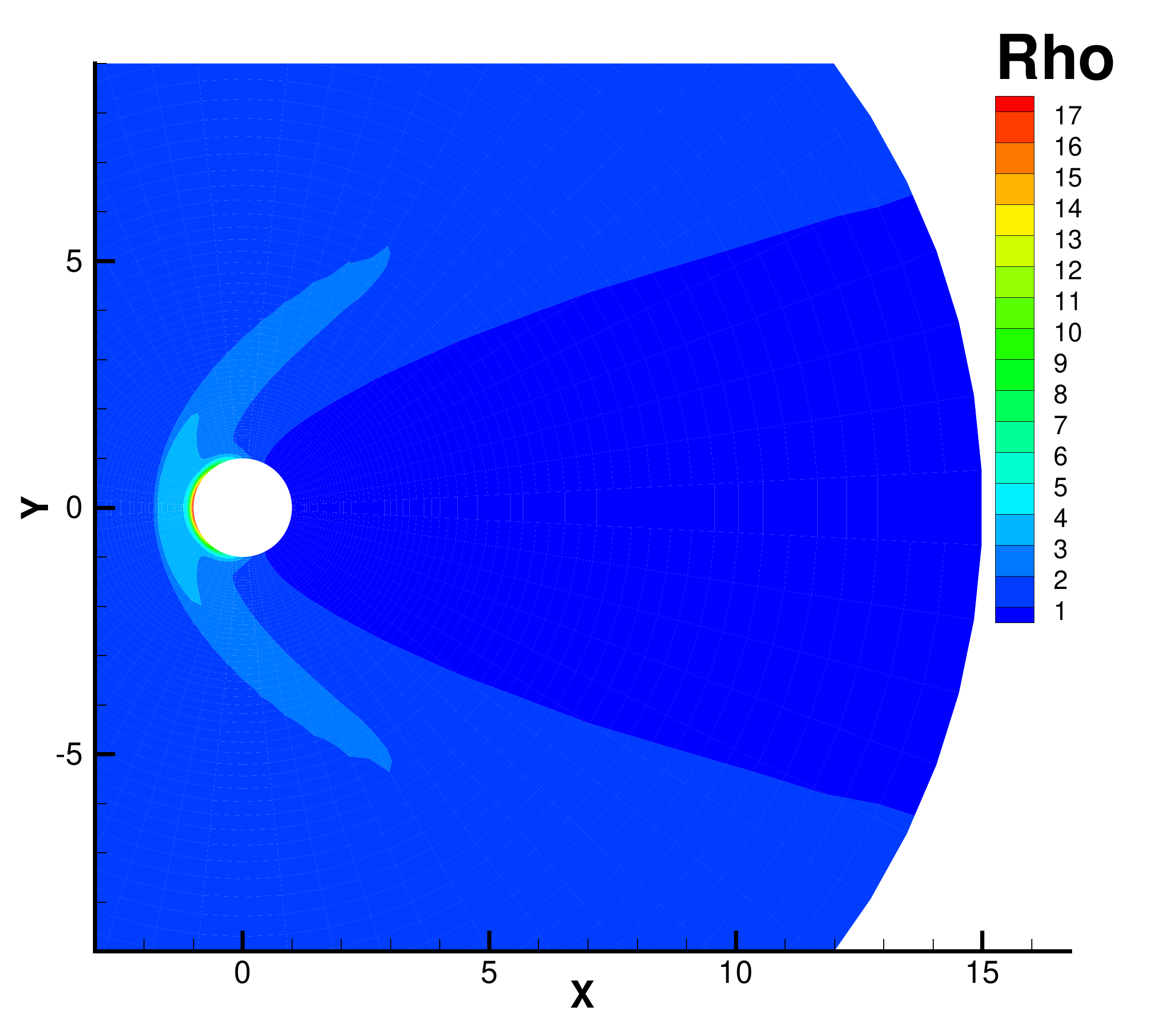}
}\hspace{0.05\textwidth}%
\subfigure[\label{Fig:cylinderKn001U} U-velocity]{
\includegraphics[width=0.45\textwidth]{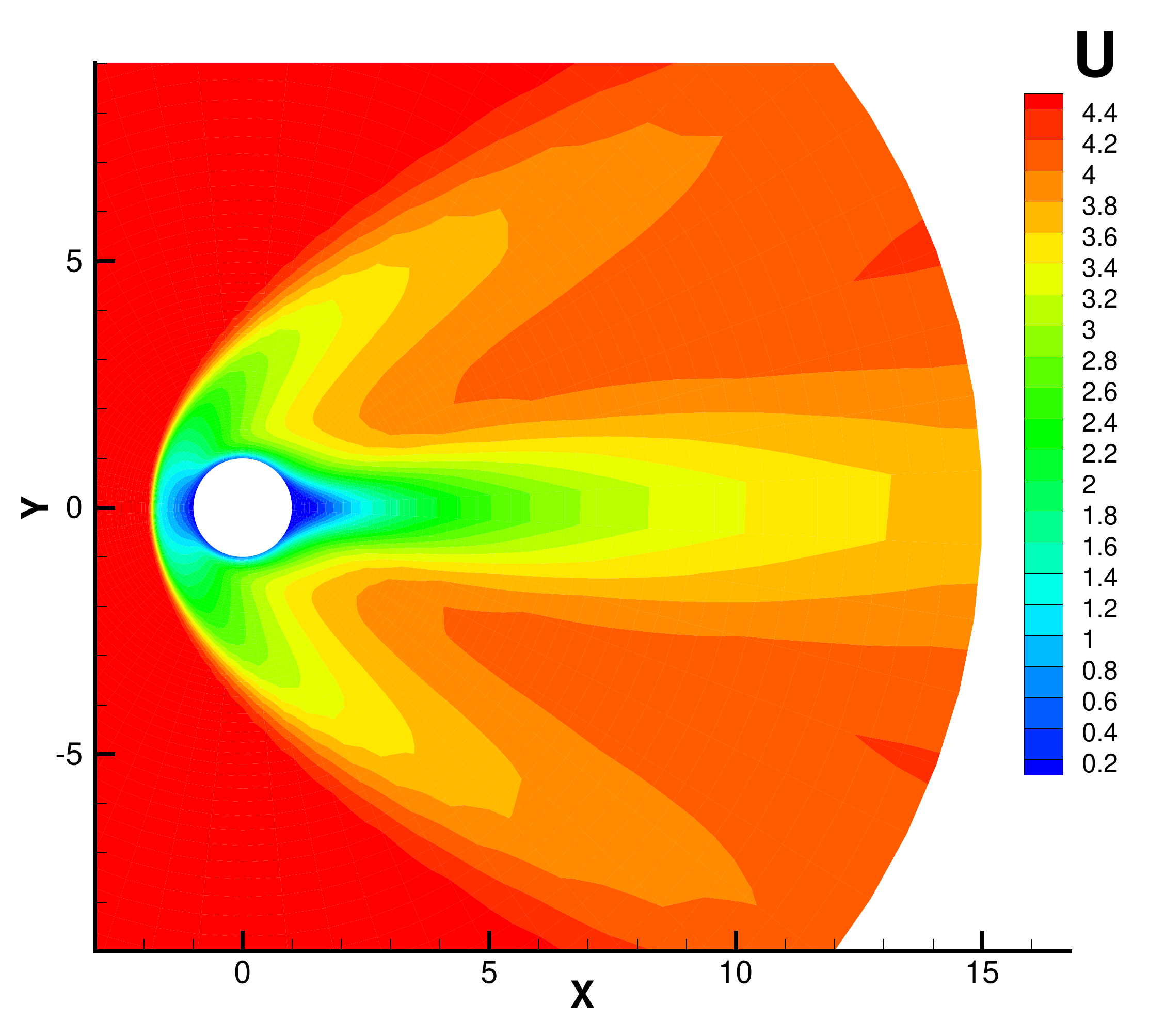}
}\\
\subfigure[\label{Fig:cylinderKn001V} V-velocity]{
\includegraphics[width=0.45\textwidth]{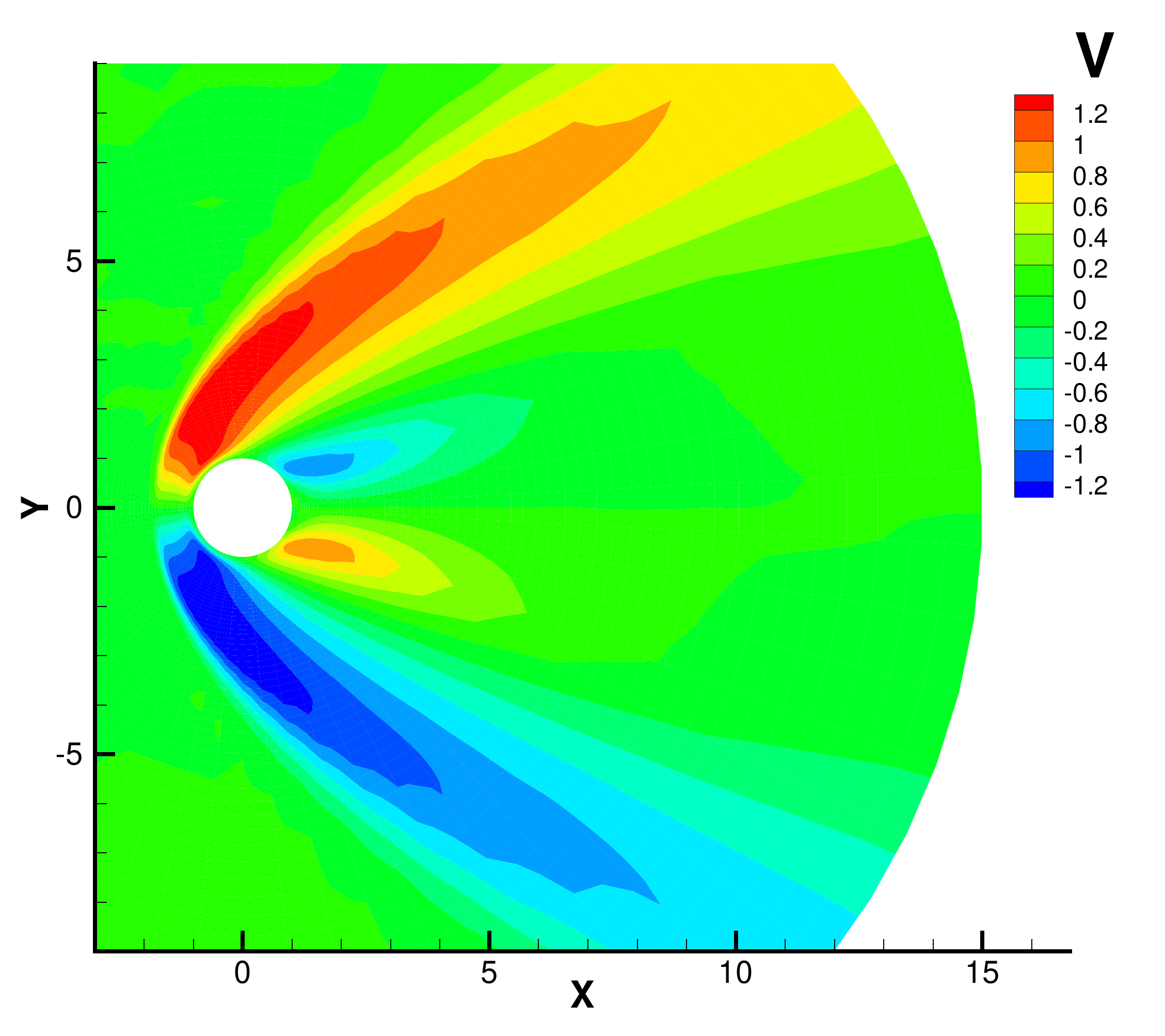}
}\hspace{0.05\textwidth}%
\subfigure[\label{Fig:cylinderKn001T} Temperature]{
\includegraphics[width=0.45\textwidth]{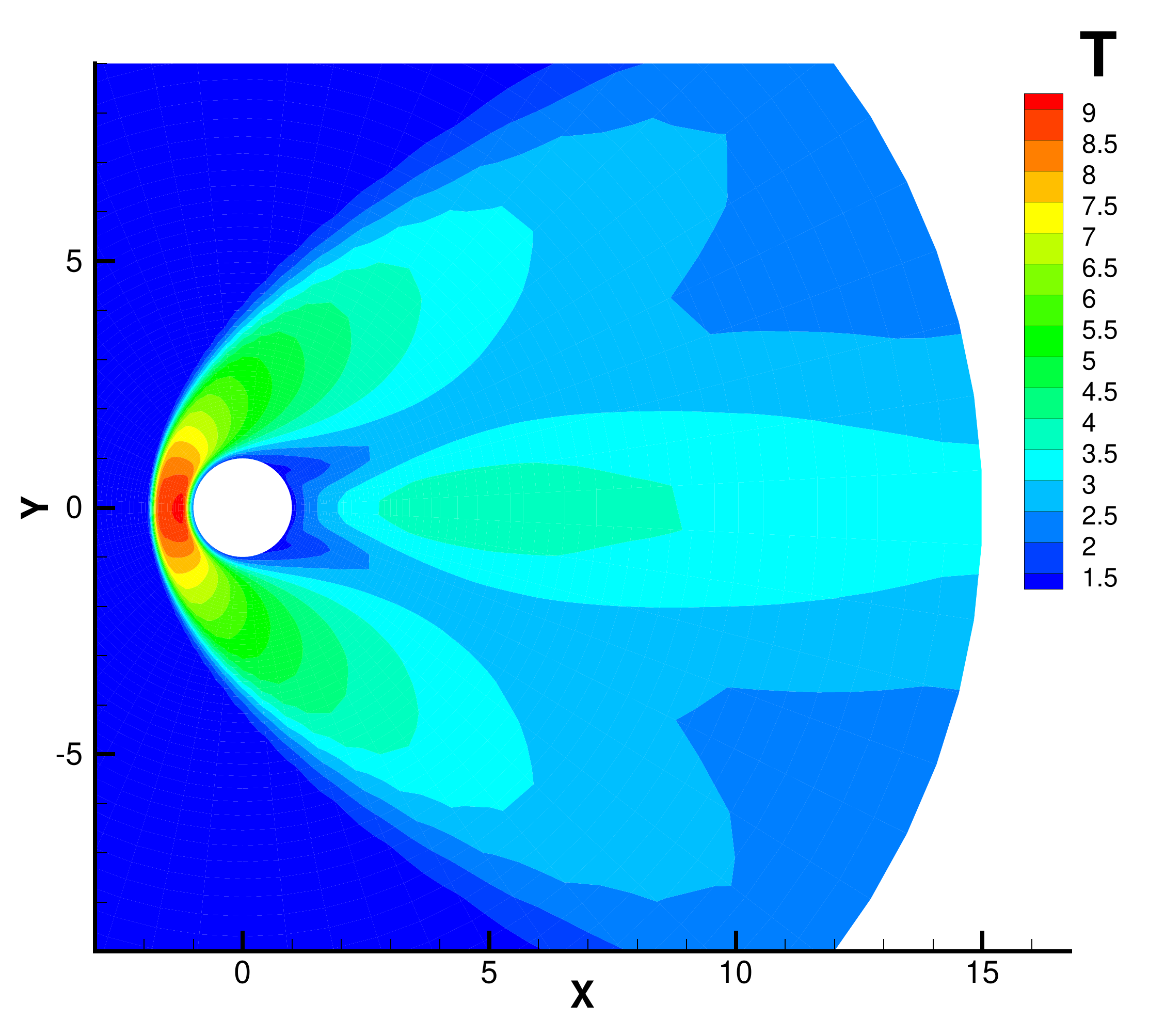}
}
\caption{\label{Fig:cylinderKn001} Ma$=5$ flow around the cylinder at Kn$=0.01$}
\end{figure}

The flow fields (The contours of density, U-velocity, V-velocity and Temperature) at Kn number 10, 1, 0.1 and 0.01 are illustrated in Fig.~\ref{Fig:cylinderKn10}, Fig.~\ref{Fig:cylinderKn1}, Fig.~\ref{Fig:cylinderKn01} and Fig.~\ref{Fig:cylinderKn001}, respectively. In the Kn=10 case where the flow is governed by free-transport mechanism, the flow is smooth and shock wave can not be identified in the flow field. When Kn number decreases, the bow shock becomes obvious, and its structure becomes clear. At Kn=0.01, the structure of bow shock is already the same with that in the continuum regime. Since as the Kn number increases, the free-transport mechanism prevails gradually, then molecules can take their information to a large distance without collision. This leads to a large regime influenced by the solid wall (cylinder). This phenomenon can be seen from the temperature contours easily (Fig.~\ref{Fig:cylinderKn001T}, Fig.~\ref{Fig:cylinderKn01T}, Fig.~\ref{Fig:cylinderKn1T} and Fig.~\ref{Fig:cylinderKn10T}).

\begin{figure}
\centering
\subfigure[\label{Fig:cchr10} Kn=10]{
\includegraphics[width=0.45\textwidth]{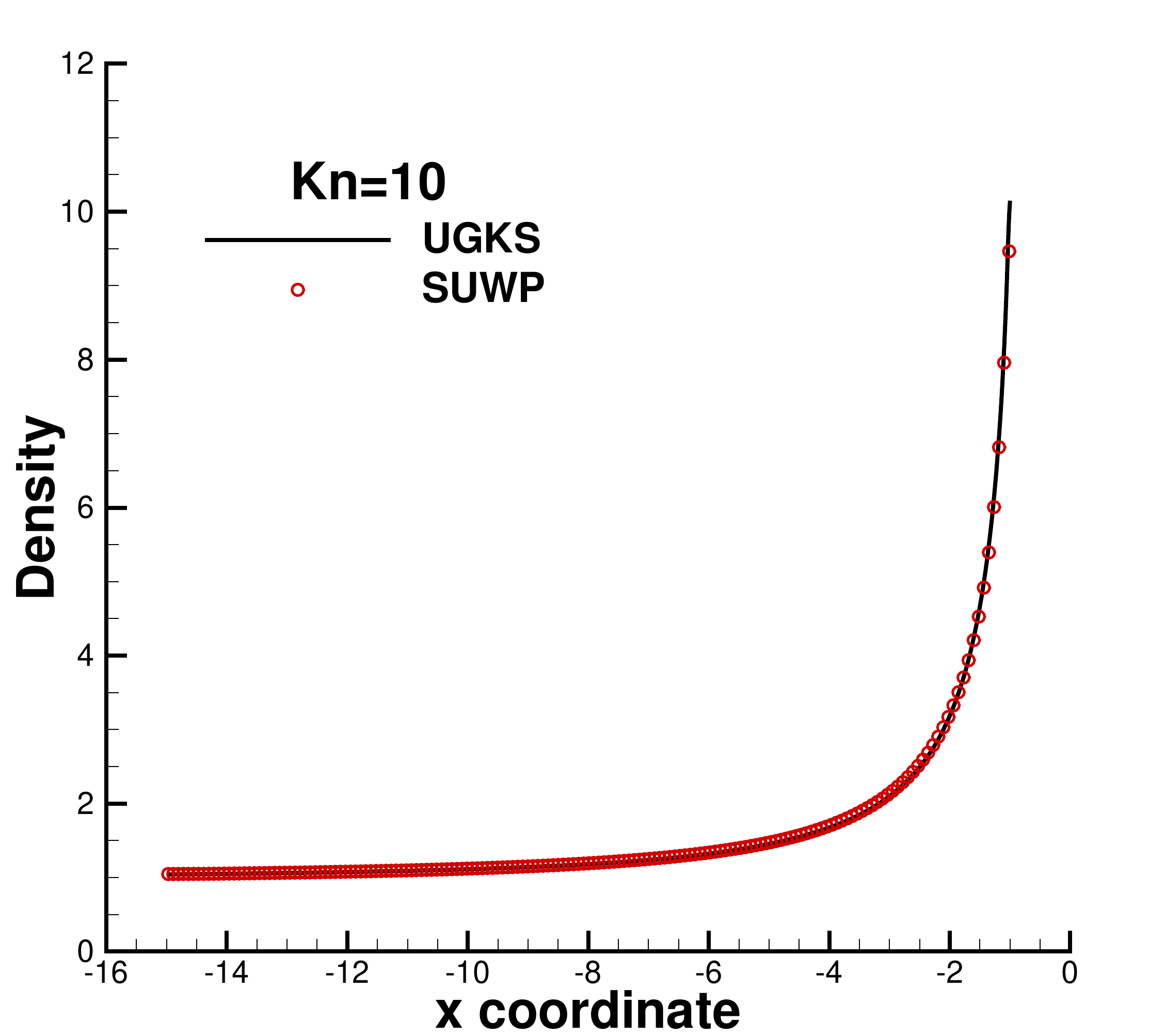}
}\hspace{0.05\textwidth}%
\subfigure[\label{Fig:cchr1} Kn=1]{
\includegraphics[width=0.45\textwidth]{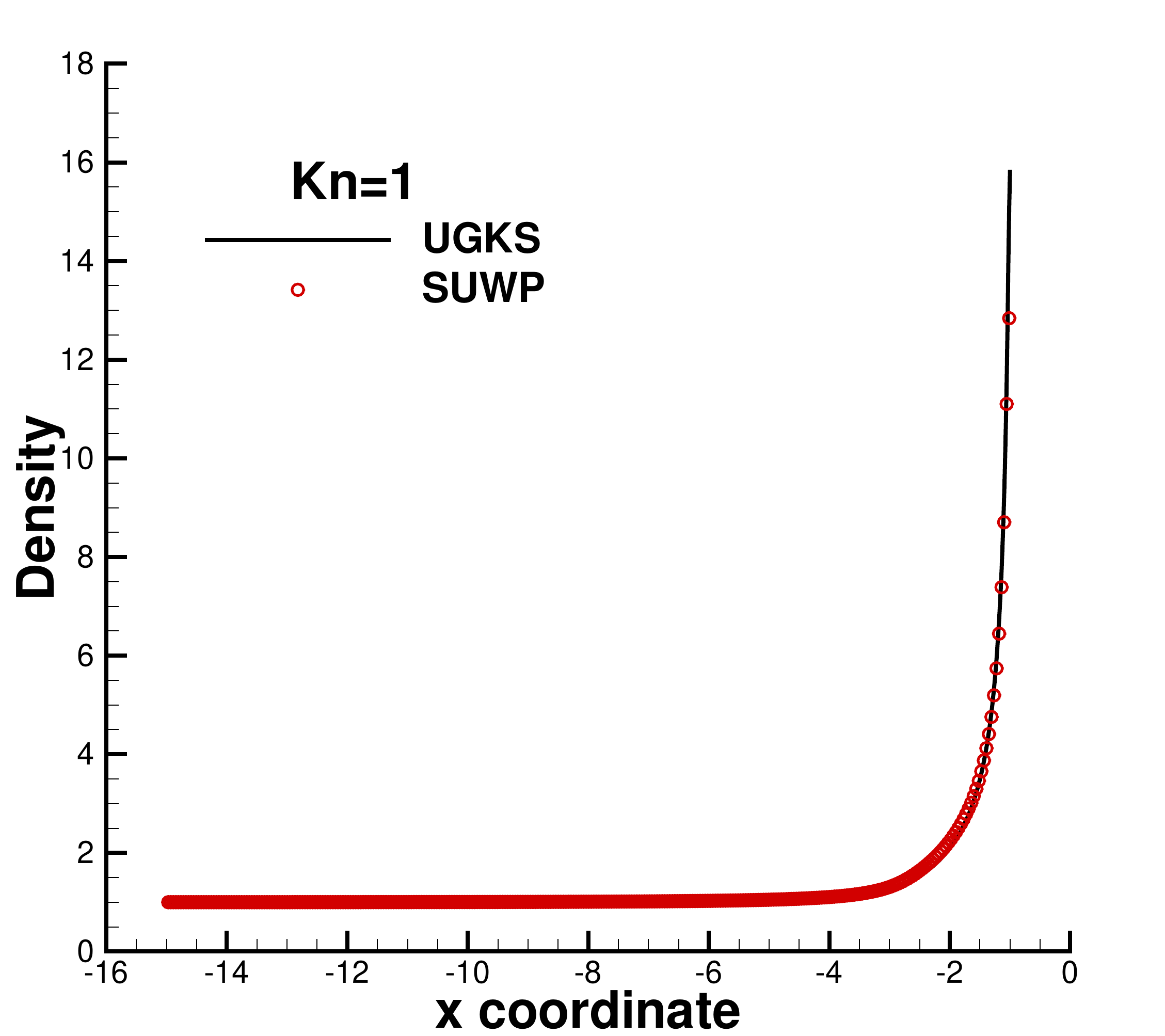}
}\\
\subfigure[\label{Fig:cchr01} Kn=0.1]{
\includegraphics[width=0.45\textwidth]{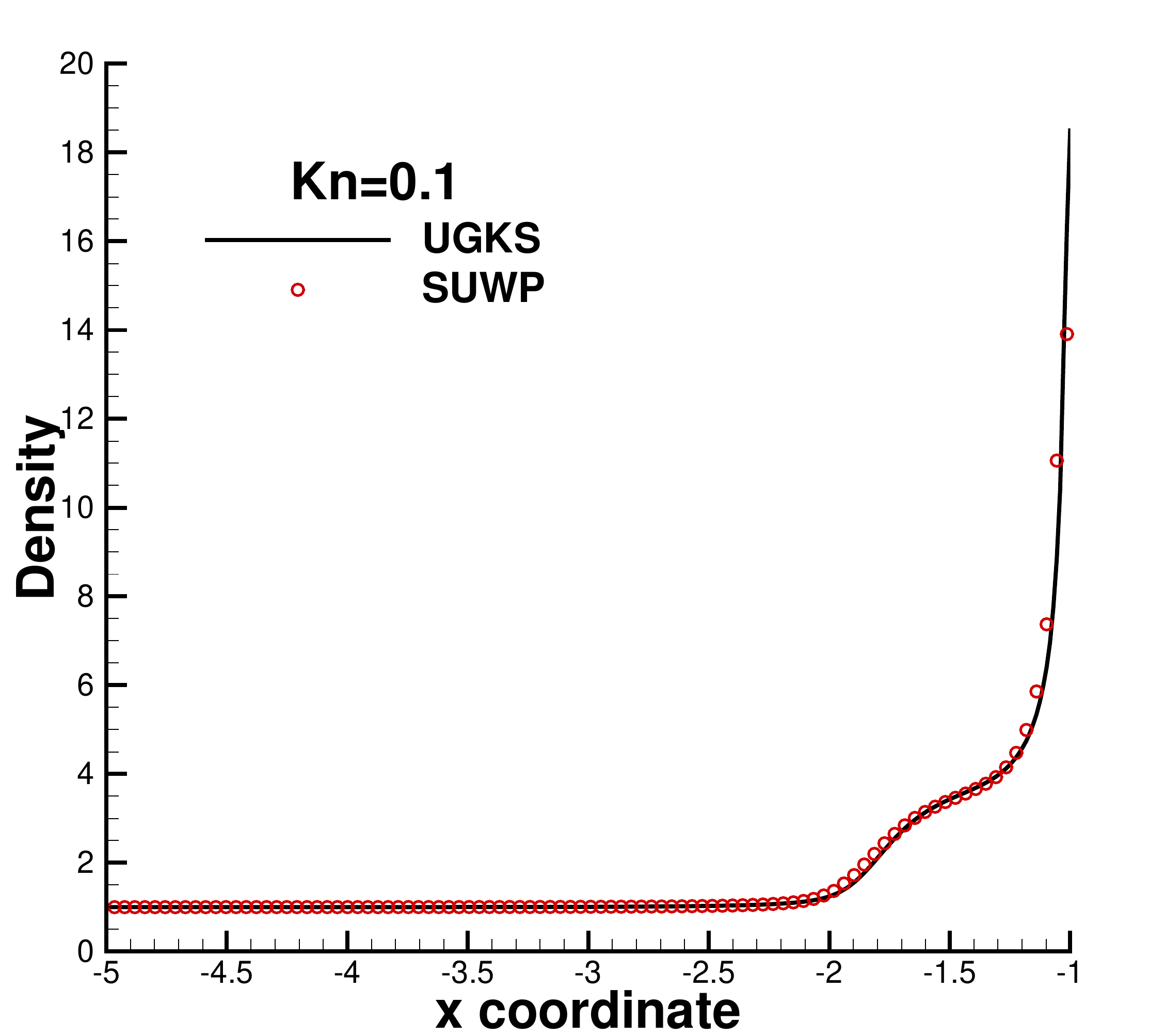}
}\hspace{0.05\textwidth}%
\subfigure[\label{Fig:cchr001} Kn=0.01]{
\includegraphics[width=0.45\textwidth]{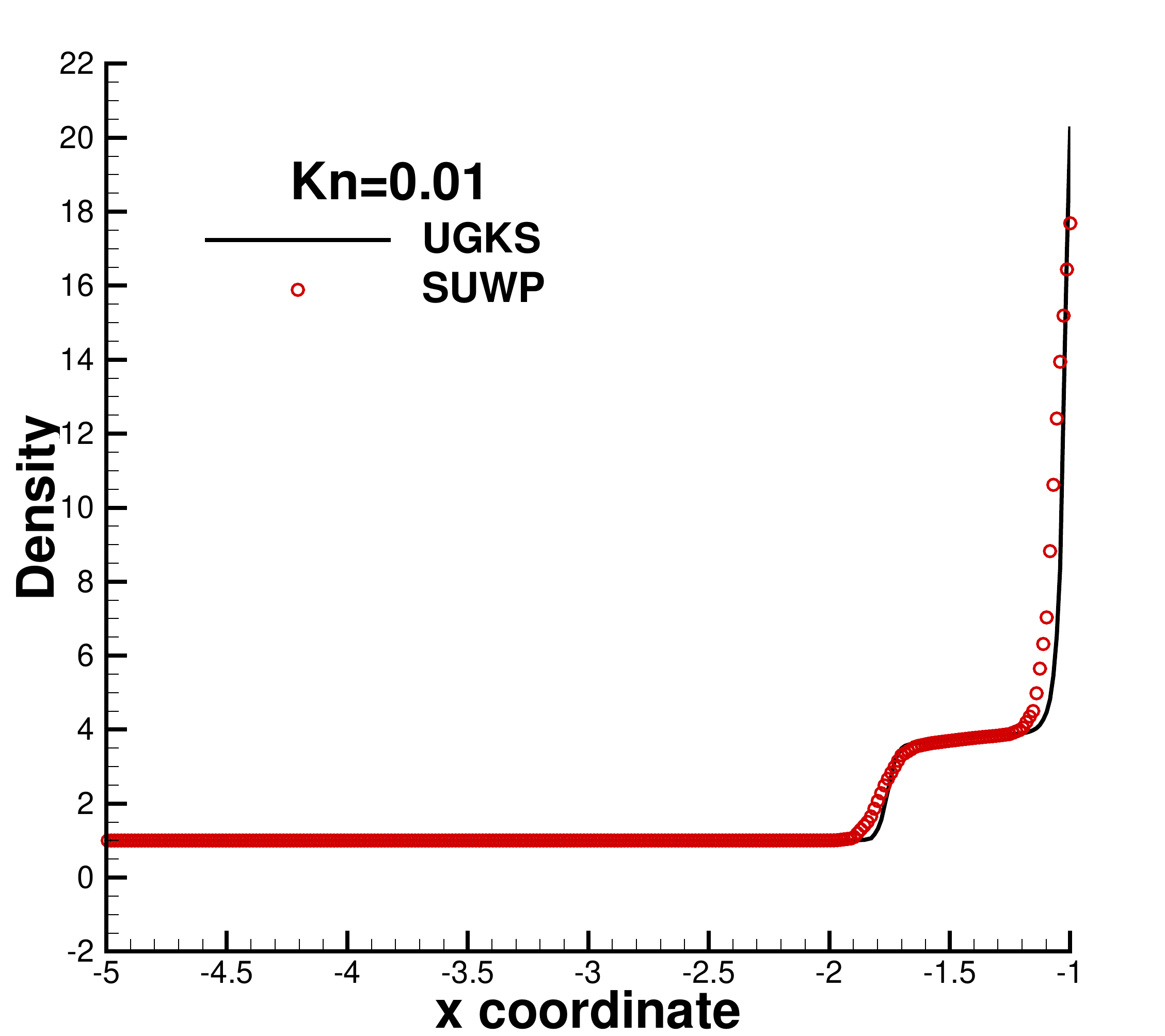}
}
\caption{\label{Fig:cchr} The density distribution along the stagnation lines of Ma$=$5 cylinder at Kn$=$ 10, 1, 0.1 and 0.01}
\end{figure}

\begin{figure}
\centering
\subfigure[\label{Fig:cchu10} Kn=10]{
\includegraphics[width=0.45\textwidth]{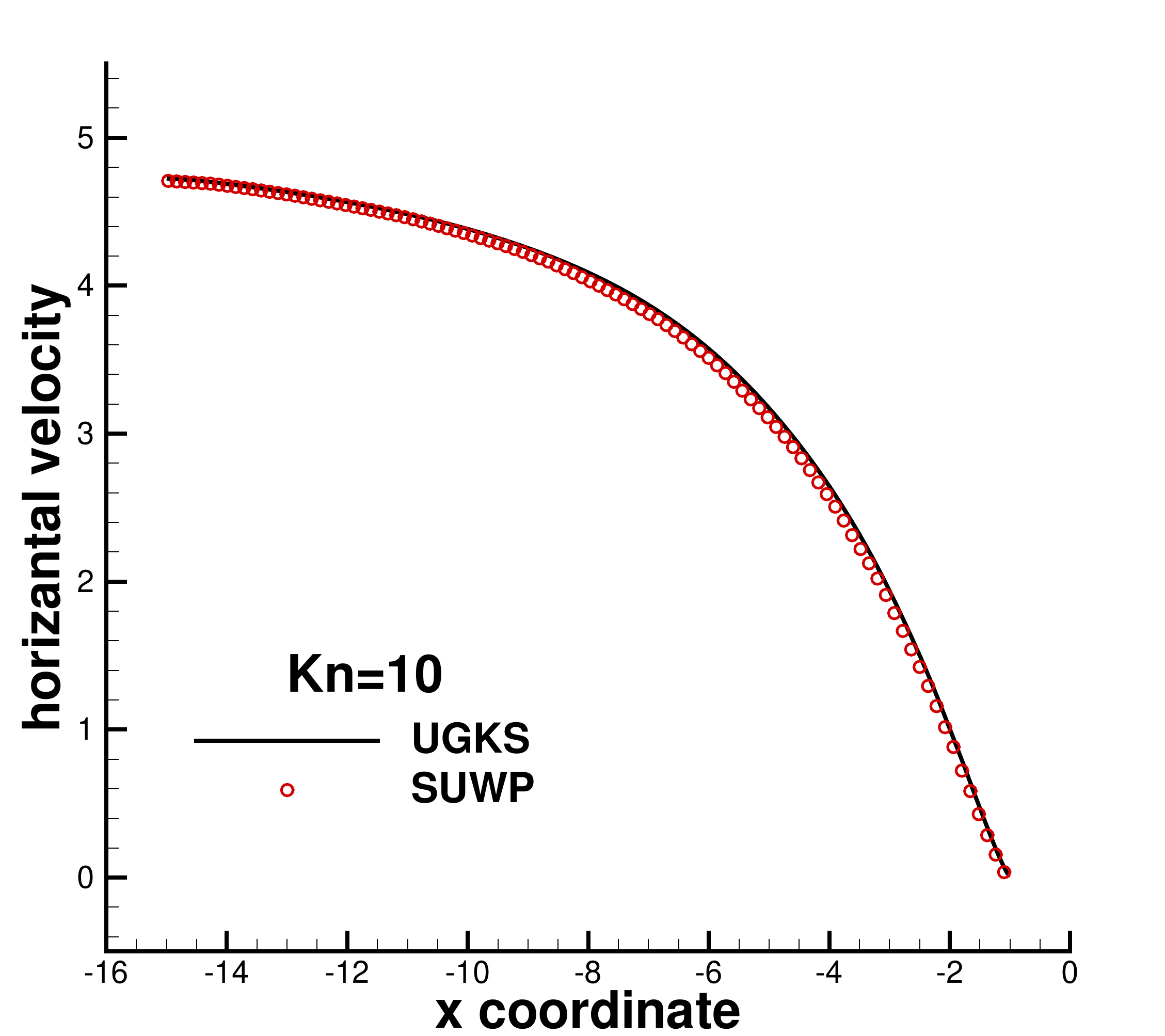}
}\hspace{0.05\textwidth}%
\subfigure[\label{Fig:cchu1} Kn=1]{
\includegraphics[width=0.45\textwidth]{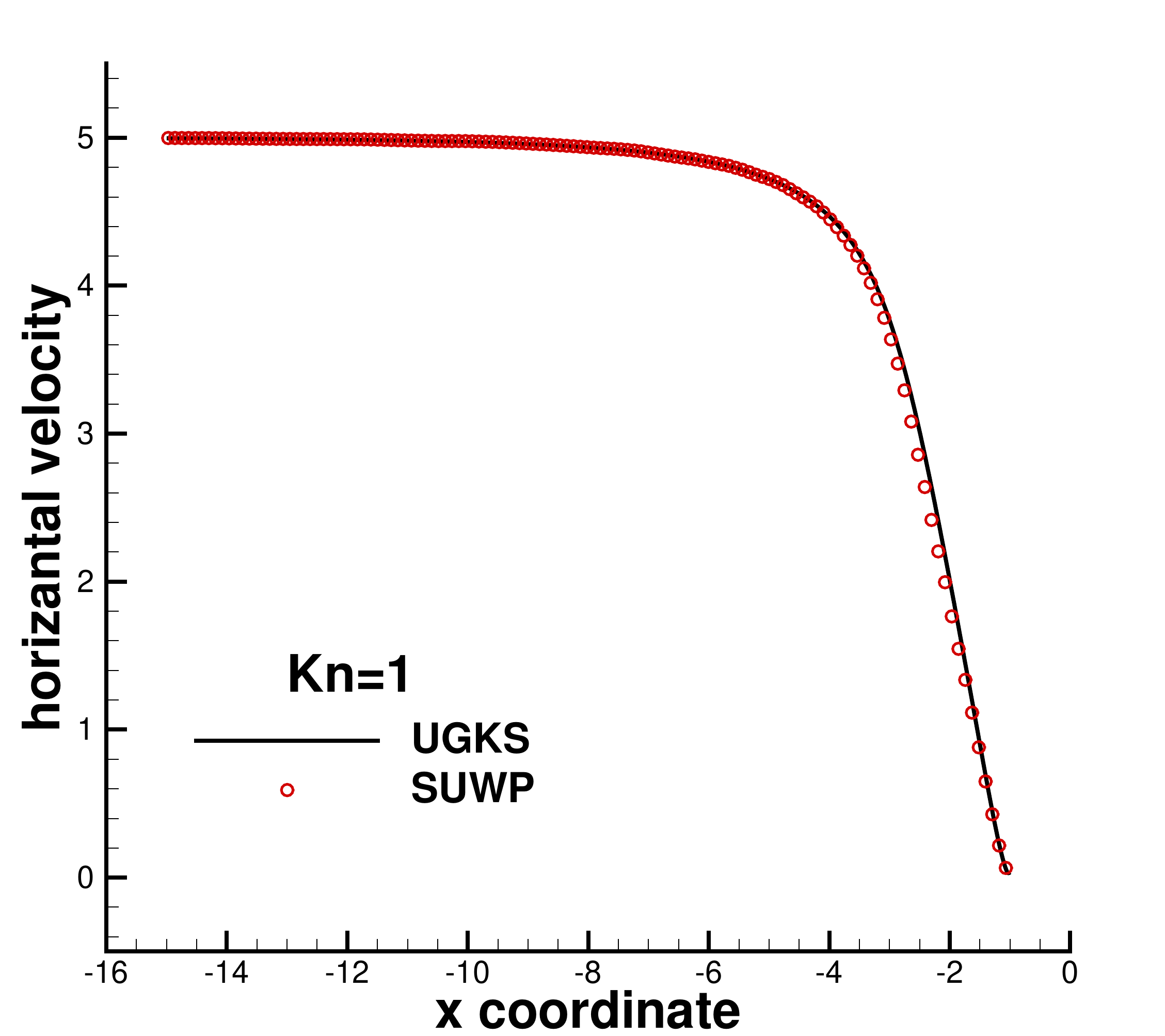}
}\\
\subfigure[\label{Fig:cchu01} Kn=0.1]{
\includegraphics[width=0.45\textwidth]{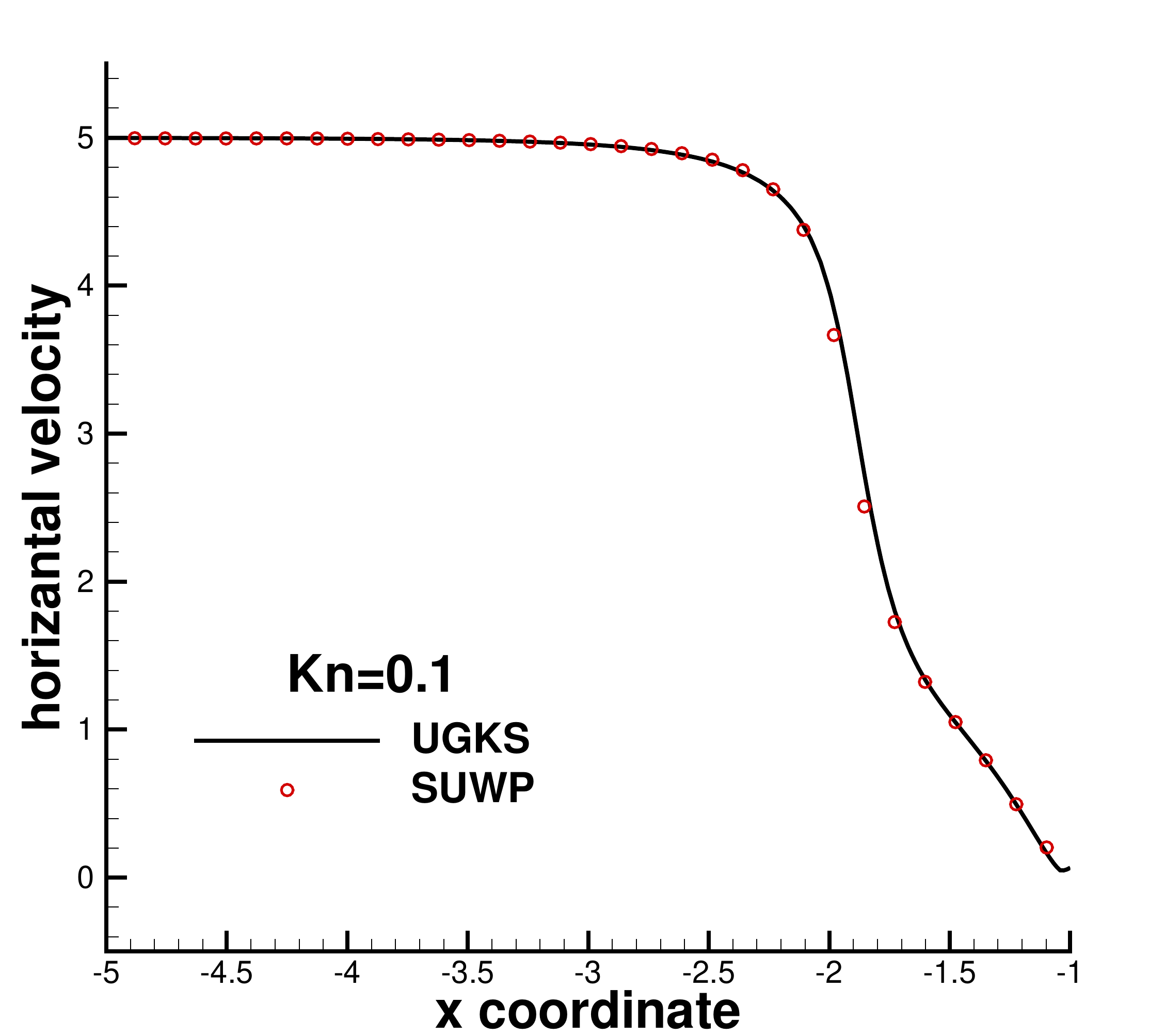}
}\hspace{0.05\textwidth}%
\subfigure[\label{Fig:cchu001} Kn=0.01]{
\includegraphics[width=0.45\textwidth]{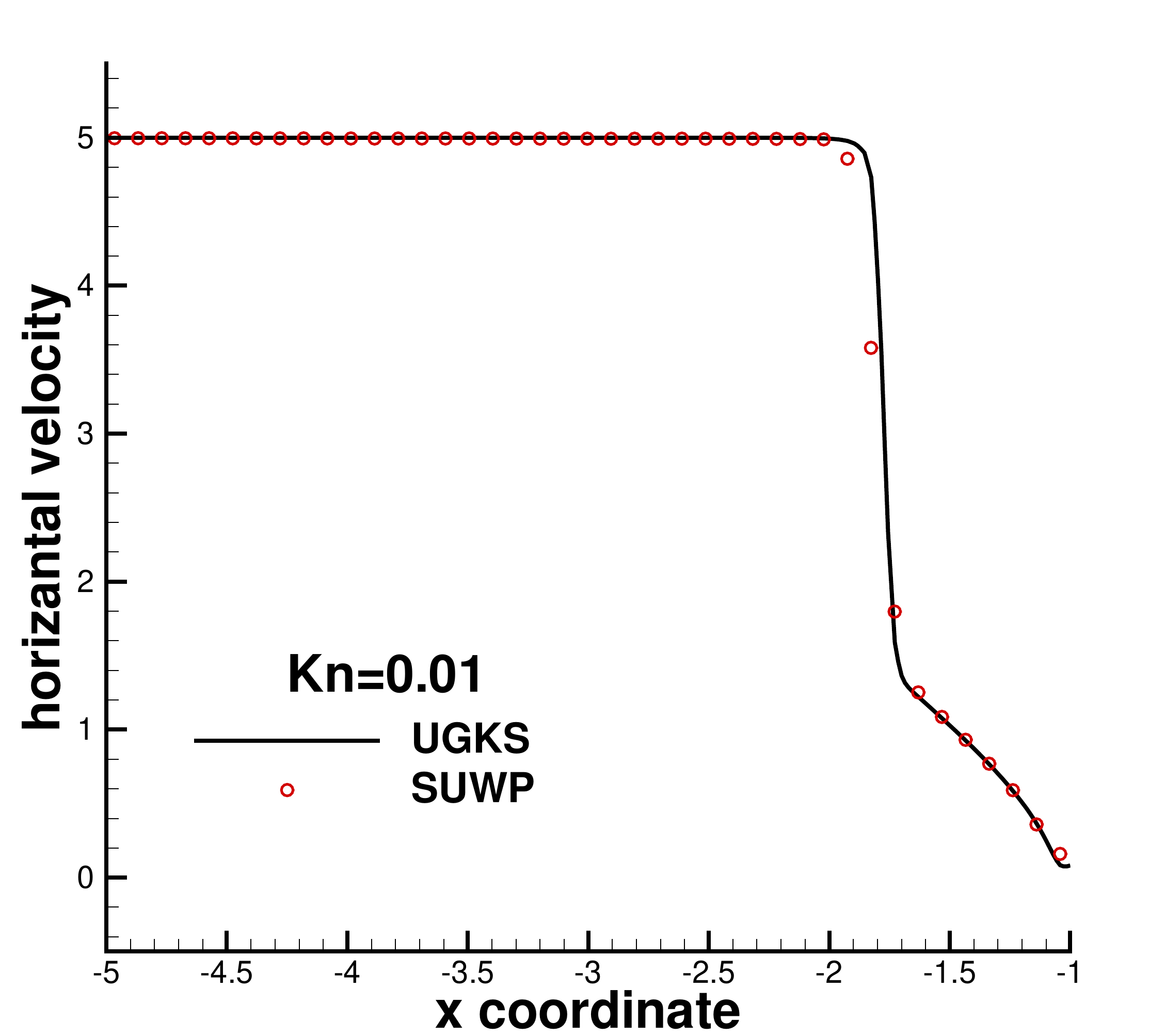}
}
\caption{\label{Fig:cchu} The u-velocity distribution along the stagnation lines of Ma$=$5 cylinder at Kn$=$ 10, 1, 0.1 and 0.01}
\end{figure}

\begin{figure}
\centering
\subfigure[\label{Fig:ccht10} Kn=10]{
\includegraphics[width=0.45\textwidth]{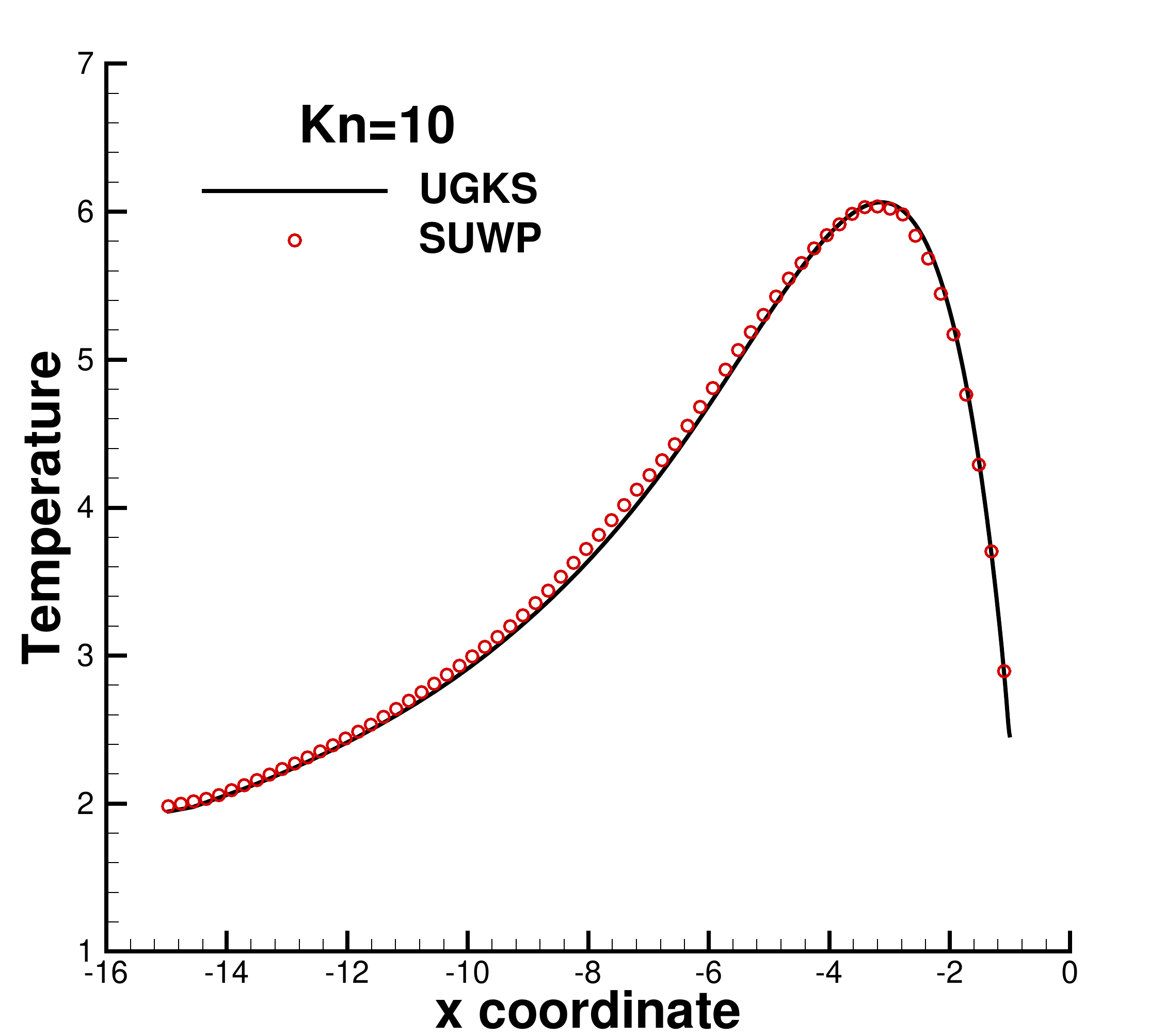}
}\hspace{0.05\textwidth}%
\subfigure[\label{Fig:ccht1} Kn=1]{
\includegraphics[width=0.45\textwidth]{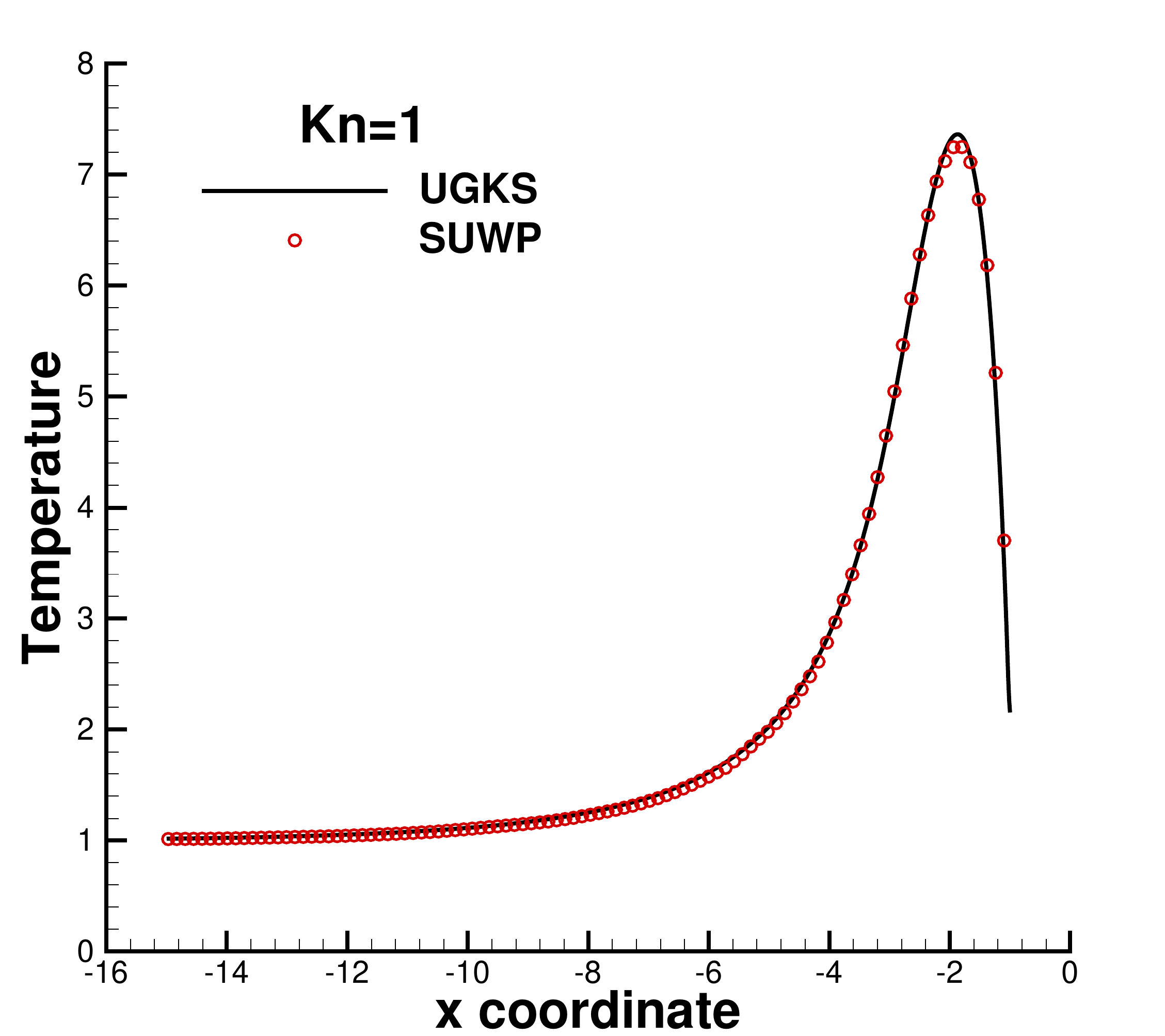}
}\\
\subfigure[\label{Fig:ccht01} Kn=0.1]{
\includegraphics[width=0.45\textwidth]{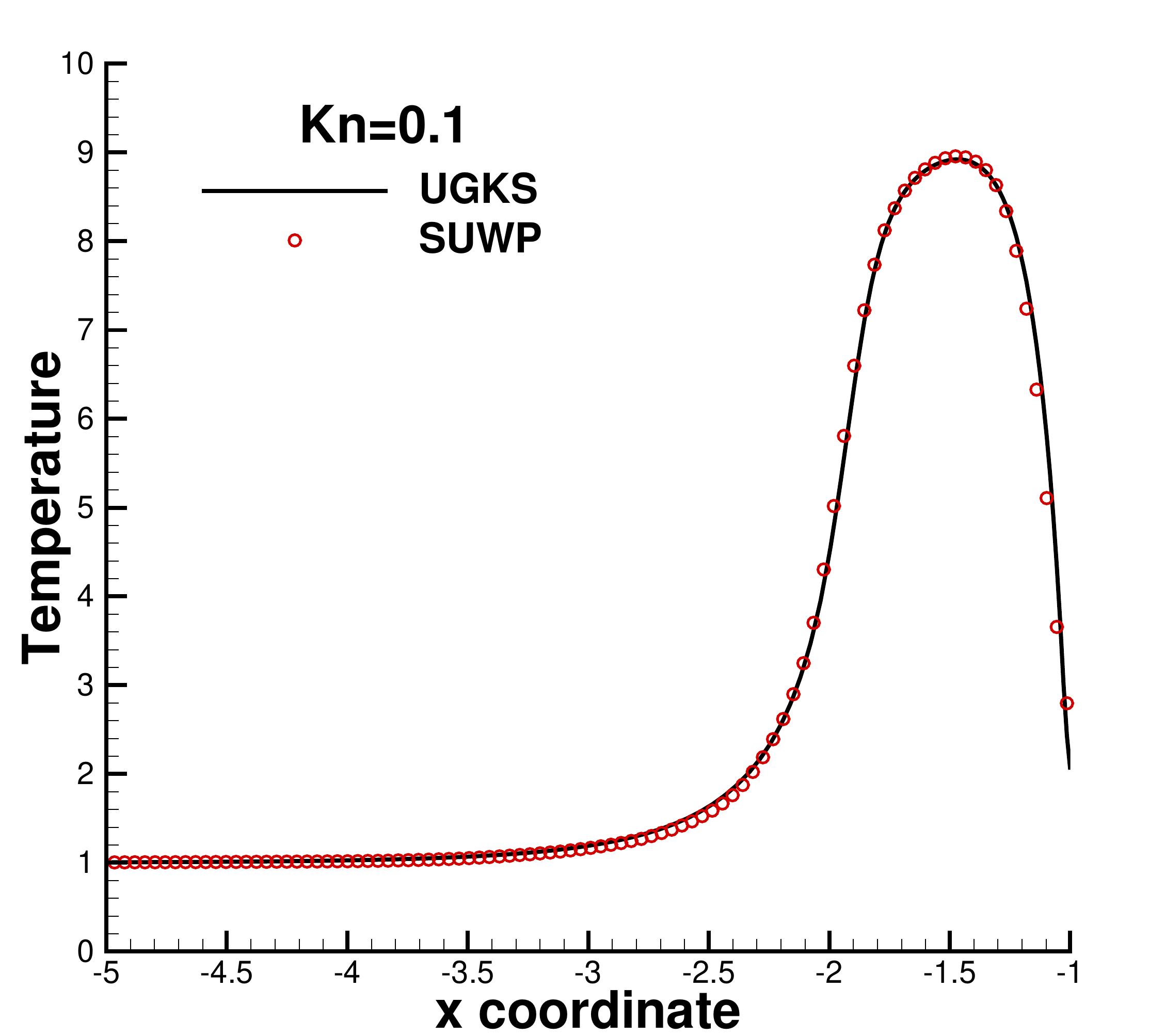}
}\hspace{0.05\textwidth}%
\subfigure[\label{Fig:ccht001} Kn=0.01]{
\includegraphics[width=0.45\textwidth]{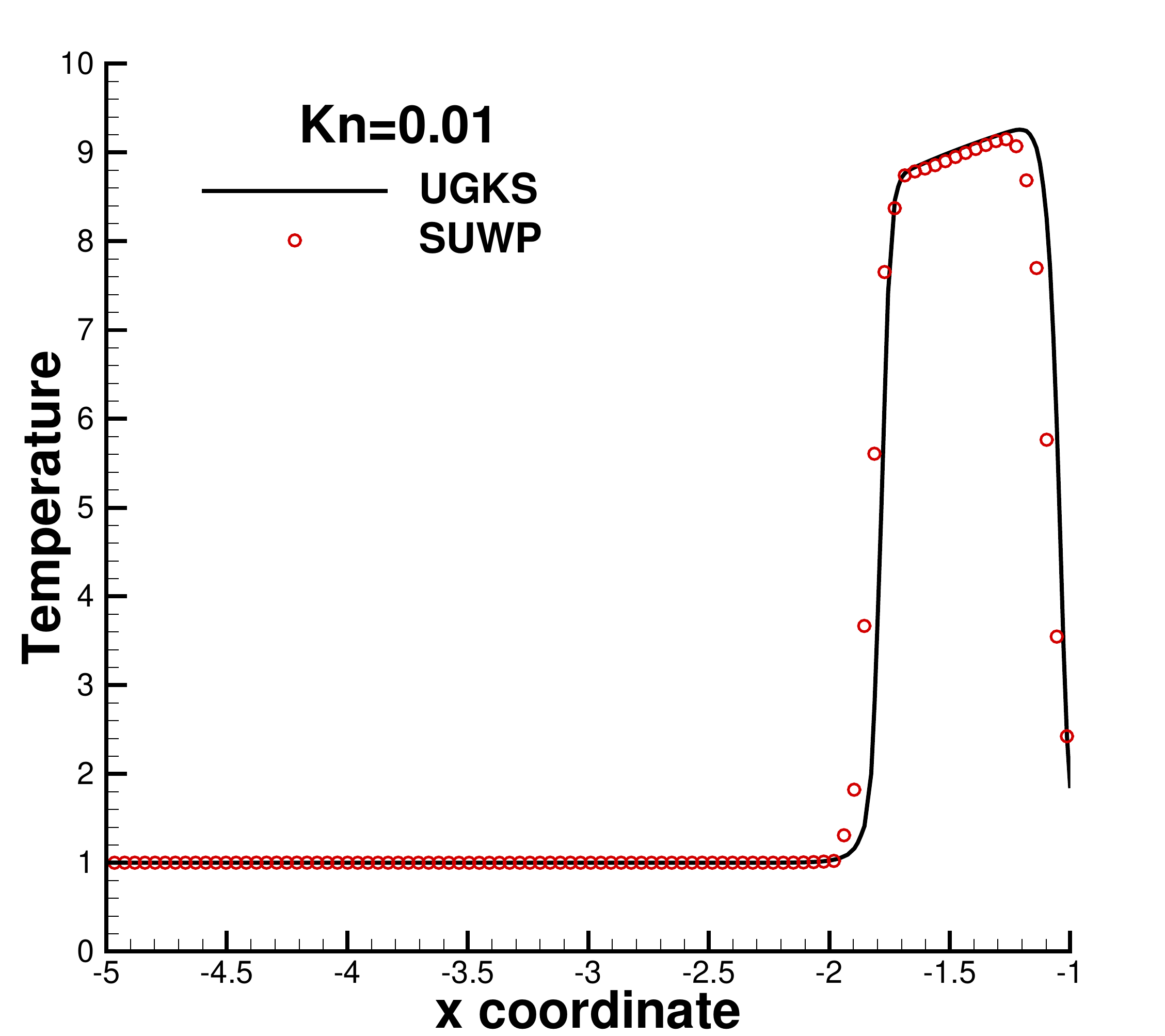}
}
\caption{\label{Fig:ccht} The temperature distribution along the stagnation lines of Ma$=$5 cylinder at Kn$=$ 10, 1, 0.1 and 0.01}
\end{figure}

The density, velocity and temperature profiles along the stagnation line are plotted in Fig.~\ref{Fig:cchr}, Fig.~\ref{Fig:cchu} and Fig.~\ref{Fig:ccht}, respectively. The results calculated by the present method are compared to those from the DUGKS code in Ref.~\cite{Chen2019Conserved}. The SUWP results match well with those from the DUGKS. At Kn=0.01, the profiles calculated by SUWP deviate slightly from the DUGKS results. That may because that the transient statistical fluctuation amplify the effect of the slop limiter and more numerical viscosity is added into the scheme. As the Kn number decreases, the profiles of the physical variables becomes thin, and same sharp structures appears in the profiles at Kn=0.01. This phenomenon is consistent with the observation of the flow field.

\begin{figure}
\centering
\includegraphics[width=0.45\textwidth]{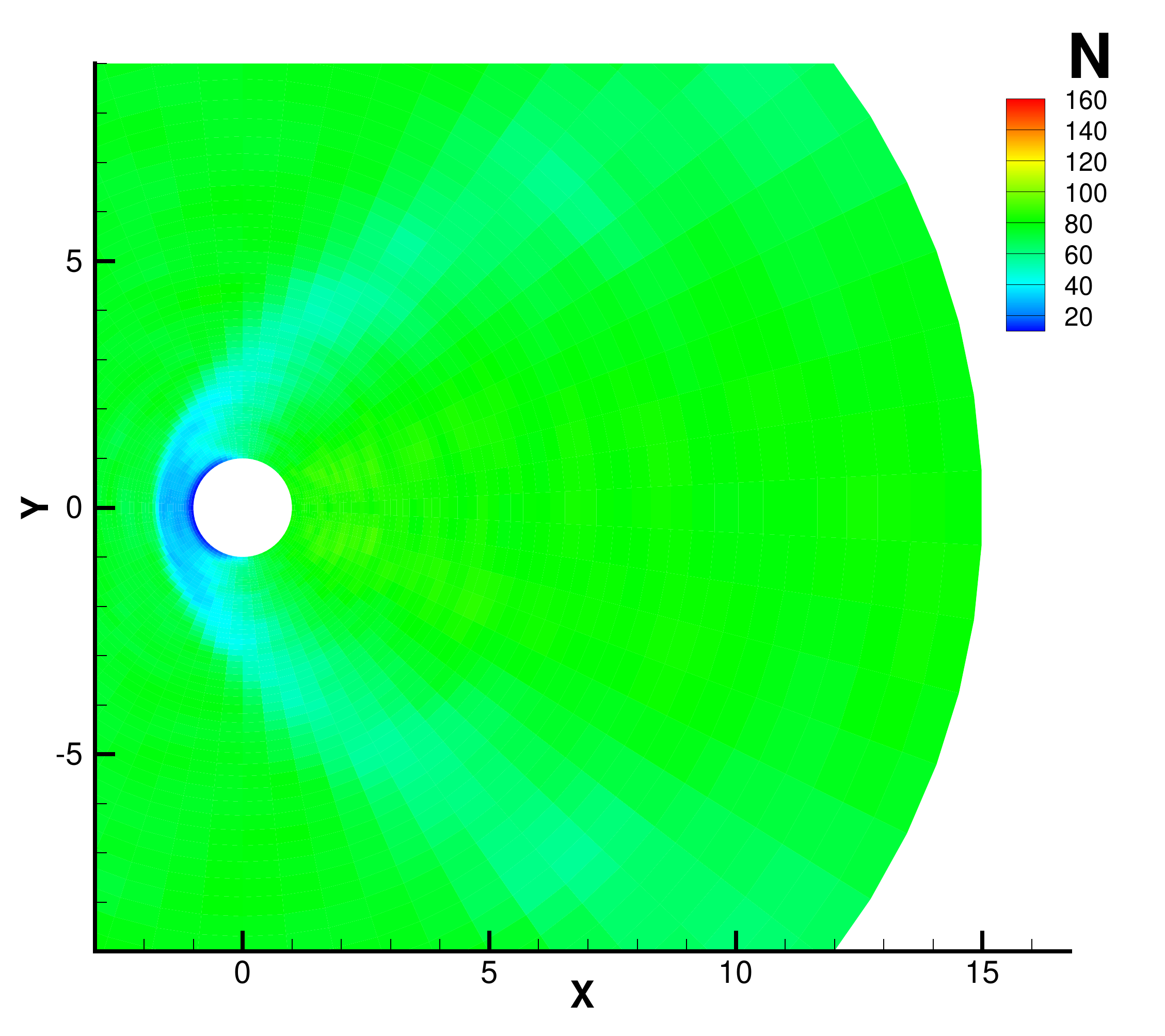}
\caption{\label{Fig:numberofmolecule} The model molecule number in cell for Ma$=5$ cylinder flow at Kn$=0.01$}
\end{figure}

Being the same with UGKWP, as cell Kn number decreases, the model molecules in cell will decrease. It is good property for computational efficiency. In Fig.~\ref{Fig:numberofmolecule}, the number of model molecules in cell are illustrated at the transient time after the flow achieving the steady state. It can been seen that the number of model molecules in the cells close to the front of cylinder are only within the range from 20 to 40.

\subsection{Viscous boundary layer}
\begin{figure}
\centering
\subfigure[\label{Fig:blafloodm} Mesh]{
\includegraphics[width=0.45\textwidth]{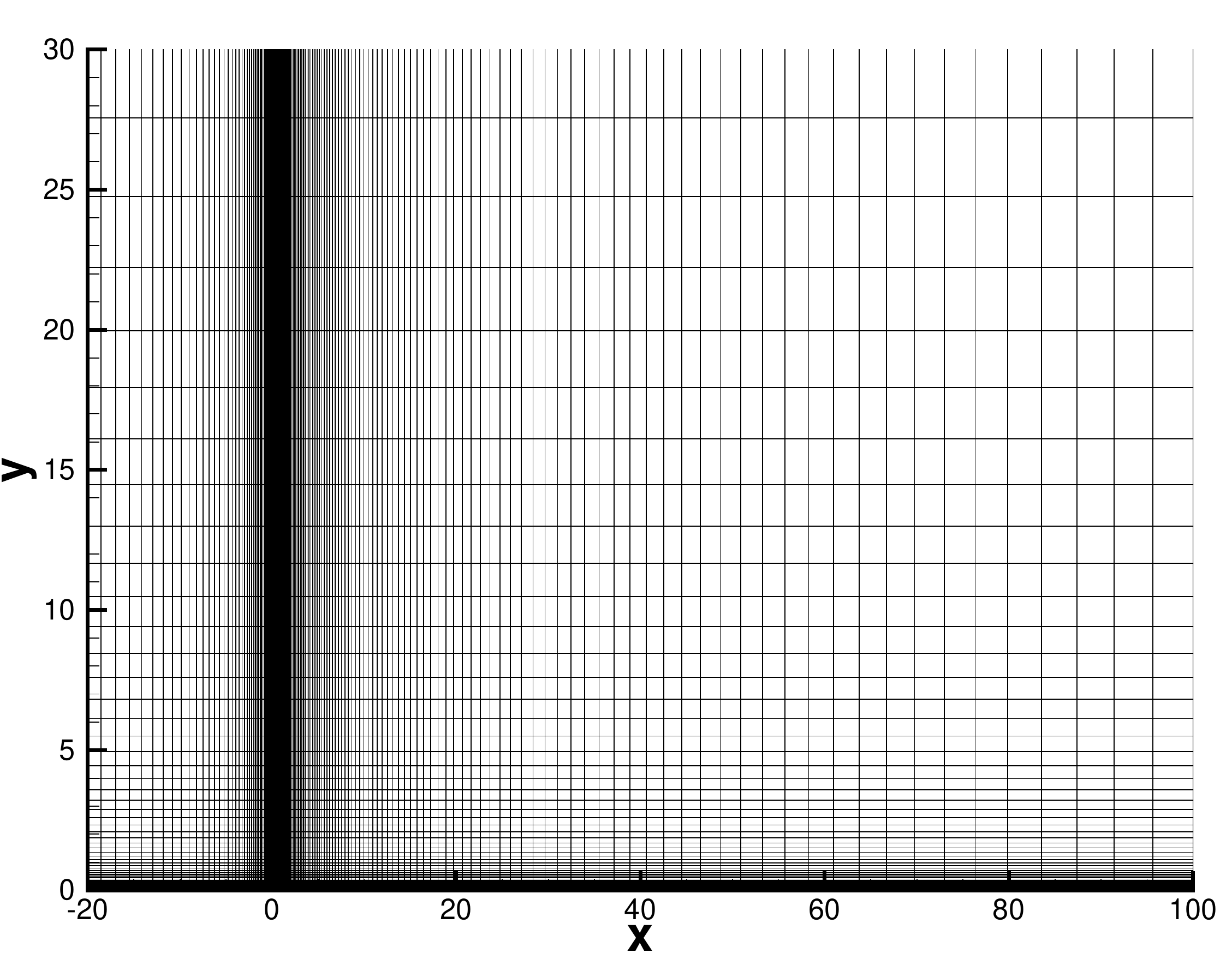}
}\hspace{0.05\textwidth}%
\subfigure[\label{Fig:blafloodr} Density]{
\includegraphics[width=0.45\textwidth]{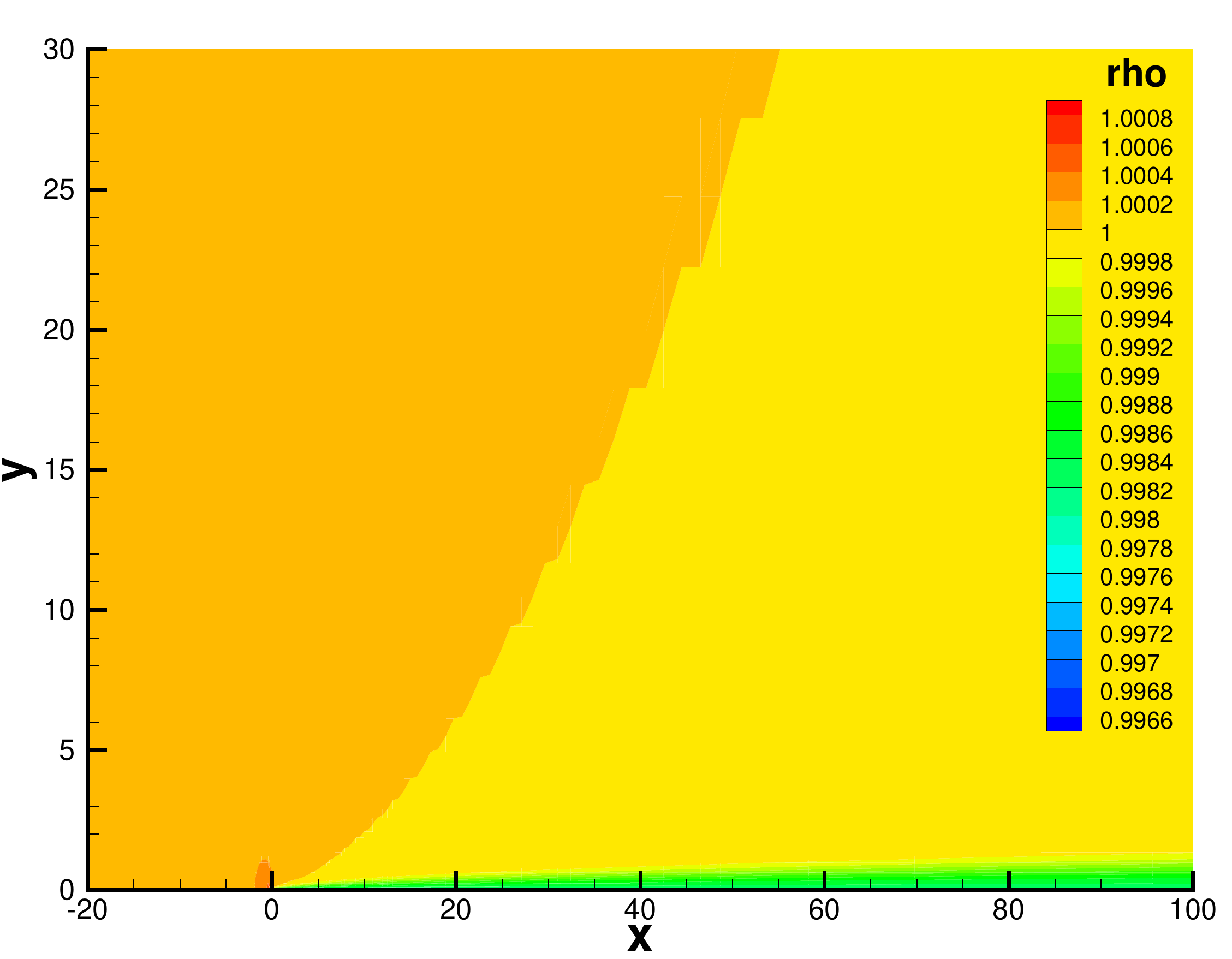}
}\\
\subfigure[\label{Fig:blafloodu} U-velocity]{
\includegraphics[width=0.45\textwidth]{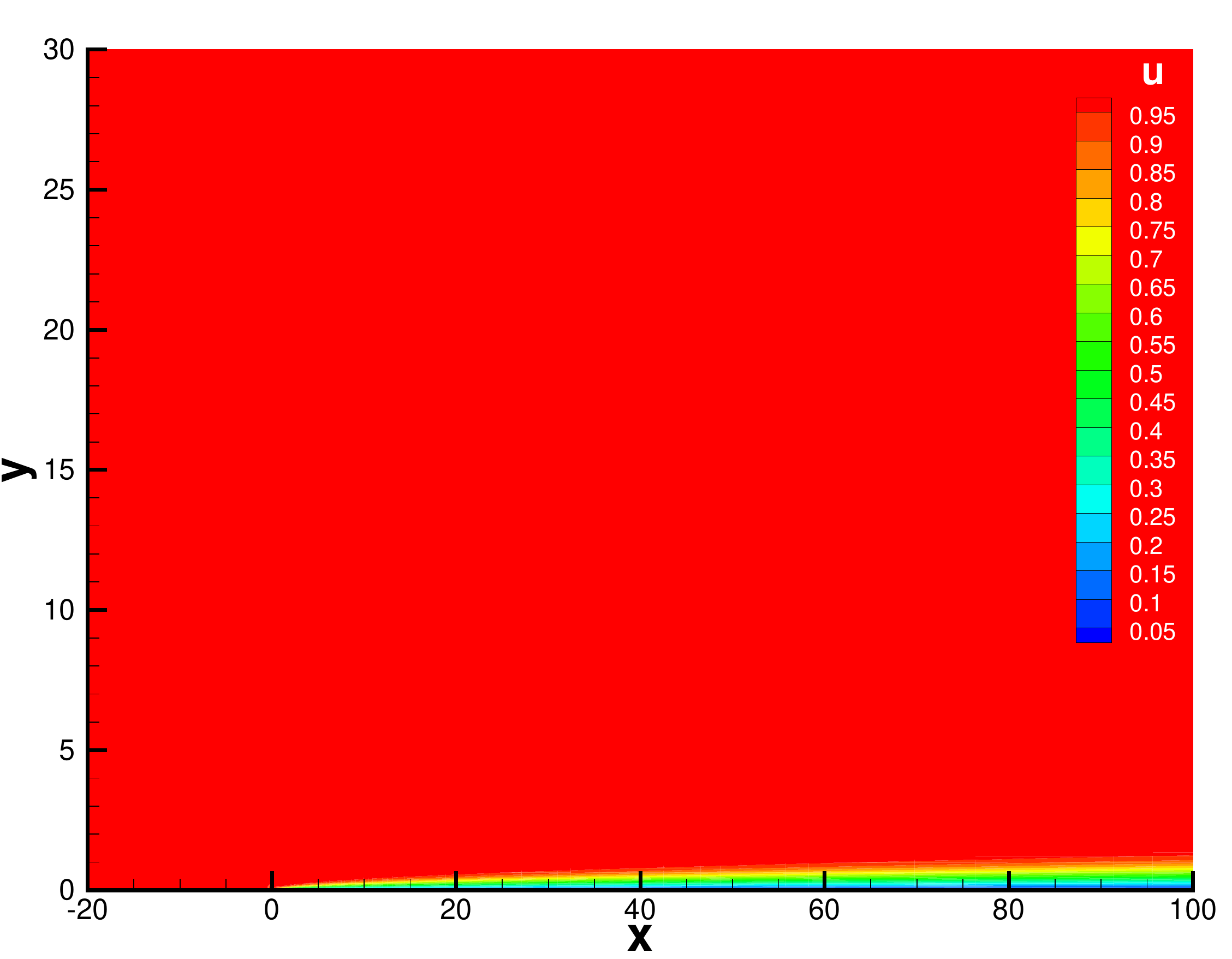}
}\hspace{0.05\textwidth}%
\subfigure[\label{Fig:blafloodv} V-velocity]{
\includegraphics[width=0.45\textwidth]{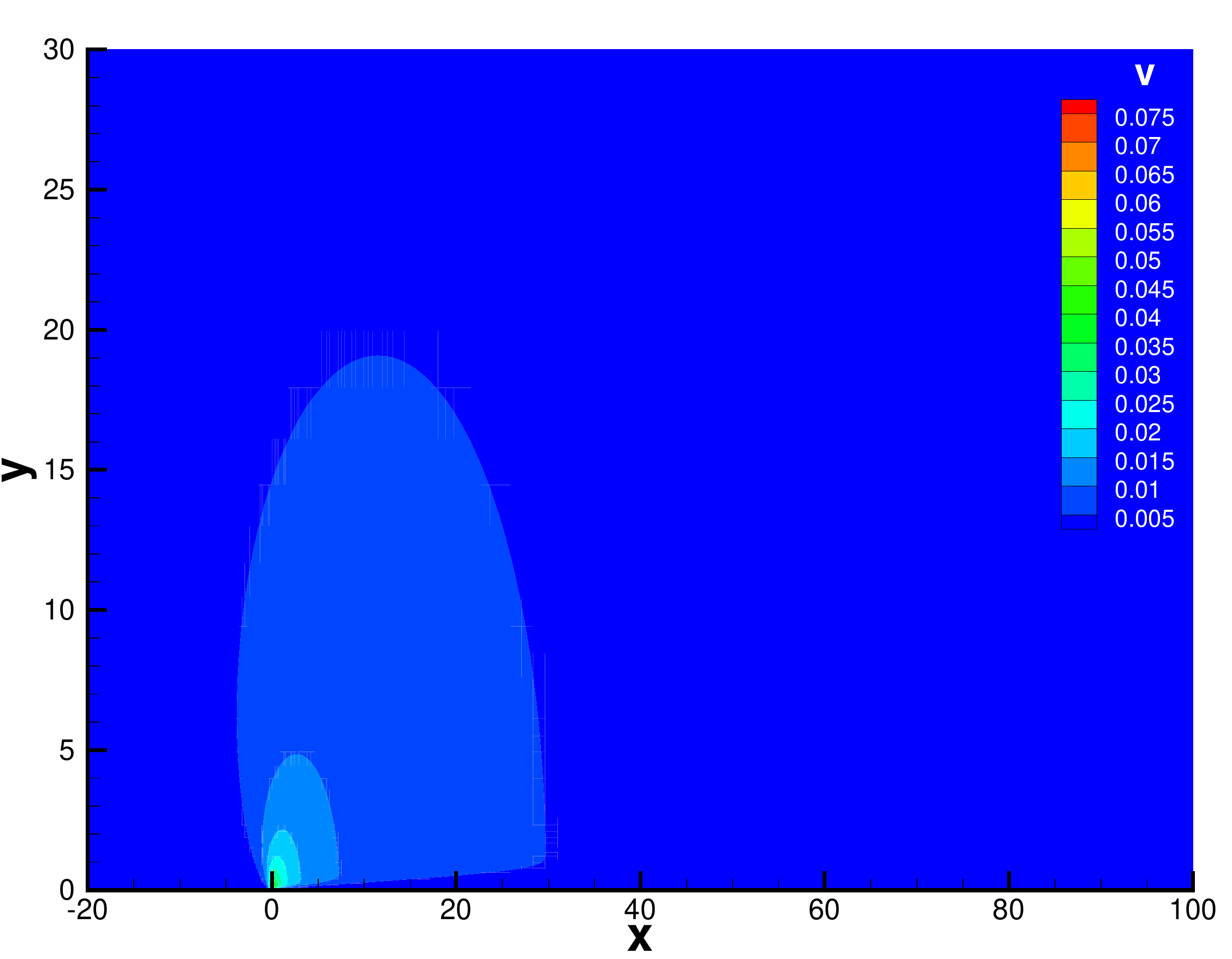}
}
\caption{\label{Fig:blaflood} The mesh and flow field of the viscous boundary layer at Ma$=0.1$ and Re$=10^{5}$}
\end{figure}

\begin{figure}
\centering
\subfigure[\label{Fig:blau} Normalized u-velocity]{
\includegraphics[width=0.45\textwidth]{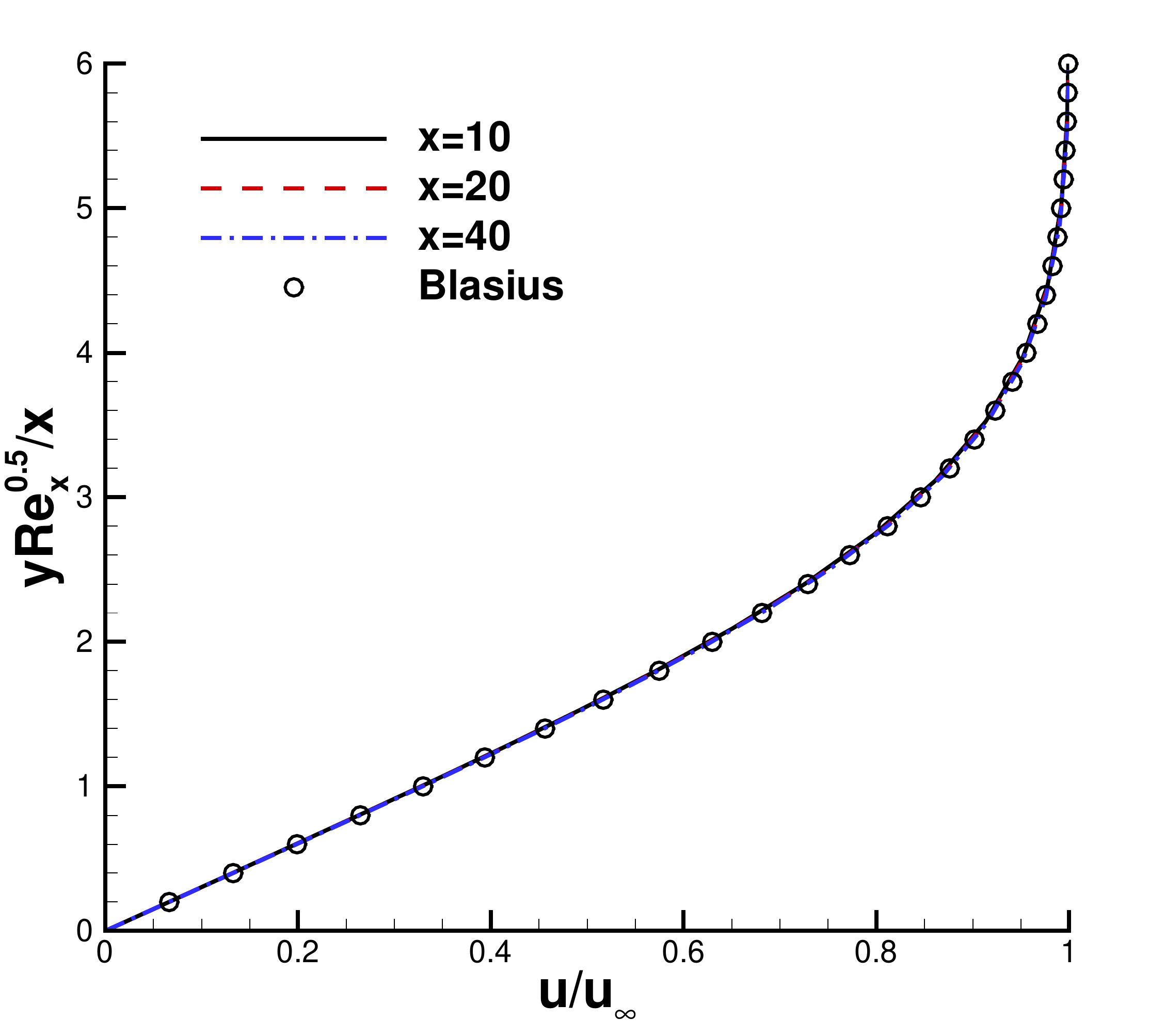}
}\hspace{0.05\textwidth}%
\subfigure[\label{Fig:blav} Normalized v-velocity]{
\includegraphics[width=0.45\textwidth]{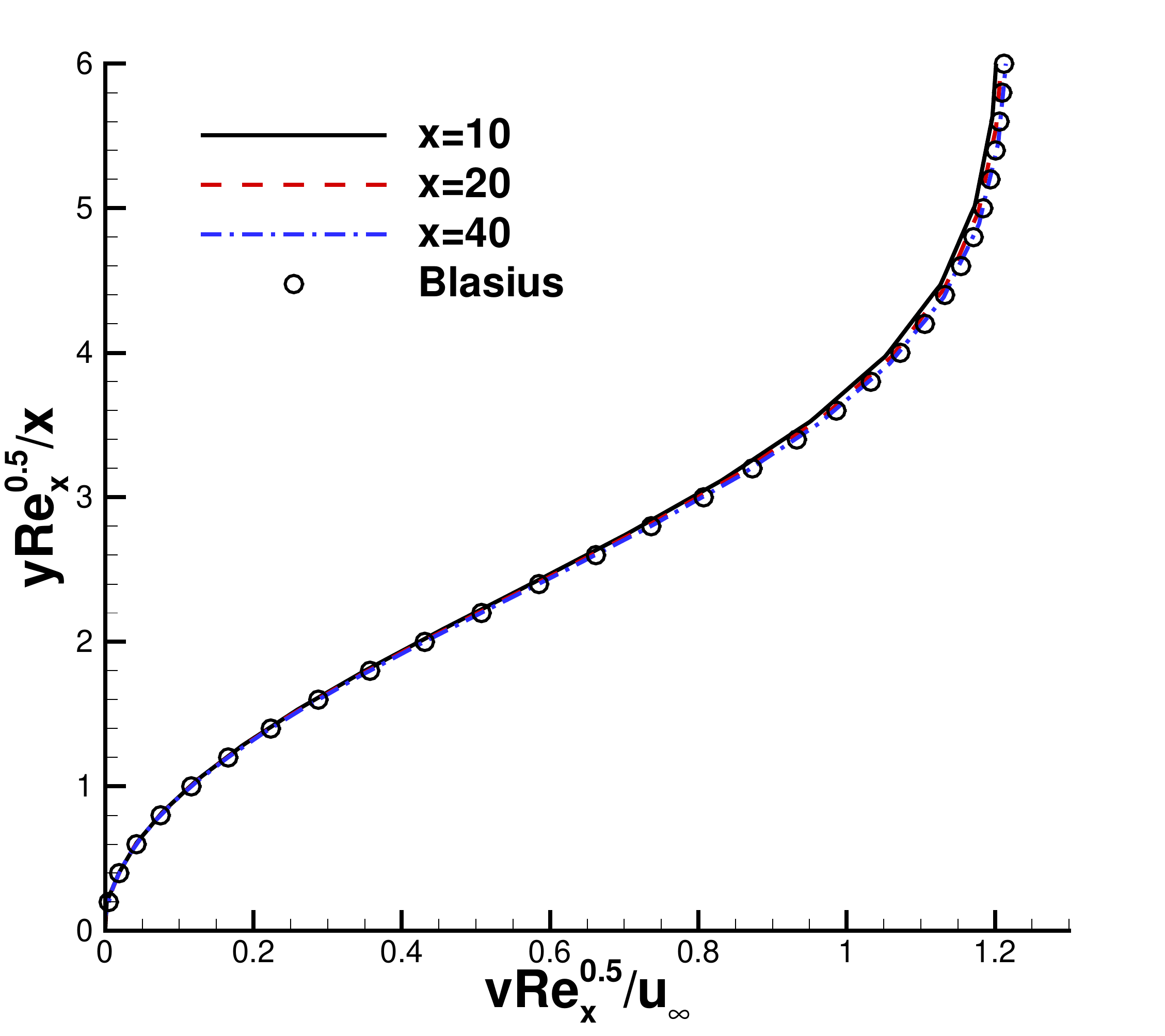}
}
\caption{\label{Fig:bla} The velocity profiles in the viscous boundary layer at Ma$=0.1$ and Re$=10^{5}$}
\end{figure}

Since the SUWP is designed that it is can be reduced to the N-S solver with a correct dissipation, the continuum flow passing a flat plate is simulated in the section. The computational domain is $\left[-50,100\right]\times\left[0,100\right]$, and rectangular mesh with $150\times65$ cells are utilized. The height of the cell adjacent to the plate is $0.02$, and its length is $0.1$. The inflow Ma number and Re number are $0.1$ and $10^{5}$, respectively. According to the inflow condition, $\Delta t/tau \gg 1$. Therefore, $w_{free}$ approaches zero and $w_{hydro}$ approaches unity. The smallest value of the total mass for the model molecules sampled from the hydrodynamic macroscopic variables is set as $10^{-4}$ of the total mass in cell. Below this value, no molecule need to be sampled (flow is in continuum regime, it does not need the free-transport mechanism). Therefore, at the initial time, there is no model molecule in the flow field. Since the density field is almost a constant in this case, there is also no model molecule needed to be sampled in the evolution process. In this case, the SUWP is reduced to a classic N-S solver. The mesh, density contours, u-velocity contours and v-velocity are illustrated in Fig.~\ref{Fig:blaflood}. As illustrated in Fig.~\ref{Fig:bla}, the u-velocity and v-velocity profiles matches well with that from the Blasius solution.

\section{Conclusions}\label{sec:DIS}
In this paper, a Quantified Model-Competition (QMC) mechanism is extracted from the integral solution of the Boltzmann-BGK model equation, and a novel Simplified Unified Wave-Particle method (SUWP) is proposed with the aid of this QMC mechanism. The validity and accuracy of the present SUWP method are verified through a series of multi-scale test cases. The SUWP combines the stochastic particle method and the continuum N-S method in the algorithm level. Both stochastic particle and N-S calculations are conducted in a single discrete cell, while their weights are quantified by the QMC mechanism. At the free-molecular limit, the SUWP is reduced to the stochastic particle method. Because the number of free molecules in a single cell is very small in the near continuum regime, the SUWP needs only a few amount of model molecules in such situations. At the continuum limit, the SUWP is reduced to the pure N-S solver completely. Since the SUWP is not strictly based on the BGK-type model equation, it is flexible and can be extended to the gas mixture and chemical reaction easily in the future research.

\section*{Acknowledgements}
The authors thank Prof. Kun Xu in Hong Kong University of Science and Technology for discussions of the UGKS method, the UGKWP method and the direct modeling of multi-scale flows. Sha Liu thanks Prof. Jun Zhang in Beihang University and Dr. Fei Fei in Huazhong University of Science and Technology for discussion of the Particle FP methods and the USP method. Sha Liu thanks Dr. Chang Liu and Dr. Yajun Zhu in Hong Kong University of Science and Technology for useful suggestions in constructing the present SUWP method. The present work is supported by National Natural Science Foundation of China (Grant No. 11702223, No. 11902266 and No. 11902264), National Numerical Wind-Tunnel Project of China (Grant No. NNW2019ZT3-A09) and 111 Project of China (Grant No. B17037).

\section*{Reference}
\bibliographystyle{elsarticle-num}
\bibliography{SUWP}
\clearpage

\clearpage

\clearpage

\end{document}